\documentclass[a4paper,11pt]{article}
\usepackage{jheppub}
\usepackage[T1]{fontenc}
\usepackage[utf8]{inputenc}
\usepackage{slashed}
\usepackage{subcaption}

\usepackage{todonotes}
\usepackage{hyperref}
\usepackage{bm}

\usepackage{tocloft}

\newcommand{\nocontentsline}[3]{}
\newcommand{\tocless}[2]{\bgroup\let\addcontentsline=\nocontentsline#1{#2}\egroup}

\newcommand{\pot}{(p)}
\def\relf{\sigma}
\newcommand{\vect}{\bm}

\usepackage{tikz,environ}

\usetikzlibrary{snakes}
\usetikzlibrary{decorations}
\usetikzlibrary{trees}
\usetikzlibrary{decorations.pathmorphing}
\usetikzlibrary{decorations.markings}
\usetikzlibrary{external}
\usetikzlibrary{intersections}
\usetikzlibrary{shapes,arrows}
\usetikzlibrary{arrows.meta}
\usetikzlibrary{calc}
\usetikzlibrary{shapes.misc}
\usetikzlibrary{decorations.text}
\usetikzlibrary{backgrounds}
\usetikzlibrary{fadings}
\usetikzlibrary{tikzmark,calc,arrows,shapes,decorations.pathreplacing}

\usetikzlibrary{shapes.geometric,positioning}

\tikzset{
	graviton/.style={decorate,line width=0.1mm, decoration={snake,amplitude=.3mm, segment length=0.8mm}},
	photon/.style={decorate, decoration={snake,amplitude=.4mm, segment length=2mm}, draw=red},
	scalar/.style={postaction={decorate},
	},
	massive/.style={postaction={decorate},
		line width=0.75mm,
	},
	massless/.style={postaction={decorate},
	},
	masslessWithDot/.style={postaction={decorate},
		decoration={
			markings,
			mark=at position 0.5 with {\fill circle (2pt);}}
	},
	massiveWithDot/.style={postaction={decorate},
		line width=0.5mm,
		decoration={
			markings,
			mark=at position 0.5 with {\fill circle (2pt);}}
	},
	massiveWithArrow/.style={postaction={decorate},
		line width=0.75mm,
		decoration={
			markings,
			mark=at position 0.5 with {\arrow{latex}}}
	},
	massiveLin/.style={postaction={decorate},
		double,
		thick,
		fill=white
	},
	massivePhi/.style={postaction={decorate},
		line width=0.75mm,
		dashed
	},
	masslessPhi/.style={postaction={decorate},
		dashed
	},
	unitaryCut/.style={postaction={draw,densely dashed,blue,thin},
		line width = 0.2cm,white
	},
	gluon/.style={decorate, draw=magenta,
		decoration={coil,amplitude=4pt, segment length=5pt}},
	partial ellipse/.style args={#1:#2:#3}{
		insert path={+ (#1:#3) arc (#1:#2:#3)}
	},
	cross/.style={cross out, draw=black, minimum size=2*(#1-\pgflinewidth), inner sep=0pt, outer sep=0pt},
	branchCut/.style={postaction={decorate},
		snake=zigzag,
		decoration = {snake=zigzag,segment length = 2mm, amplitude = 2mm}	
	}
	cross/.default={1pt}
}

\colorlet{mred}{black!30!red}
\colorlet{mgreen}{black!30!green}
\colorlet{mblue}{black!30!blue}
\colorlet{morange}{blue!70!red}

\newcommand{\threePtPhoton}{
\begin{tikzpicture}[scale=1]
		\coordinate (e1) at (-0.5,1.);
		\coordinate (e2) at (.5,1.);
		\coordinate (v1) at (0,.8);
		\coordinate (e3) at (0,0);	
		
		\draw[massive] (e1) -- (v1) -- (e2) ;
		\draw[photon] (v1) -- (e3) ;

       		 \node[above left] at (e1) {$p_a$};
        		\node[above right] at (e2) {$p_b$};
\end{tikzpicture}}

\newcommand{\comptonAmp}{
\begin{tikzpicture}[scale=1]
		\coordinate (e1) at (-0.7,1.);
		\coordinate (e4) at (.7,1.);
		\coordinate (v1) at (0,.8);
		\coordinate (e2) at (-0.5,0);	
		\coordinate (e3) at (0.5,0);
		
		\draw[massive] (e1) -- (v1) -- (e4) ;
		\draw[photon] (v1) -- (e2) ;
		\draw[photon] (v1) -- (e3) ;
		
		\draw[fill=lightgray, opacity=1] (v1) circle (0.25);

       		 \node[above left] at (e1) {$p_1$};
		 \node[below left] at (e2) {$p_2,\varepsilon_2$};
		  \node[below right] at (e3) {$p_3,\varepsilon_3$};
        		 \node[above right] at (e4) {$p_4$};
\end{tikzpicture}}

\newcommand{\threePhotonAmp}{
\begin{tikzpicture}[scale=1]
		\coordinate (e1) at (-0.7,1.);
		\coordinate (e5) at (.7,1.);
		\coordinate (v1) at (0,.8);
		\coordinate (e2) at (-0.6,0);	
		\coordinate (e3) at (0,-.2);
		\coordinate (e4) at (0.6,0);
		
		\draw[massive] (e1) -- (v1) -- (e5) ;
		\draw[photon] (v1) -- (e2) ;
		\draw[photon] (v1) -- (e3) ;
		\draw[photon] (v1) -- (e4) ;
		
		\draw[fill=lightgray, opacity=1] (v1) circle (0.25);

       		 \node[above left] at (e1) {$p_1$};
		 \node[below left] at (e2) {$p_2,\varepsilon_2$};
		 \node[below] at (e3) {$p_3,\varepsilon_3$};
		  \node[below right] at (e4) {$p_4,\varepsilon_4$};
        		 \node[above right] at (e5) {$p_5$};
\end{tikzpicture}}

\newcommand{\treetPhoton}{
\begin{tikzpicture}[scale=1]
		\coordinate (e1) at (-0.5,0.);
		\coordinate (e2) at (.5,0.);
		\coordinate (e3) at (.5,-1.);
		\coordinate (e4) at (-0.5,-1.);	
		
		\coordinate (v1) at (0,0);
		\coordinate (v2) at (0,-1.);
		
		\draw[massive] (e1) -- (v1) -- (e2) ;
		\draw[massive] (e3) -- (v2) -- (e4) ;
		
		\draw[photon] (v1) -- (v2) ;

        		\node[below left]  at (e4) {$p_1$};
        		\node[above left]  at (e1) {$p_2$};
        		\node[above right] at (e2) {$p_3$};
        		\node[below right] at (e3) {$p_4$};
\end{tikzpicture}}

\newcommand{\bubbleCut}{
\begin{tikzpicture}[scale=1]
		\coordinate (e1) at (-0.5,0.);
		\coordinate (e2) at (.5,0.);
		\coordinate (e3) at (.5,-1.);
		\coordinate (e4) at (-0.5,-1.);	
		
		\coordinate (v1) at (0,0);
		\coordinate (v2) at (0,-1.);
		
		\draw[massive] (e1) -- (v1) -- (e2) ;
		\draw[massive] (e3) -- (v2) -- (e4) ;
		
		\draw[photon] (v1)  to[out=-25,in=25] (v2);
		\draw[photon] (v1)  to[out=180+25,in=180-25] (v2);
		
		\draw[fill=lightgray, opacity=1] (v1) circle (0.2);
		\draw[fill=lightgray, opacity=1] (v2) circle (0.2);

        		\node[below left]  at (e4) {$p_1$};
        		\node[above left]  at (e1) {$p_2$};
        		\node[above right] at (e2) {$p_3$};
        		\node[below right] at (e3) {$p_4$};
\end{tikzpicture}}

\newcommand{\bboxLabel}{
\begin{tikzpicture}[scale=1]
		\coordinate (e1) at (-0.5,0.);
		\coordinate (e2) at (1.5,0.);
		\coordinate (e3) at (1.5,-1.);
		\coordinate (e4) at (-0.5,-1.);	
		
		\coordinate (v1) at (0,0);
		\coordinate (v2) at (1,0);
		\coordinate (v3) at (0,-1);
		\coordinate (v4) at (1,-1);
		
		\coordinate (v5) at (.5,-1);
		\coordinate (v6) at (.5,0);
		\coordinate (v7) at (0,-.5);
		\coordinate (v8) at (1,-.5);
		
		\draw[massive] (e1) -- (v1) ;
		\draw[massive] (v1) -- (v2) ;
		\draw[massive] (v2) -- (e2) ;
		\draw[massive] (e4) -- (v3) ;
		\draw[massive] (v3) -- (v4) ;
		\draw[massive] (v4) -- (e3) ;
		
		\draw[massless] (v1) -- (v3) ;
		\draw[massless] (v2) -- (v4) ;

        \node[below left]  at (e4) {$p_1$};
        \node[above left]  at (e1) {$p_2$};
        \node[above right] at (e2) {$p_3$};
        \node[below right] at (e3) {$p_4$};
        
        \node[below]  at (v5) {$D_1$};
        \node[above]  at (v6) {$D_2$};
        \node[left] at (v7) {$D_3$};
        \node[right] at (v8) {$D_4$};
\end{tikzpicture}}

\newcommand{\bbox}{
\begin{tikzpicture}[scale=1]
		\coordinate (e1) at (-0.5,0.);
		\coordinate (e2) at (1.5,0.);
		\coordinate (e3) at (1.5,-1.);
		\coordinate (e4) at (-0.5,-1.);	
		
		\coordinate (v1) at (0,0);
		\coordinate (v2) at (1,0);
		\coordinate (v3) at (0,-1);
		\coordinate (v4) at (1,-1);
		
		\draw[massive] (e1) -- (v1) ;
		\draw[massive] (v1) -- (v2) ;
		\draw[massive] (v2) -- (e2) ;
		\draw[massive] (e4) -- (v3) ;
		\draw[massive] (v3) -- (v4) ;
		\draw[massive] (v4) -- (e3) ;
		
		\draw[massless] (v1) -- (v3) ;
		\draw[massless] (v2) -- (v4) ;

        \node[below left]  at (e4) {$p_1$};
        \node[above left]  at (e1) {$p_2$};
        \node[above right] at (e2) {$p_3$};
        \node[below right] at (e3) {$p_4$};
\end{tikzpicture}}

\newcommand{\bxbox}{
\begin{tikzpicture}[scale=1]
		\coordinate (e1) at (-0.5,0.);
		\coordinate (e2) at (1.5,0.);
		\coordinate (e3) at (1.5,-1.);
		\coordinate (e4) at (-0.5,-1.);	
		
		\coordinate (v1) at (0,0);
		\coordinate (v2) at (1,0);
		\coordinate (v3) at (0,-1);
		\coordinate (v4) at (1,-1);
		
		\draw[massless] (v2) -- (v3) ;
	    \draw[line width = 0.2cm,white ] (v1) -- (v4) ;
		\draw[massless] (v1) -- (v4) ;

		\draw[massive] (e1) -- (v1) ;
		\draw[massive] (v1) -- (v2) ;
		\draw[massive] (v2) -- (e2) ;
		\draw[massive] (e4) -- (v3) ;
		\draw[massive] (v3) -- (v4) ;
		\draw[massive] (v4) -- (e3) ;
		
		\node[below left]  at (e4) {$p_1$};
        \node[above left]  at (e1) {$p_2$};
        \node[above right] at (e2) {$p_3$};
        \node[below right] at (e3) {$p_4$};
\end{tikzpicture}}

\newcommand{\softkinA}{
\begin{tikzpicture}
        \coordinate (e1) at (-1,-1);	
        \coordinate (e2) at (-1, 1);
        \coordinate (e3) at ( 1, 1);
        \coordinate (e4) at ( 1,-1);
      
        \coordinate (v1) at (0,0);
      
        \draw[massiveWithArrow] (-.75,-.75) -- (-1,-1);
        \draw[massive] (e1) -- (v1);
        \draw[massiveWithArrow] (-.75,.75) -- (-1,1);
        \draw[massive] (e2) -- (v1);
        \draw[massiveWithArrow] (.75,-.75) -- (1,-1);
        \draw[massive] (e4) -- (v1);
        \draw[massiveWithArrow] (.75,.75) -- (1,1);
        \draw[massive] (e3) -- (v1);
      
        \draw[fill=lightgray, opacity=1] (0,0) circle (0.65);
      
        \node[below left]  at (e1) {$p_1$};
        \node[above left]  at (e2) {$p_2$};
        \node[above right] at (e3) {$p_3=-p_2-q$};
        \node[below right] at (e4) {$p_4=-p_1+q$};
\end{tikzpicture}}

\newcommand{\softkinB}{
\begin{tikzpicture}
        \coordinate (e1) at (-1,-1);	
        \coordinate (e2) at (-1, 1);
        \coordinate (e3) at ( 1, 1);
        \coordinate (e4) at ( 1,-1);
      
        \coordinate (v1) at (0,0);
      
        \draw[massiveWithArrow] (e1) -- (v1);
        \draw[massiveWithArrow] (e2) -- (v1);
        \draw[massiveWithArrow] (.75,-.75) -- (1,-1);
        \draw[massive] (e4) -- (v1);
        \draw[massiveWithArrow] (.75,.75) -- (1,1);
        \draw[massive] (e3) -- (v1);
      
        \draw[fill=lightgray, opacity=1] (0,0) circle (0.65);
      
        \node[below left]  at (e1) {$\overline{p}_1 - \frac{q}{2}$};
        \node[above left]  at (e2) {$\overline{p}_2 + \frac{q}{2}$};
        \node[above right] at (e3) {$\overline{p}_2 - \frac{q}{2}$};
        \node[below right] at (e4) {$\overline{p}_1 + \frac{q}{2}$};
\end{tikzpicture}}

\newcommand{\kmocvirtual}{
\begin{tikzpicture}
        \coordinate (e1) at (-1,-1);	
        \coordinate (e2) at (-1, 1);
        \coordinate (e3) at ( 1, 1);
        \coordinate (e4) at ( 1,-1);
      
        \coordinate (v1) at (0,0);
      
        \draw[massive] (e1) -- (v1);
        \draw[massive] (e2) -- (v1);
        \draw[massive] (e4) -- (v1);
        \draw[massive] (e3) -- (v1);
      
        \draw[fill=lightgray, opacity=1] (0,0) circle (0.65);
        \node at (v1) {$\mathcal{A}$};
      
        \node[below left]  at (e1) {$p_1$};
        \node[above left]  at (e2) {$p_2$};
        \node[above right] at (e3) {$p_3$};
        \node[below right] at (e4) {$p_4$};
\end{tikzpicture}}

\newcommand{\kmocreal}[2]{
\begin{tikzpicture}
        \coordinate (e1) at (-2.25,-1);	
	    	\coordinate (e2) at (-2.25, 1);
	    	\coordinate (e3) at ( 2.25, 1);
	    	\coordinate (e4) at ( 2.25,-1);
    
	    	\coordinate (v1) at (-1.25, 0);
	    	\coordinate (v2) at ( 1.25, 0);
	    	
	    	\coordinate (m1) at (0, 1.00);
	    	\coordinate (m2) at (0,-1.00);
    
	    	\draw[massive] (e1) -- (v1);
	    	\draw[massive] (e2) -- (v1);
	    	\draw[massive] (e3) -- (v2);
	    	\draw[massive] (e4) -- (v2);
    
	    	\draw[massive, name path=rung1] (v1) to[out=90, in=90](v2);
	    	\draw[massive, name path=rung2] (v1) to[out=-90, in=-90](v2);
	    	\draw[photon, name path=rung3, line width=0.25mm] (v1) to (v2);
	    	\draw[line width=0.25mm, dashed, name path=rung4, white] (0, 1) to (0, -1);
	    	
            	\draw[fill=lightgray, opacity=1] (v1) circle (0.65);
           	 \draw[fill=lightgray, opacity=1] (v2) circle (0.65);
	    	\node  at (v1) {$\mathcal{A}$};
	    	\node  at (v2) {$\mathcal{A}^*$};
    
	    	\fill[name intersections={of=rung1 and rung4, name=i,total=\t},white] foreach \s in {1,...,\t}{(i-\s) circle (5pt)};
	    	\fill[name intersections={of=rung2 and rung4, name=i,total=\t},white] foreach \s in {1,...,\t}{(i-\s) circle (5pt)};
	    	\fill[name intersections={of=rung3 and rung4, name=i,total=\t},white] foreach \s in {1,...,\t}{(i-\s) circle (5pt)};
    
	    	\draw[line width=0.3mm, dashed, name path=rung4, blue] (m1) to (m2);
    
	        \node[below left]  at (e1) {$p_1$};
	        \node[above left]  at (e2) {$p_2$};
	        \node[above right] at (e3) {$p_3$};
	        \node[below right] at (e4) {$p_4$};
	    	\node[fill=white, opacity=1]        at (0,0) {$\ell_X$};
	    	\node[above, fill=white, opacity=1] at (m1) {#1}; %$\ell_2-p_2$
	    	\node[below, fill=white, opacity=1] at (m2) {#2}; %$\ell_1-p_1$
\end{tikzpicture}}

\newcommand{\kmocvirtualtreenolab}{
\begin{tikzpicture}
                \coordinate (e1) at (-1,-1);	
	     	\coordinate (e2) at (-1, 1);
	     	\coordinate (e3) at ( 1, 1);
	     	\coordinate (e4) at ( 1,-1);
	     	
	     	\coordinate (v1) at (0,0);
     
	     	\draw[massive] (e1) -- (v1);
	     	\draw[massive] (e2) -- (v1);
	     	\draw[massive] (e4) -- (v1);
	     	\draw[massive] (e3) -- (v1);
	     	
            \draw[fill=lightgray, opacity=1] (0,0) circle (0.65);
\end{tikzpicture}}

\newcommand{\kmocvirtualnlonolab}{
\begin{tikzpicture}
                \coordinate (e1) at (-1,-1);	
	     	\coordinate (e2) at (-1, 1);
	     	\coordinate (e3) at ( 1, 1);
	     	\coordinate (e4) at ( 1,-1);
	     	
	     	\coordinate (v1) at (0,0);
     
	     	\draw[massive] (e1) -- (v1);
	     	\draw[massive] (e2) -- (v1);
	     	\draw[massive] (e4) -- (v1);
	     	\draw[massive] (e3) -- (v1);
	     	
            \draw[fill=lightgray, opacity=1] (0,0) circle (0.65);
            \draw[fill=white, opacity=1] (0,0) circle (0.35);
\end{tikzpicture}}

\newcommand{\kmocvirtualnnlonolab}{
\begin{tikzpicture}
                \coordinate (e1) at (-1,-1);	
	     	\coordinate (e2) at (-1, 1);
	     	\coordinate (e3) at ( 1, 1);
	     	\coordinate (e4) at ( 1,-1);
	     	
	     	\coordinate (v1) at (0,0);
     
	     	\draw[massive] (e1) -- (v1);
	     	\draw[massive] (e2) -- (v1);
	     	\draw[massive] (e4) -- (v1);
	     	\draw[massive] (e3) -- (v1);
	     	
            \draw[fill=lightgray, opacity=1] (0,0) circle (0.65);
            \draw[fill=white, opacity=1] (-0.25,0) ellipse (0.2 and 0.4);
            \draw[fill=white, opacity=1] ( 0.25,0) ellipse (0.2 and 0.4);
\end{tikzpicture}}

\newcommand{\wcut}{
\begin{tikzpicture}
		\coordinate (e1) at (-0.75, 0.25);
		\coordinate (e2) at ( 2.75, 0.25);
		\coordinate (e3) at ( 2.00,-1.25);
		\coordinate (e4) at ( 0.00,-1.25);	
		
		\coordinate (v1) at (0,0);
		\coordinate (v2) at (1,0);
		\coordinate (v3) at (2,0);
		\coordinate (v4) at (1,-1);
		
		\draw[massive] (e1) -- (v1);
		\draw[massive] (v1) -- (v2);
		\draw[massive] (v2) -- (v3);
		\draw[massive] (v3) -- (e2);
		\draw[massive] (e3) -- (v4);
		\draw[massive] (e4) -- (v4);
		\draw[photon] (v1) -- (v4);
		\draw[photon] (v2) -- (v4);
		\draw[photon] (v3) -- (v4);

		\draw[fill=lightgray, opacity=1] (v1) circle (0.2);
                \draw[fill=lightgray, opacity=1] (v2) circle (0.2);
                \draw[fill=lightgray, opacity=1] (v3) circle (0.2);
                \draw[fill=lightgray, opacity=1] (v4) ellipse (0.4 and 0.2);
		\end{tikzpicture}
		}
\newcommand{\ncut}{
\begin{tikzpicture}
		\coordinate (e1) at (-0.75, 0.25);
		\coordinate (e2) at ( 1.75, 0.25);
		\coordinate (e3) at ( 1.75,-1.25);
		\coordinate (e4) at (-0.75,-1.25);	
		
		\coordinate (v1) at (0,0);
		\coordinate (v2) at (0,-1);
		\coordinate (v3) at (1,0);
		\coordinate (v4) at (1,-1);

		\draw[massive] (e1) -- (v1);
		\draw[massive]  (v1) -- (v3);
		\draw[massive] (e2) -- (v3);
		
		\draw[massive] (e4) -- (v2);
		\draw[massive]  (v2) -- (v4);
		\draw[massive] (e3) -- (v4);
		
		\draw[photon] (v1) -- (v2);
		\draw[photon] (v3) -- (v4);
		\draw[photon] (v1) --  (v4);

                \draw[fill=lightgray, opacity=1] (v1) circle (0.2);
                \draw[fill=lightgray, opacity=1] (v2) circle (0.2);
                \draw[fill=lightgray, opacity=1] (v3) circle (0.2);
                \draw[fill=lightgray, opacity=1] (v4) circle (0.2);
		\end{tikzpicture}
}

\newcommand{\boxbubcut}{
\begin{tikzpicture}
		\coordinate (e1) at (-0.75, 0.25);
		\coordinate (e2) at ( 1.75, 0.25);
		\coordinate (e3) at ( 1.75,-1.25);
		\coordinate (e4) at (-0.75,-1.25);	
		
		\coordinate (v1) at (0,0);
		\coordinate (v2) at (0,-1);
		\coordinate (v3) at (1,0);
		\coordinate (v4) at (1,-1);

		\draw[massive] (e1) -- (v1);
		\draw[massive]  (v1) -- (v3);
		\draw[massive] (e2) -- (v3);
		
		\draw[massive] (e4) -- (v2);
		\draw[massive]  (v2) -- (v4);
		\draw[massive] (e3) -- (v4);
		
		\draw[photon] (v1) -- (v2);
		\draw[photon] (v3)  to[out=-25,in=25] (v4);
		\draw[photon] (v3)  to[out=180+25,in=180-25] (v4);

                \draw[fill=lightgray, opacity=1] (v1) circle (0.2);
                \draw[fill=lightgray, opacity=1] (v2) circle (0.2);
                \draw[fill=lightgray, opacity=1] (v3) circle (0.2);
                \draw[fill=lightgray, opacity=1] (v4) circle (0.2);
		\end{tikzpicture}
}

\newcommand{\threecomptoncut}{
\begin{tikzpicture}
		\coordinate (e1) at (-0.75, 0.25);
		\coordinate (e2) at ( 1.75, 0.25);
		\coordinate (e3) at ( 1.50,-1.25);
		\coordinate (e4) at (-0.50,-1.25);	
		
		\coordinate (v1) at (0,0);
		\coordinate (v2) at (0.5,-1);
		\coordinate (v3) at (1,0);

		\draw[massive] (e1) -- (v1);
		\draw[massive]  (v1) -- (v3);
		\draw[massive] (e2) -- (v3);
		
		\draw[massive] (e4) -- (v2);
		\draw[massive] (e3) -- (v2);
		
		\draw[photon] (v1) -- (v2);
		\draw[photon] (v3) -- (v2);
		\draw[photon] (v1) to[out=100,in=80]  (v3);

        \draw[fill=lightgray, opacity=1] (v1) circle (0.2);
        \draw[fill=lightgray, opacity=1] (v2) ellipse (0.4 and 0.2);
        \draw[fill=lightgray, opacity=1] (v3) circle (0.2);	
\end{tikzpicture}}

\newcommand{\graphone}{
		\centering	%H
		\tikzsetnextfilename{graph_1}
		\begin{tikzpicture}
		\coordinate (e1) at (-0.5, 0);
		\coordinate (e2) at ( 2.5, 0);
		\coordinate (e3) at ( 2.5,-1);
		\coordinate (e4) at (-0.5,-1);	
		
		\coordinate (v1) at (0,0);
		\coordinate (v2) at (1,0);
		\coordinate (v3) at (2,0);
		\coordinate (v4) at (2,-1);
		\coordinate (v5) at (1,-1);
		\coordinate (v6) at (0,-1);
		
		\draw[massive] (e1) -- (v1);
		\draw[massive] (v1) -- (v2);
		\draw[massive] (v2) -- (v3);
		\draw[massive] (v3) -- (e2);
		\draw[massive] (e4) -- (v6);
		\draw[massive] (v6) -- (v5);
		\draw[massive] (v5) -- (v4);
		\draw[massive] (v4) -- (e3);
		
		\draw[massless] (v1) -- (v6);
		\draw[massless] (v2) -- (v5);
		\draw[massless] (v3) -- (v4);
		\end{tikzpicture}
}
\newcommand{\graphfour}{
		\tikzsetnextfilename{graph_4}
          	\begin{tikzpicture}
		\coordinate (e1) at (-0.5, 0);
		\coordinate (e2) at ( 2.5, 0);
		\coordinate (e3) at ( 2.5,-1);
		\coordinate (e4) at (-0.5,-1);

		\coordinate (t1) at (0,0);
		\coordinate (t2) at (1,0);
		\coordinate (t3) at (2,0);
		\coordinate (b1) at (0,-1);
		\coordinate (b2) at (1,-1);
		\coordinate (b3) at (2,-1);
		
		\draw[massive] (e1) -- (t1) -- (t2)-- (t3) -- (e2);
		\draw[massive] (e4) -- (b1) -- (b2)-- (b3) -- (e3);
		
		\draw[name path=rung1,massless] (t1) -- (b1);
		\draw[name path=rung2,massless] (t2) -- (b3);
		\draw[name path=rung3,massless] (t3) -- (b2);
		
		\fill[name intersections={of=rung2 and rung3, name=i,total=\t},white] foreach \s in {1,...,\t}{(i-\s) circle (5pt)};
		\draw[massless] (t3) -- (b2);
		\end{tikzpicture}
}
\newcommand{\graphnine}{
		\tikzsetnextfilename{graph_9}
		\begin{tikzpicture}
		\coordinate (e1) at (-1, 0);
		\coordinate (e2) at ( 2, 0);
		\coordinate (e3) at ( 2,-1);
		\coordinate (e4) at (-1,-1);

		\coordinate (v1) at (0,0);
		\coordinate (v2) at (0,-1);
		\coordinate (v3) at (1,0);
		\coordinate (v4) at (1,-1);
		\coordinate (v5) at (-0.5,0);
		\coordinate (v6) at (1.5,0);
		
		\draw[massive] (e1) -- (v1);
		\draw[massive]  (v1) -- (v3);
		\draw[massive] (e2) -- (v3);

		\draw[massless] (1.5, 0) arc (0:180:1cm and 0.75cm);
		
		\draw[massive] (e4) -- (v2);
		\draw[massive]  (v2) -- (v4);
		\draw[massive] (e3) -- (v4);
		
		\draw[massless] (v1) -- (v2);
		\draw[massless] (v3) -- (v4);
		\end{tikzpicture}
}
\newcommand{\grapheleven}{
		\tikzsetnextfilename{graph_11}
		\begin{tikzpicture}
		\coordinate (e1) at (-1, 0);
		\coordinate (e2) at ( 1.5, 0);
		\coordinate (e3) at ( 1.5,-1);
		\coordinate (e4) at (-1,-1);

		\coordinate (v1) at (0,0);
		\coordinate (v2) at (0,-1);
		\coordinate (v3) at (1,0);
		\coordinate (v4) at (1,-1);
		\coordinate (v5) at (-0.5,0);
		\coordinate (v6) at (0.5,0);
		
		\draw[massive] (e1) -- (v1);
		\draw[massive]  (v1) -- (v3);
		\draw[massive] (e2) -- (v3);
		
		\draw[massive] (e4) -- (v2);
		\draw[massive]  (v2) -- (v4);
		\draw[massive] (e3) -- (v4);
		
		\draw[massless] (v1) -- (v2);
		\draw[massless] (v3) -- (v4);
		\draw[massless] (0.5, 0) arc (0:180:0.5cm and 0.5cm);
		\end{tikzpicture}
}
\newcommand{\graphthirteen}{
		\tikzsetnextfilename{graph_13}
		\begin{tikzpicture}
		\coordinate (e1) at (-1, 0);
		\coordinate (e2) at ( 2, 0);
		\coordinate (e3) at ( 2,-1);
		\coordinate (e4) at (-1,-1);

		\coordinate (v1) at (-0.5,0);
		\coordinate (v2) at (-0.5,-1);
		\coordinate (v3) at (1.5,0);
		\coordinate (v4) at (1.5,-1);
		\coordinate (v5) at (0,0);
		\coordinate (v6) at (1,0);

		\draw[massive] (e1) -- (v1);
		\draw[massive]  (v1) -- (v3);
		\draw[massive] (e2) -- (v3);
		
		\draw[massive] (e4) -- (v2);
		\draw[massive]  (v2) -- (v4);
		\draw[massive] (e3) -- (v4);
		
		\draw[massless] (v1) -- (v2);
		\draw[massless] (v3) -- (v4);
		\draw[massless] (1, 0) arc (0:180:0.5cm and 0.5cm);
		\end{tikzpicture}
}
\graphicspath{{figures/}}
\usepackage[export]{adjustbox}

\newcommand{\calA}{\mathcal{A}}
\newcommand{\calO}{\mathcal{O}}
\newcommand{\calN}{\mathcal{N}}

\newcommand{\calI}{\mathcal{I}}

\newcommand{\nn}{\nonumber}
\renewcommand{\imath}{\mathrm{i}}

\newcommand{\mb}{{\overline{m}}}

\def\RT{\mathrm{III}}

\def\EulerGamma{\gamma_{E}}
\def\relfbar{y}
\def\sqrtmQSq{\sqrt{-q^2}}

\def\sect#1{Sec.~{\ref{#1}}}

\begin{document}

\title{Scalar QED as a toy model for higher-order effects in classical gravitational scattering}

\author[1]{Zvi Bern,}
\affiliation[1]{Mani L. Bhaumik Institute for Theoretical Physics,\\
UCLA Department of Physics and Astronomy, Los Angeles, CA 90095, USA}
\emailAdd{bern@physics.ucla.edu}

\author[1]{Juan Pablo Gatica,}
\emailAdd{jpgatica3541@g.ucla.edu}

\author[1]{Enrico Herrmann,}
\emailAdd{eh10@g.ucla.edu}

\author[2]{Andres Luna,}
\affiliation[2]{Niels Bohr International Academy $\&$ Discovery Center, Niels Bohr Institute, \\
Blegdamsvej 17, DK- 2100, Copenhagen {\O}, DENMARK}
\emailAdd{andres.luna@nbi.ku.dk}

\author[3]{Mao Zeng}
\affiliation[3]{Higgs Centre for Theoretical Physics, University of Edinburgh,\\
James Clerk Maxwell Building, Edinburgh EH9 3FD, UK}
\emailAdd{mzeng@ed.ac.uk}

%================================================
\abstract{
  Quantum Electrodynamics (QED) serves as a useful toy model for classical
  observables in gravitational two-body systems with reduced complexity due to the
  linearity of QED.  We investigate scattering observables in scalar
  QED at the sixth order in the charges (two-loop order) in a
  classical regime analogous to the post-Minkowskian expansion in
  General Relativity.  We employ modern scattering amplitude
  tools and extract classical observables by both eikonal methods and the
  formalism of Kosower, Maybee, and O'Connell (KMOC).  In
  addition, we provide a simplified approach to extracting the
  radial action beyond the conservative sector.}

%\preprint{XXX}
%================================================

\maketitle

%==============================================================
%
\vspace{1cm}
\section{Introduction}
%
%==============================================================

The landmark detection of gravitational
waves~\cite{Abbott:2016blz,TheLIGOScientific:2017qsa} has opened a
remarkable new window into the Universe that promises major new
advances into black holes, neutron stars, and perhaps even provides new
insights into fundamental physics. The recent experimental progress has
inspired efforts to develop new theoretical tools for predicting
gravitational-wave signals that meet the precision
challenges of current and future detectors~\cite{Punturo:2010zz,
  Dwyer:2014fpa, LISA:2017pwj, Reitze:2019iox}.  A variety of
complementary tools are being used, including the effective one-body
(EOB) formalism~\cite{Buonanno:1998gg}, numerical
relativity~\cite{Pretorius:2005gq, Campanelli:2005dd, Baker:2005vv},
the self-force formalism~\cite{Mino:1996nk,Quinn:1996am}, as well as
perturbative methods such as the post-Newtonian~(PN)
expansion~\cite{PNDroste,Einstein:1938yz}, the effective field theory (EFT)
known as nonrelativistic general relativity (NRGR)~\cite{Goldberger:2004jt,
  Kol:2007rx, Kol:2007bc,  Gilmore:2008gq, Foffa:2011ub, Foffa:2016rgu,
  Porto:2017dgs,Foffa:2019hrb,Blumlein:2019zku, Foffa:2019rdf,
  Foffa:2019yfl, Blumlein:2020pog}, as well as the post-Minkowskian~(PM)
expansion~\cite{Bertotti:1956, Kerr:1959zlt, Bertotti:1960wuq,
  Portilla:1979xx, Westpfahl:1979gu, Bel:1981be, Westpfahl:1985tsl,
  Ledvinka:2008tk, Damour:2017zjx, Bjerrum-Bohr:2013bxa,
  Bjerrum-Bohr:2018xdl, Cheung:2018wkq, Kosower:2018adc,
  Maybee:2019jus, Bern:2019nnu, Bern:2019crd, Antonelli:2019ytb,
  KoemansCollado:2019ggb, Cristofoli:2020uzm}.  Information from
various approaches can be combined into state-of-the-art results and
provide nontrivial cross-checks, see e.g.~Refs.~\cite{Bern:2019crd,
  Blumlein:2020znm, Cheung:2020gyp,Bini:2020wpo, Bini:2020nsb,
  Bini:2019nra, Siemonsen:2019dsu, Antonelli:2020aeb, Bern:2021dqo, Bini:2021gat, Blumlein:2021txe, Bern:2021yeh}.

The post-Minkowskian approach is a weak-field expansion in
Newton's constant, $G$, and has risen in prominence in recent years.
It has the advantage of maintaining Lorentz invariance and gives results with exact
relativistic velocity dependence. The scattering amplitude framework
for post-Minkowskian calculations~\cite{Neill:2013wsa,
  Bjerrum-Bohr:2018xdl, Cheung:2018wkq, Kosower:2018adc, Bern:2019nnu,
  Bern:2019crd,Cristofoli:2021vyo} naturally meshes with this
covariant approach.  Calculations of scattering amplitudes have
advanced enormously, making this a natural framework for
state-of-the-art post-Minkowskian calculations.  The modern amplitude
tools include the unitarity
method~\cite{Bern:1994zx,Bern:1994cg,Britto:2004nc} which constructs
loop-level scattering amplitude integrands from lower-order
gauge-invariant on-shell data, as well as the double copy which relates
gauge and gravity
theories~\cite{KLT,Bern:2008qj,Bern:2010ue,Bern:2012uf,Bern:2019prr}.
Furthermore, amplitude methods also incorporate powerful integration
procedures~\cite{Parra-Martinez:2020dzs}, originally developed for particle-collider physics applications, such
as integration by parts (IBP)~\cite{Tkachov:1981wb,
  Chetyrkin:1981qh, Laporta:2001dd}, differential
equations~\cite{Kotikov:1990kg, Bern:1992em, Remiddi:1997ny, Gehrmann:1999as,
  Henn:2013pwa, Henn:2014qga}, and reverse
unitarity~\cite{Anastasiou:2002yz, Anastasiou:2002qz,
  Anastasiou:2003yy, Anastasiou:2015yha}.

Combining techniques based on scattering amplitudes with those of
effective field theory (EFT), two-body effective Hamiltonians have
been derived in Refs.~\cite{Neill:2013wsa,Cheung:2018wkq,Bern:2019nnu,
  Bern:2019crd} that straightforwardly determine the conservative
classical dynamics of bound orbits via their equations of motion,
whenever nonlocalities associated with the tail
effect~\cite{Bonnor:1959, BonnorRotenberg:1966, Thorne:1980ru,
  Blanchet:1987wq, Blanchet:1992br, Blanchet:1993ec} are absent.
Such Hamiltonians can be imported into the EOB
framework~\cite{Damour:2017zjx, Antonelli:2019ytb} used by LIGO for
constructing gravitational-wave templates.  One can alternatively
obtain bound-state physical observables from the ones of hyperbolic
scattering processes via appropriate analytic
continuation~\cite{Kalin:2019rwq,Kalin:2019inp,Bini:2020hmy,Saketh:2021sri}.
Cases involving the tail effect are more subtle~\cite{Bini:2017wfr,
  Cho:2021arx}.

Scattering amplitudes are the natural realm to describe the hyperbolic
motion of classical objects from the asymptotic past to the asymptotic
future. This idea has been implemented in the work by Kosower, Maybee,
and O'Connell (KMOC)~\cite{Kosower:2018adc} whose approach allows us
to extract classical observables directly from scattering amplitudes
and what are essentially unitarity cuts.  Alternatively, in the
classical limit, appropriately defined finite parts of the scattering
amplitudes can be directly connected to the scattering angle or the
isotropic gauge two-body Hamiltonian~\cite{Bern:2019crd,
  Kalin:2019rwq}.  The scattering amplitude can also be interpreted
directly in terms of the radial action~\cite{Bern:2021dqo,
  Damgaard:2021ipf, Kol:2021jjc}.  A related but distinct approach
based on the eikonal phase~\cite{Glauber:1956Lecture} (for more recent
examples see e.g. Refs.~\cite{Amati:1990xe, Laenen:2008gt,
  Akhoury:2013yua,Bern:2020gjj,delaCruz:2021gjp}) provides a natural way to extract the
classical scattering angle from amplitudes. In this paper we will use
a variety of these approaches to extract classical observables in both
the conservative and radiative sectors.

The usefulness of the scattering amplitude framework has been
demonstrated through the first construction of the conservative
two-body Hamiltonian at $\calO(G^3)$~\cite{Bern:2019nnu,Bern:2019crd}, 
as well as new results at
$O(G^4)$~\cite{Bern:2021dqo,Bern:2021yeh}. There have also been a
variety of new results for spin~\cite{Vaidya:2014kza, Vines:2017hyw,
  Guevara:2017csg, Vines:2018gqi, Guevara:2018wpp, Chung:2018kqs,
  Guevara:2019fsj, Chung:2019duq, Damgaard:2019lfh, Aoude:2020onz,
  Bern:2020buy, Kosmopoulos:2021zoq, Guevara:2020xjx, Levi:2020uwu, Levi:2020kvb,
  Chen:2021qkk}, tidal effects~\cite{Brandhuber:2019qpg,
  Huber:2019ugz, Cheung:2020sdj, Kalin:2020lmz, Haddad:2020que,
  Aoude:2020ygw, AccettulliHuber:2020oou, Huber:2020xny,
  Cheung:2020gbf, Bern:2020uwk}, and
waveforms~\cite{Cristofoli:2021vyo}.

In carrying out such calculations, it is useful to analyze simpler
models compared to Einstein gravity that eliminate unnecessary
complications.  As a recent example, the authors of
Ref.~\cite{DiVecchia:2020ymx} analyzed $\calN{=} 8$ supergravity to
demonstrate the cancellation of mass singularities between
conservative and radiative contributions to the classical scattering
angle, leading to a complete
resolution~\cite{Damour:2020tta}. Likewise, gauge theory, especially
electrodynamics, served as a toy model for gravity in the context of
two-body dynamics for many
decades~\cite{Westpfahl:1985tsl,Buonanno:2000qq}.  While being a
linear theory, it captures some of the technical difficulties
encountered with general relativity (GR) at high orders of
perturbation theory.  Another reason for studying corresponding
quantities in gauge theories, especially in non-abelian cases, is the
double-copy relation between gauge and gravity
theories~\cite{KLT,Bern:2008qj,Bern:2010ue}.

Recently, the conservative and radiative dynamics in classical
relativistic scattering was obtained by Saketh, Vines, Steinhoff, and
Buonanno in scalar electrodynamics to the sixth order in the
charges or the third order in the fine-structure constant~\cite{Saketh:2021sri}.
This was accomplished by the direct iteration of the equations
of motion.  Here, we compare to these results using scattering
amplitudes based approaches, finding full agreement.
Using amplitude methods, we evaluate the angle including
radiative effects in three distinct ways.  First, we use the
Kosower-Maybee-O'Connell formalism to obtain the impulse on two
massive charged scalar particles scattering at large impact parameter
$b$ from which we extract the scattering angle.  As an alternative, we
extract the eikonal phase from the scattering amplitude which allows
us to determine the scattering angle.  Finally, using recent
observations on the connection of the scattering amplitude in the
classical limit to the radial action~\cite{Bern:2021dqo} (see
also~\cite{Damgaard:2021ipf}), we present a simple prescription for
extracting from the scattering amplitude a radial action that
determines the scattering angle, including radiative effects.
Although the system is not conservative (we include radiation reaction
effects in the scattering angle), we find that the scattering angle
obtained by differentiating this generalized radial action matches the
previous results. All three of these approaches for extracting the
classical scattering angle match, and agree with the result from the
classical approach of Ref.~\cite{Saketh:2021sri}.

As previously discussed in Ref.~\cite{Saketh:2021sri}, in
electromagnetism we encounter a mass singularity in the scattering
angle, which does not cancel between potential and radiative
contributions, as it does in the corresponding $\calO(G^3)$
calculation in gravity~\cite{DiVecchia:2020ymx,Damour:2020tta}. In
fact, the singularity is power divergent for $m \rightarrow 0$,
similar to the situation in the conservative sector of gravity at
$\calO (G^4)$~\cite{Bern:2021dqo, Bern:2021yeh}.  From our perspective, we interpret this singularity as
a breakdown of the classical expansion which requires $ m^2 |b|^2 {\gg}
1$.  Another interesting feature of the amplitudes based approaches is
that the Abraham-Lorentz-Dirac (ALD) force~\cite{lorentz:1892theorie,
  abraham:1912theorie, Dirac:1938a} that appears in more tradiational
methods~\cite{Westpfahl:1985tsl} is automatically built in and does
not require any special treatment~\cite{Kosower:2018adc}.

For the conservative sector, we also extracted a two-body Hamiltonian
valid through the sixth order in the charges analogous to the $\calO(G^3)$ 
isotropic-gauge Hamiltonian of the gravitational case.  This is
obtained from the mapping between infrared-finite parts of the
amplitude to the coefficients in the two-body potential~\cite{Bern:2019crd,
Kalin:2019rwq}.  As for any standard Hamiltonian it can be directly
applied to the bound state case. 

The paper is organized as follows.  In \sect{sec:methods} we briefly
review the methods used here.  Then in \sect{sec:conservative} we
evaluate the conservative contributions to the two-particle scattering through the sixth
order in the charges. In \sect{sec:radiative} we include radiative
corrections to the scattering angle and impulse and also compute the radiated momentum. We give our conclusions
in \sect{sec:conclusions}. All our results are available in computer-readable form 
in the ancillary file attached to this article.

%==============================================================
%
%\newpage
\section{Review of Methods}
\label{sec:methods}
%
%==============================================================

The present section briefly summarizes and reviews the main technical ingredients that are required to obtain classical scattering observables in (scalar) QED up to two-loop order ($\calO(\alpha^3)$ where $\alpha = e^2/4\pi$), corresponding to the sixth order in the charges, from various scattering amplitude based frameworks. Readers only interested in the final results may skip this section on a first reading. The remainder of this section is structured as follows: We first review the kinematic parametrization tailored towards the classical expansion of quantum scattering amplitudes in Subsection \ref{subsec:kinematics}, before outlining the generalized unitarity framework to determine the amplitude integrands in Subsection \ref{subsec:integrands}. In Subsection \ref{subsec:integration}, we telegraphically sketch the applicability of modern collider-physics based integration tools to compute precision-level classical observables with the help of integration-by-parts reduction to a minimal set of master integrals and their evaluation by differential equation methods. Subsection \ref{subsec:soft_radial_action_subtraction_scheme} introduces a new concept that allows us to define a radial action in the presence of soft-region radiation effects. In particular, we find an efficient computational scheme that allows us to compute the \emph{soft radial action} by a well-motivated modification of the boundary conditions for the soft-region master integrals. Finally, in Subsection \ref{subsec:KMOC_general}, we briefly summarize the Kosower, Maybee, and O'Connell (KMOC) formalism which allows us to extract the classical electromagnetic impulse, the radiated momentum, and the classical scattering angle up to $\calO(\alpha^3)$.

%================================================
%
\subsection{Classical limit of quantum scattering amplitudes--soft and potential region}
\label{subsec:kinematics}
%
%================================================
%
\begin{figure}[b!]
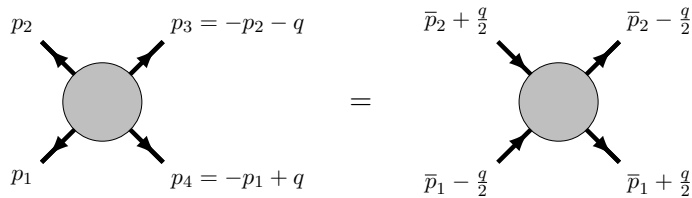

\centering
    $\vcenter{\hbox{\scalebox{0.8}{\softkinA}}} 
    \quad
    =
    \quad
    \vcenter{\hbox{\scalebox{0.8}{\softkinB}}}$
    \caption{\label{fig:soft_kinematics}Parametrization of external kinematics.}
\end{figure}
We compute classical observables for the relativistic scattering of two point-charges, in what might be called the ``post-Lorentzian'' (PL) expansion. This regime is in direct correspondence to the post-Minkowskian expansion in gravity. 

\begin{table}[h!]
\centering
\begin{tabular}{c|c|c}
 &  GR & QED \\
\hline
quantum: 				& $\lambda_c \sim \frac{\hbar}{m}$ 	& $\lambda_c \sim \frac{\hbar}{m}$ 	\\[5pt]
classical particle size:	& $r_S = G m$				       	& $r_Q = \frac{e^2 q^2_i}{4 \pi\, m} $ 		\\[5pt]
particle separation:		& $b$						& $b$
\end{tabular}
\caption{
\label{tab:gr_qed_comparison_scales}
Comparison between relevant length scales in the post-Minkowskian (PM) expansion in GR and the post-Lorentzian (PL) regime in (scalar) QED in units where $c=1$ and the fundamental charge $e$ is measured in units where $ \epsilon_0 =1$. }
\end{table}

The relevant length scales, summarized in Table \ref{tab:gr_qed_comparison_scales}, are the Compton wavelength $\lambda_c$, related to Planck's constant $\hbar$ and the particle mass scale $m$, the typical classical particle size $r_S$ or $r_Q$ (Schwarzschild radius or classical charge radius), as well as the inter-particle separation $b$. The classical PM expansion corresponds to the following hierarchy of scales: $\lambda_c \ll r_S \ll b$, and similarly $\lambda_c  \ll r_Q  \ll b$ in the PL case. The classical limit posits that the individual particle size is much bigger than the Compton wavelength and forces us into a regime of large charges: $r_S/\lambda_c \gg 1 \leftrightarrow G$ $m^2/\hbar \gg 1$, or $r_Q/\lambda_c \gg1  \leftrightarrow e^2 q^2_i/\hbar \gg 1$, where $q_i$ is the electric charge of the classical object in units of $e$. The large (macroscopic) charge regime makes intuitive sense from the point of view that classical physics should arise from the quantum theory in the limit of large quantum numbers. This seems to suggest that we are outside the traditional range of validity of perturbation theory. This, however, is a premature conclusion, because the PM or PL regime amounts to an expansion in terms of the small ratio $r_S/b\ll 1$ or $r_Q/b\ll1$ which leads to a well-defined perturbative expansion. For an especially nice discussion of the relevant scales in the gravitational context, see e.g.~Ref.~\cite{DiVecchia:2021bdo}. 

In order to simplify our discussion, we focus on the scattering of spinless, structureless objects described by massive scalar fields. In the PL approximation, the above hierarchy of scales is converted into momentum space as follows: it is assumed that the masses of the scalars are very heavy and that the momentum transfer $|q|\sim 1/|b|$ is small in the classical limit, $(-q^2)\ll m^2_i $, in complete analogy to gravitational scattering. In order to extract classical physics, we utilize special kinematic variables that facilitate the classical $\hbar\to0$ or equivalently \emph{soft} (small |q|) expansion\footnote{From now on, we work in natural units and set $\hbar=1$ unless stated otherwise.} in the context of the method of regions~\cite{Beneke:1997zp}. These variables have previously appeared in e.g. Ref.~\cite{Parra-Martinez:2020dzs} and are summarized in Fig.~\ref{fig:soft_kinematics}, 
\begin{align}
    p_1 = - \left(\overline{p}_1 - \frac{q}{2}\right)\,, \ 
    p_2 = - \left(\overline{p}_2 + \frac{q}{2}\right)\,, \ 
    p_3 =   \left(\overline{p}_2 - \frac{q}{2}\right)\,, \ 
    p_4 =   \left(\overline{p}_1 + \frac{q}{2}\right)\,.
    \label{eq:soft_kinematics}
\end{align}
The new vectors $\overline{p}_i$ are orthogonal to the momentum transfer $q$, $\overline{p}_i \cdot q = 0\,, $ which directly follows from the on-shell conditions $p^2_1 = p^2_4 = m^2_1$ and $p^2_2 = p^2_3 = m^2_2$. For later convenience, we also introduce `soft-masses' $\overline{m}_i$ defined by
\begin{align}
 \overline{m}^2_i = \overline{p}^2_i = m^2_i - \frac{q^2}{4} 
 \quad \rightarrow \quad 
 \overline{m}_i =  m_i + \frac{(-q^2)}{8 m_i} + \calO(q^4) \,.
\end{align}
Notably, in the specialized barred variables, $s{=}(\overline{p}_1+\overline{p}_2)^2 {=}(p_1 + p_2)^2 $ the physical scattering region $s{>}(m_1+m_2)^2,\, q^2{<}0$ remains the same. Following earlier conventions~\cite{Parra-Martinez:2020dzs}, we define the soft four-velocities of the two black holes $u^\mu_i = \overline{p}^\mu_i/|\overline{p}_i|$, such that $u_i^2=1$, and 
\begin{align}
\label{eq:y_and_x_def}
 y \equiv u_1\cdot u_2 = \frac{1+x^2}{2x}  
   = \sigma - (-q^2)\frac{\left(m_1^2{+}m_2^2\right) \sigma +2 m_1 m_2}{8 \, m_1^2 m_2^2}+ \calO(q^4) \,.
\end{align}
For physical scattering in the $s$-channel we have $y>1$ . Often, it will prove advantageous to change variables to $x$ in the range $0<x<1$ in order to rationalize the naturally appearing square-root $\sqrt{y^2-1} = \frac{1-x^2}{2x}$. 

Note that the soft velocities $u_i$ coincide with the classical four velocities of the massive scalars only up to corrections of $\calO(q)$. The KMOC setup directly targets physical observables where this difference is immaterial. However, for the eikonal computations, the $\calO(q)$ corrections do matter and one has to carefully track them. In the classical limit (without restricting to the conservative sector), we are interested in the soft expansion of loop amplitudes with the hierarchy of scales given by $|\ell | \sim |q| \ll |\overline{p}_i|, m , \sqrt{s}$.  Here, $\ell$ schematically represents arbitrary combinations of photon momenta of the form $(\ell_1, \ell_2, \ell_1 \pm \ell_2, \ldots )$ and typical photon propagators take the form $\frac{1}{\ell^2}\,,\ \frac{1}{(\ell-q)^2}$. These have a homogeneous $|q|$-scaling and do not require any further expansions. Distinctly, matter propagators do have a non-trivial $|q|$ expansion expressed via dimensionless velocity variables $u_i$
\begin{align}
    \frac{1}{(\ell-p_i)^2-m^2_i} = \frac{1}{\ell^2 - 2\, \ell \cdot p_i}
    =
    \frac{1}{2 u_i \cdot \ell}\,\frac{1}{m_i} 
    - \frac{\ell^2 \mp \ell \cdot q}{(2 u_i \cdot \ell)^2}\,\frac{1}{m^2_i} 
    + \cdots\,. \label{eq:softExpansionMatterProp}
\end{align}
Each order in the expansion is homogeneous in $|q|$ and the mass dependence factorizes. The matter propagators effectively ``eikonalize'' and the soft expansion to higher orders in $|q|$ can lead to raised propagator powers.

To focus on conservative dynamics one would perform a further expansion where the temporal part of any photon line is suppressed by an additional power of the formally small velocity $v$, related to $y\approx \sigma = \frac{1}{\sqrt{1-v^2}}$ to signal instantaneous interactions. These \emph{potential region} expansions have been described in great detail elsewhere~\cite{Bern:2019crd,Bern:2019nnu,Parra-Martinez:2020dzs} and we refrain from repeating them here for the sake of brevity. 

%================================================
%
\subsection{Generalized Unitarity and scalar QED scattering amplitudes up to $\mathcal{O}(\alpha^3)$}
\label{subsec:integrands}
%
%================================================

As we are going to review in the following subsections, a number of novel approaches to the classical two-body problem in gravity and electromagnetism involve the (classical limit of) quantum scattering amplitudes. These enter either in the EFT matching calculation to a classical two-body potential, in the eikonal approach to classical scattering, or in the KMOC framework that expresses classical physical observables (e.g. the impulse or the radiate momentum) in terms of scattering amplitudes and weighted cross-section-like objects. Therefore, it is crucial to have at our disposal compact expressions for the relevant (classical parts) of the higher-loop scattering amplitudes in the theories under consideration. Recent years have seen enormous advances in our ability to obtain analytic results for quantum scattering amplitudes via modern on-shell methods. On one hand, this progress enhanced our ability to compute phenomenologically relevant collider physics processes in quantum chromodynamics (QCD) and the Standard Model. On the other hand, in simplified toy theories such as maximally supersymmetric Yang-Mills theory or in supersymmetric gravity theories, similar computations were crucial to shed light on a number of impressive theoretical insights into the deeper structures of quantum field theory.  A chief ingredient in many of these calculations is an efficient way to obtain a scattering amplitude \emph{integrand}, i.e. an expression of the amplitude before loop integration. Generalized unitarity~\cite{Bern:1994zx,Bern:1994cg,Britto:2004nc} is based on the factorization of amplitudes into simpler gauge-invariant on-shell building blocks which allows to export the simplicity of tree-amplitudes to loop-calculations. In the context of classical gravitational dynamics, these methods have been recently used~\cite{Herrmann:2021tct} to obtain the radiated momentum and the impulse at $\calO(G^3)$ in general relativity from the KMOC setup and from eikonal considerations~\cite{DiVecchia:2021bdo}.  Since these methods have been comprehensively documented elsewhere in the context of general relativity~\cite{Bern:2019crd,Bern:2019nnu,Parra-Martinez:2020dzs,Herrmann:2021tct}, we are only giving a telegraphic account of the main ingredients of our QED calculation.

%================================================
%
\subsubsection{Tree-level amplitudes in scalar QED}
\label{subsubsec:treeAmps}
%
%================================================
The main building blocks in the derivation of loop integrands via generalized unitarity are on-shell tree-level amplitudes out of which unitarity cuts are built. Later, these products of tree-level amplitudes are compared to the unitarity cuts of a putative ansatz of Feynman-like loop integrals in order to fix the free coefficients in the ansatz by solving a linear system of equations.  We are interested in the scattering of two massive charged scalars in scalar QED that have charges $e q_{1,2}$ and masses $m_{1,2}$, respectively and interact via the exchange of $U(1)$ gauge bosons, i.e. photons. The Lagrangian for the system is 
\begin{align}
\mathcal{L} = -\frac{1}{4} F_{\mu \nu} F^{\mu \nu} + \sum^2_{i=1}  \Big[(D_\mu \phi_i)^\dagger (D^\mu \phi_i) - m^2_i \phi^\dagger_i  \phi^{\phantom{\dagger}}_i \Big]\,,
\end{align}
where the covariant derivative $D^\mu = \partial^\mu - \imath\, e\, q_i\,  A^\mu$ contains the photon field $A^\mu(x)$ and the appropriate electric charge $q_i$ (in multiples of the fundamental charge\footnote{In the following, we often trade the square of elementary charges for the coupling constant $\alpha = e^2/(4\pi)$.} $e$) of the scalar $\phi_i$. The $U(1)$ field strength is $F_{\mu \nu} = \partial_\mu A_\nu - \partial_\nu A_\mu$. 

The basic input is the three-point coupling between the scalars and a photon in an all-outgoing convention for the particle momenta
\begin{align}
 \vcenter{\hbox{\scalebox{1}{\threePtPhoton}}} =- \imath \, e\, q_i (p_a - p_b)^\mu\,,
\end{align}
from which we can build the tree-level scattering amplitude between the two charged scalars due to photon exchange
\begin{align}
\label{eq:4s_tree_amp_soft_vars}
\mathcal{A}^{\text{tree}}_4(p_1,p_2,p_3,p_4)=
 \vcenter{\hbox{\scalebox{1}{\treetPhoton}}} = - \frac{4 e^2\,  q_1 q_2 \,\mb_1 \mb_2\, y}{-q^2}\,,
\end{align}
written in terms of the soft-kinematics of Eq.~\eqref{eq:soft_kinematics}. For higher-order calculations, we also require the Compton amplitude for the tree-level scattering of two scalars with two photons
\begin{align}
\label{eq:compton_amp}
\begin{split}
\hspace{-.5cm}
 \vcenter{\hbox{\scalebox{1}{\comptonAmp}}} 
 \hspace{-.8cm}
 & = \frac{-2 e^2 q^2_i }{(2 p_1\cdot p_2) (2p_1 \cdot p_3)} \Big[
 2 p_1{\cdot} F_2 {\cdot}  F_3 {\cdot}  p_4 + 2 p_1 {\cdot}  F_3 {\cdot}  F_2 {\cdot}  p_4 + \frac{1}{2} (p_1 {+} p_4)^2 F_2 {\cdot}  F_3
 \Big] 
 \\[-10pt]
 & = 
 -2 e^2 q^2_i \Big[ 
 \varepsilon_2\cdot \varepsilon_3 
 - \frac{\varepsilon_2\cdot p_1 (\varepsilon_3\cdot p_1+\varepsilon_3\cdot p_2)}{p_1\cdot p_2}
 - \frac{\varepsilon_3\cdot p_1 (\varepsilon_2\cdot p_1+\varepsilon_2\cdot p_2)}{p_1\cdot p_3}
 \Big]\,,
 \end{split}
\end{align}
where we have introduced the linearized field-strengths $F^{\mu \nu}_i = \varepsilon^\mu_i p^\nu_i - \varepsilon^\nu_i p^\mu_i $. From the first line of Eq.~(\ref{eq:compton_amp}) it is clear that we have expressed the amplitude in terms of gauge-invariant building blocks that manifestly vanish when $\varepsilon_i \to p_i$, so that physical state sums that appear in the cut sewing procedure of generalized unitarity can be performed by the simple substitution (see the discussion in Ref.~\cite{Kosmopoulos:2020pcd})
\begin{align}
\sum_{\lambda} \varepsilon^{\ast \mu}_{i,\lambda}(k)  \varepsilon^{\nu}_{i,\lambda}(-k) \to \eta^{\mu\nu}\,,
\label{StateSumSimp}
\end{align}
where $\lambda$ denotes the physical polarizations. To get to the compact expression on the second line of Eq.~(\ref{eq:compton_amp}), we have used momentum conservation to eliminate $p_4$ and the transversality condition $\varepsilon_i\cdot p_i = 0$ of the polarization vectors.

For the two-loop computation, we also require the amplitude between three photons and two massive scalars 
\begin{align}
\label{eq:3photon_amp}
%\begin{split}
\hspace{-.5cm}
 \vcenter{\hbox{\scalebox{1}{\threePhotonAmp}}} 
 \hspace{-1cm}
  =
-2 i e^3q^3_i \Bigg[& 
 \frac{(\varepsilon_3{\cdot} \varepsilon_4) (p_1{\cdot} F_2{\cdot} p_5) }{(p_1{\cdot} p_2)(p_2{\cdot} p_5)}  
{-}\frac{\varepsilon_2{\cdot} p_1}{p_1{\cdot} p_2}  
\left[ 
 	 \frac{(\varepsilon_3{\cdot} p_5)\,  \varepsilon_4{\cdot}( p_3{+} p_5)}{p_3{\cdot} p_5}
 {+} 	\frac{(\varepsilon_4{\cdot} p_5)\, 	 \varepsilon_3{\cdot} (p_4{+} p_5)}{p_4{\cdot} p_5}
 \right]
 \nn \\[-15pt]
 &
 \hspace{2cm}
+ (2\leftrightarrow 3) + (2\leftrightarrow 4)
\qquad
 \Bigg]\,.
%\end{split}
\end{align}
One can check that the representation of the amplitude satisfies generalized gauge invariance~\cite{Kosmopoulos:2020pcd} for each of the photon lines.  This 
property is defined to be that longitudinal states automatically decouple without the need to impose physical state conditions on other legs.  The net effect
is that state sums simplify as described in Eq.~\eqref{StateSumSimp}.

%================================================
%
\subsubsection{One-loop integrand in scalar QED}
%
%================================================
Equipped with the tree-level building blocks, we follow the generalized-unitarity framework~\cite{Bern:1994zx,Bern:1994cg,Britto:2004nc} to write an ansatz of Feynman-like graphs with associated numerators dictated by the power-counting of scalar QED. At one loop, we can write the full integrand in terms of box, triangle, and bubble topologies. However, sometimes it is convenient to re-absorb contributions from topologies with fewer propagators (`contact terms') into the definition of the box numerator by multiplying the contact terms by appropriate powers of inverse propagators $D_i$. 
\begin{align}
\label{eq:box_skeleton}
 \vcenter{\hbox{\scalebox{1}{\bboxLabel}}} = \int \!\! \frac{d^D\ell}{(2\pi)^D} \frac{1}{D_1 D_2 D_3 D_4}
\end{align}
where the inverse propagators $D_i$ are
\begin{align}
D_1 = (\ell-p_1)^2{-}m^2_1\,, \ 
D_2 = (\ell+p_2)^2{-}m^2_2\,,\ 
D_3 = \ell^2\,, \ 
D_4 = (\ell-q)^2\,.
\end{align}
We employ the graphical notation in which thin lines denote massless propagators and thick lines denote the propagators of the massive particles. (We do not graphically distinguish particles of mass $m_1$ and $m_2$ that are always associated to the external momenta $p_1,p_4$ and $p_2,p_3$, respectively.) As has been advocated in e.g.~Ref.~\cite{Bourjaily:2017wjl}, it is advantageous to directly express the numerator ansatz in terms of a basis of inverse propagators and irreducible elements (absent at one-loop). The power-counting of scalar QED dictates, that the numerator of the box integral should have a mass-scaling like $(p_i\cdot p_j)^2$. To build the ansatz, we write the numerator in terms of the following external Lorentz-products  
\begin{align}
\label{eq:1loop_ansatz_vars}
%\text{variables}=
\{p^2_1, p^2_2, s, -q^2 \} \cup \{ D_1, D_2, D_3, D_4\}\,.
\end{align}
From the power-counting of QED discussed above, we know that our numerator ansatz is quadratic in the variables of Eq.~(\ref{eq:1loop_ansatz_vars}), so that
\begin{align}
\label{eq:box_num_ansatz}
n^{\text{ansatz}}_{\text{box}} = a_1 (p^2_1)^2 + \cdots + a_{63} D_3 D_4 + a_{64} D^2_4 \,.
\end{align}
Every numerator basis element that is proportional to one of the inverse propagators $D_i$ corresponds to a contact term, so that we do not have to list these topologies separately. In a first step, we impose diagram symmetries of the scalar graph in Eq.~(\ref{eq:box_skeleton}) which reduces the number of unknown coefficients $a_i$ and ensures that we only have to determine the numerator for this single graph. 

In order to find the desired integrand, we subsequently compare the cut of the ansatz against the field theory result as determined by the product of tree-level amplitudes summed over the exchanged on-shell states that can cross the cut. Since we are interested in the classical, long-range interactions between the heavy scalar particles mediated by photon exchange, we never need to consider contact (i.e.~short distance) interactions between the scalars. In order to obtain the relevant classical and quantum terms (required for the two-loop eikonal calculation in sections \ref{subsec:eikonal_phase_cons}, \ref{subsec:eikonal_phase_rad}) of the one-loop amplitude, it suffices to match the two-particle bubble-cut
\begin{align}
\label{eq:bubble_cut}
\begin{split}
\hspace{-.5cm}
 \vcenter{\hbox{\scalebox{1}{\bubbleCut}}} \,.
\end{split}
\end{align}
Matching the above field theory cut (i.e.~the product of two Compton amplitudes of Eq.~\eqref{eq:compton_amp}) with our basis ansatz requires relabeling the basic box integrand of Eq.~\eqref{eq:box_skeleton} with the associated numerator \eqref{eq:box_num_ansatz}. Solving the cut equations and dropping all terms proportional to inverse propagators that correspond to pinches of photon lines (which would correspond to short-distance contact interactions that are irrelevant for the classical physics of interest), we find 
\begin{align}
\label{eq:1loop_box_num}
n_{\text{box}} = -4 e^4 q^2_1 q^2_2 \left(4( p_1\cdot p_2)^2 - p^2_2\, D_1 - p^2_1\, D_2 - (D^2_1 + D^2_2) + \frac{1}{4}(D_s-2) D_1 D_2 \right)\,.
\end{align}
The terms proportional to $D_1 D_2$ correspond to a bubble integral that is only relevant for the quantum subtraction for the two-loop eikonal analysis. The triangle topologies are included through the numerators proportional to $D_1$ and $D_2$, respectively. The result \eqref{eq:1loop_box_num} is written in terms of the state-counting parameter $D_s = \eta^\mu_{\ \, \mu}$ and we will work in the scheme where we set $D_s=4$ and write the one-loop amplitude as a sum of a box and cross-box,
\begin{align}
\label{eq:one-loop-QED_integrand} 
\mathcal{A}^{(1)}_4(p_1,p_2,p_3,p_4)=  
   n_{\text{box}} \hspace{-.5cm}\vcenter{\hbox{\scalebox{1}{ \bbox}}} 
+ \quad n_{\text{x-box}}\hspace{-.5cm}\vcenter{\hbox{\scalebox{1}{ \bxbox}}}\,,
\end{align}
where the numerator for the second box, $n_{\text{x-box}}$, is obtained from Eq.~(\ref{eq:1loop_box_num}) by crossing $p_1\leftrightarrow p_4$ and (the immaterial) $q_1 \to - q_1$.

%================================================
%
\subsubsection{Two-loop integrand in scalar QED}
%
%================================================

\begin{figure}[ht!]
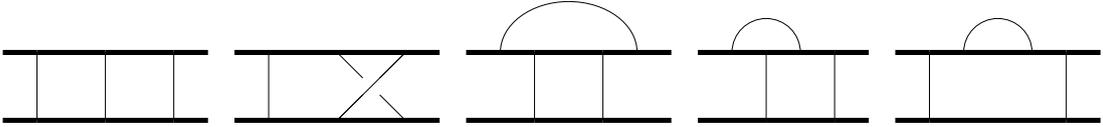

    \centering
    \scalebox{0.9}{\graphone}\!\!\!
    \scalebox{0.9}{\graphfour}\!\!\!
    \scalebox{0.9}{\graphnine}\!\!\!
    \scalebox{0.9}{\grapheleven}\!\!\!
    \scalebox{0.9}{\graphthirteen}\!\!\!
    \caption{Diagrams with cubic vertices relevant for classical $\calO(\alpha^3)$ observables in scalar QED. The first two graphs appear in the conservative sector and the three ``mushroom'' graphs are only relevant for radiative effects. The diagrams split into different gauge-invariant subsectors. The III and IX graphs (corresponding to the first two diagrams) are proportional to $q^3_1 q^3_2$, whereas the mushroom graphs are proportional to $q^2_1 q^4_2$ (and  $q^4_1 q^2_2$ for the flipped graphs not explicitly drawn).}
    \label{fig:cubic_graphs}
\end{figure}
\begin{figure}[b!]
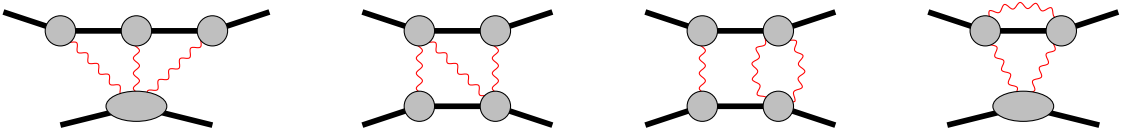

    \centering
    \raisebox{7pt}{\scalebox{1}{\wcut}}
    \qquad
    \raisebox{7pt}{\scalebox{1}{\ncut}}
    \qquad    
    \raisebox{7pt}{\scalebox{1}{\boxbubcut}}
    \qquad
    \raisebox{7pt}{\scalebox{1}{\threecomptoncut}}
    \caption{Spanning set of unitarity cuts relevant for the classical dynamics at $\calO(\alpha^3)$.}
    \label{fig:spanningcuts}
\end{figure}

At two-loops, the cut construction proceeds in a fashion similar to the previous one-loop analysis. We start from the relevant graphs with cubic vertices, summarized in Fig.~\ref{fig:cubic_graphs} and write down a numerator ansatz for each diagram consistent with QED power counting: each trivalent vertex is associated with one power of momentum in the numerator. We then impose the diagram symmetries of the graphs on the respective numerator ansatz. To determine the pieces of the scalar QED two-loop amplitudes relevant for classical physics, both in the eikonal and KMOC approach, we fix the numerators by matching against the spanning sets of cuts depicted in Fig.~\ref{fig:spanningcuts} built out of products of the tree-level amplitudes from section \ref{subsubsec:treeAmps}.

%================================================
%
\subsection{Soft and potential region expansion, IBP, and differential equations}
\label{subsec:integration}
%
%================================================
%
With the relevant one- and two-loop integrands at hand, we
directly follow similar computations that have been performed in the
gravitational
setting~\cite{Bern:2019crd,Bern:2019nnu,Parra-Martinez:2020dzs,Herrmann:2021lqe,Herrmann:2021tct}. 
In particular, we expand the scalar QED integrands of
Subsection \ref{subsec:integrands} in either the soft or potential
region. We take advantage of the technology developed in
Refs.~\cite{Parra-Martinez:2020dzs,Herrmann:2021tct} and import the
explicit values of all soft master integrals supplied in the ancillary
files of Ref.~\cite{Herrmann:2021tct} (for an alternative computation
of the soft master integrals, see Ref.~\cite{DiVecchia:2021bdo}). We
will not review these steps in any detail and refer the interested
reader to the original references. Briefly, there are two main steps
involved in order to obtain integrated results. The first is to start
from the initial integrands and expand them in small $|q|$ which
leaves graviton propagators unaffected and linearizes (eikonalizes)
all matter propagators. This step is related to the kinematic
discussion in Subsection~\ref{subsec:kinematics}. To obtain conservative
physics (i.e. the potential region in the language of the method of
regions~\cite{Beneke:1997zp}), one further expands the propagators in
a formal small velocity parameter $v$, where the graviton energy
component is suppressed by an extra factor of $v$ compared to the
spatial components. This signals instantaneous interactions in the
Fourier-conjugate time domain. In either case, upon expanding the
integrand in the desired kinematic region of interest, one
subsequently reduces all resulting integrals to a basic set of
so-called master integrals.  One can then solve for the values of the
remaining master integrals using modern differential equation
methods~\cite{Kotikov:1990kg, Bern:1992em, Remiddi:1997ny,
  Gehrmann:1999as,  Henn:2013pwa, Henn:2014qga}. For the KMOC setup, besides the virtual two-loop
integrals, one also needs to have access to certain cut-integrals
whose computation was significantly simplified using reverse
unitarity~\cite{Anastasiou:2002yz,Anastasiou:2002qz,Anastasiou:2003yy}---a
well-known tool from collider physics computations (see
e.g.~Ref.~\cite{Anastasiou:2015yha}). In this work, we leverage the
fact that all relevant integrals have been computed to the order
required for our work and we essentially re-use the integration
pipeline that already has been successfully implemented for GR both in
the soft~\cite{Herrmann:2021lqe,Herrmann:2021tct} and potential
region~\cite{Bern:2019crd,Bern:2019nnu,Parra-Martinez:2020dzs}.

%================================================
%
\subsection{Soft radial action and master integral subtraction of classically divergent terms}
\label{subsec:soft_radial_action_subtraction_scheme}
%
%================================================
%
In the discussion so far, we mainly focused on the computation of scattering amplitudes in the so-called \emph{soft region} where we only assume that the momentum transfer $-q^2\ll m^2_i,\, s $ is much smaller than the masses or the energy of the scattering process. As we will explain in most of the remainder of our work, these amplitudes then enter either the eikonal formalism or the KMOC framework in order to extract the relevant classical observables from the scattering amplitudes (or certain combinations of scattering amplitudes). Crucially, inherent in both the eikonal or the KMOC formalism is the fact that amplitudes not only have classical contributions but, at higher orders in perturbation theory, also involve classically divergent (`super-classical') terms that are more singular and have to cancel for classically well-defined observables.\footnote{Similar statements also hold in the EFT matching approach for the conservative two-body problem where classically divergent terms correspond to iterations of lower-order potentials, see e.g.~Ref.~\cite{Bern:2021dqo} and references therein.} In light of this discussion, one might wonder, whether or not there exists a formalism that directly targets the classical terms directly, without the need to compute the classically divergent terms directly and avoid problems at higher perturbative order of the form $\hbar/\hbar$ where the classically divergent terms interfere with quantum contributions to yield a naively classical result. In the conservative sector,~\cite{Bern:2021dqo} advocated for an EFT based approach with a particular subtraction scheme of classical iterations that allowed the definition and computation of the \emph{radial action} $I_r$ directly from the relation (Eq.~(2) of~Ref.~\cite{Bern:2021dqo})
\begin{align}
  \imath \, \calA(\mathbf{q}) =  \int\limits_J \left[e^{\imath\, I_r(J)} -1\right]\,,
  \label{eq:radialExp}
\end{align}
which looks very similar to the eikonal exponentiation, but differs in important details~\cite{Bern:2021dqo}.

As is well-known from classical physics, see e.g.~Ref.~\cite{Landau:1976mech}, the radial action is an important quantity, associated with the classical Hamilton-Jacobi equation for the system, from which to extract relevant classical observables. (See e.g.\!~recent work in the probe-limit~\cite{Kol:2021jjc}.) Subsequently, the authors of Ref.~\cite{Damgaard:2021ipf} argued for a related exponential representation of the S-matrix, $S = e^{\imath\, \hat{N}}$, where one calculates its phase, $\hat{N}$, from which one can extract classical observables (including radiation) due to a relation to the WKB approximation. This has been assembled into a computational framework in Ref.~\cite{Bjerrum-Bohr:2021wwt} where the radial action has been tied to certain \emph{velocity cuts} (see also Ref.~\cite{Brandhuber:2021eyq} for related work). 

Similarly to the eikonal approach, the radial action $I_r(J)$ (and also the phase of the S-matrix $\hat{N}$) has a perturbative expansion $I_r(J) =  I^{(0)}_r(J) + I^{(1)}_r(J) +  I^{(2)}_r(J) +\cdots $ which leads to classical iterations from expanding the exponential to higher orders in the small coupling constant. In this section, we describe an approach to calculate the classical radial action at a given loop order $I^{(L)}_r(J) $ without the need of explicit classically divergent subtractions. Our new setup is closest in philosophy to that of Ref.~\cite{Bern:2021dqo}. In particular, we are going to find a prescription that is implemented at the level of boundary conditions for soft master integrals which manifestly eliminates classically divergent contributions, and allows us to define a soft radial action. By explicit calculation, we show that our prescription works up to two-loop order. Its study to higher orders in perturbation theory is an interesting open problem left to future work. 

Formally, the perturbative expansion of Eq.~\eqref{eq:radialExp} to a
given loop-order $L$ still entails the subtraction of nontrivial
exponentiation terms involving two or more lower-order
\emph{iterations} $I^{(L'{<}L)}_r$ to isolate $I^{(L)}_r$ itself. The
aforementioned reference~\cite{Bern:2021dqo} writes such iterations as
$(D{-}1)$-dimensional integrals with linearized
propagators. Meanwhile, the boundary conditions for soft master
integrals near the static limit, when considering only the potential
region, are also given by $(D{-}1)$-dimensional integrals where the
``divergent'' part involves linearized
propagators~\cite{Parra-Martinez:2020dzs}. Therefore, our strategy is
to drop integrals with linearized propagators from the boundary
conditions and solving the differential equations for the soft
integrals subject to the modified ``finite'' boundary conditions.

\begin{figure}
  \centering
  \includegraphics[width=0.6\textwidth]{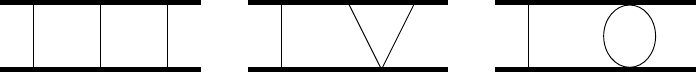}
  \caption{Master integral topologies related to iteration of radial action or EFT potential, responsible for classically divergent terms in the amplitude. The IX diagram corresponding to the second one in Fig.~\ref{fig:cubic_graphs} does not appear since it is not divergent in the potential region, so receives no subtraction in a conservative calculation. The subtraction in the soft region is designed to be identical to the subtraction in the potential region---this works as long as we restrict to the real part where the only divergences comes from $\calA_{{\rm tree}}^3$.}
  \label{fig:iterationMasters}
\end{figure}

The set of diagram topologies for master integrals related to the above ``iterations'' is shown in Fig.~\ref{fig:iterationMasters}.
It turns out that only the first graph in Fig.~\ref{fig:iterationMasters}, called the III diagram, is relevant for the classically divergent terms in the real part of the two-loop amplitude. In the notation of Eq.~(4.70) of~\cite{Parra-Martinez:2020dzs}, the top-level soft master integral for the III diagram is
\begin{equation}
  f_{\text{III}, 7} = \epsilon^4 (y^2-1) (-q^2) G_{1,1,1,1, 1,1,1, 0,0} \, ,
\end{equation}
where $G_{1,1,1,1, 1,1,1, 0,0}$ is the scalar double-box integral with a unit numerator, multiplied by additional prefactors are included. In the Euclidean region  we have ${-}1<x<0$, and $y = (1+x^2) / (2x) < -1$, with the value of the master integral given in Ref.~\cite{Parra-Martinez:2020dzs} as
\begin{align}
  f_{\text{III}, 7} &= -\frac 1 2 \epsilon^2 \log^2 (-x) + \frac 1 {12} \epsilon^3
  \big[-24 \text{Li}_3(x)-24 \text{Li}_3(-x)+12 \text{Li}_2(x) \log (-x) \nonumber \\
  &\qquad +12 \text{Li}_2(-x) \log (-x)+2 \log ^3(-x)+\pi ^2 \log (-x)+6 \zeta_3 \big] + \calO(\epsilon^4) \, .
\end{align}
By analytic continuation, the value of integral in the Lorentzian region $0<x<1$, $y > 1$ is obtained from the above formula with an infinitesimal positive imaginary part given to $y$, or an infinitesimal negative imaginary part given to $x$,
\begin{equation}
  f_{\text{III}, 7} = -\frac 1 2 \epsilon^2 \left[\log (x) + i \pi \right]^2 + \calO(\epsilon^3) \, , \label{eq:resultSoft}
\end{equation}
where we omitted the analytic continuation result at $\calO(\epsilon^3)$ which is needed for calculating the amplitude but not relevant for the discussion here.

When evaluated in the potential region, the integral cannot be analytically continued between positive and negative values of $x$, and we directly give the value of the integral for the Lorentzian region $0<x<1$,  $y = (1{+}x^2) / (2x) > 1$,
\begin{equation}
  f_{\text{III}, 7}^{(\text{p})} = \frac{\epsilon^2 \pi^2} {2} + 0 \cdot \epsilon^3 + \calO(\epsilon^4) \, . \label{eq:resultPot}
\end{equation}
Compared with Eq.~\eqref{eq:resultSoft}, in the $\calO(\epsilon^2)$ term only the $\pi^2$ part survives, and the  $\calO(\epsilon^3)$ term has become genuinely zero, and not an omission.

There are 7 pure master integrals for the soft-expanded III diagram in the even-in-$|q|$ sector. In the potential region, the boundary condition near the static limit $y=1$ is given in terms of $(3-2\epsilon)$-dimensional integrals in Eqs.~(A.9)-(A.11) of Ref.~\cite{Parra-Martinez:2020dzs}. In particular, for the top-level master integral near the static limit,
\begin{equation}
  f_{\RT,7}^{\pot} \big|_{\relfbar=1} = \pi \epsilon ^4  (-q^2) \!\int\!\frac{\mathrm{d}^{D-1}\vect{\ell}_1\mathrm{d}^{D-1}\vect{\ell}_2 \left( e^{\EulerGamma\epsilon}\right)^2}
  {(\imath\pi^{(D-1)/2})^2 \, \vect{\ell}_1^{\,2}\vect{\ell}_2^{\,2}(\vect{\ell}_1+\vect{\ell}_2-\vect{q})^2(2\ell_1^z)(-2\ell_2^z)} \, ,
  \label{eq:IIIboundary}
\end{equation}
which is precisely of the form of a $(D-1)$-dimensional integral involving linearized propagators. The superscript $(p)$ in the equation above indicates that only the potential region is considered. Now we implement a subtraction scheme similar to the 4PM potential-region calculation, by dropping $(3-2\epsilon)$-dimensional integrals involving linearized propagators arising from iterations of lower-loop potentials. This is equivalent to keeping Eqs.~(A.9) and (A.10) in the reference while changing the RHS of Eq.~(A.11), reproduced in Eq.~\eqref{eq:IIIboundary}, to zero. Solving differential equations with the altered boundary conditions, Eq.~\eqref{eq:resultPot} becomes
\begin{equation}
  f_{\text{III}, 7}^{\text{(p),subtracted}} =  0 \cdot \epsilon^2 + 0 \cdot \epsilon^3 + \calO(\epsilon^4) \, , \label{eq:resultPotSubtracted}
\end{equation}
i.e. vanishes until $\calO(\epsilon^4)$, which is beyond the order of $\epsilon$ needed in the classical calculation.

The boundary values for the master integrals in the soft region are decomposed into the sum of their values in the potential region and their soft-region corrections. We perform the same subtraction for the potential-region part, while keeping the soft-region corrections unchanged. Since the solutions to the homogeneous system of differential equations have multi-linear dependence on the boundary conditions, we have
\begin{equation}
  f_{\text{III}, 7}^{\text{subtracted}} = f_{\text{III}, 7} + \left( f_{\text{III}, 7}^{\text{(p),subtracted}} - f_{\text{III}, 7}^{\pot} \right) \, .
\end{equation}
Explicitly,
\begin{equation}
  f_{\text{III}, 7}^{\text{subtracted}} = -\frac 1 2 \epsilon^2 \left[\log (x)^2 + 2 i \pi \log(x) \right] + \calO(\epsilon^3) \, , \label{eq:resultSoftSubtracted}
\end{equation}
where in the $\calO(\epsilon^2)$ term, the $\pi^2$ part has been removed, and the omitted $\calO(\epsilon^3)$ term (also needed for assembling the amplitude) is completely unchanged.

If we calculate the two-loop amplitude in the soft expansion using the subtracted value Eq.~\eqref{eq:resultSoftSubtracted} for the top-level III master integral and unsubtracted original results for all other master integrals, the real part of the result directly gives the radial action after Fourier transform.

%================================================
%
\subsection{KMOC framework for classical conservative and radiative observables}
\label{subsec:KMOC_general}
%
%================================================
%
In this part of our review, we schematically recall aspects of the KMOC framework~\cite{Kosower:2018adc} as presented in Ref.~\cite{Herrmann:2021tct}. We only introduce the relevant final formulae. For further details, the interested reader is encouraged to consult Refs.~\cite{Kosower:2018adc} or~\cite{Herrmann:2021tct} directly. 

In the KMOC~\cite{Kosower:2018adc} approach one first sets up a quantum mechanical Gedanken experiment for the scattering of two wavepackets representing massive particles from which the classical limit is carefully taken in order to extract the classical observables of interest. In the quantum setup, we can measure the change of some observable $\Delta O$ (corresponding to some quantum operator $\mathbb{O}$) following the time-evolution of states from the asymptotic past to the asymptotic future. In the asymptotic past, the wavepackets are represented by $|{\text{in}} \rangle$, an \emph{in} quantum state constructed from the superposition of two-particle momentum eigenstates $|p_1,p_2\rangle_{\text{in}}$ with wavefunctions $\phi_i(p_i)$. For the case of interest to us, these states are well separated by an impact parameter $b^\mu$\,\footnote{The impact parameter $b^\mu$ is distinct from the eikonal impact parameter $b_e$ that will appear later.}
\begin{align}
\label{eq:wavefuncs}
 | {\text{in}} \rangle = \int \mathrm{d}\Phi_2(p_1,p_2) \, 
					\phi_1(p_1)\phi_2(p_2) e^{\imath \, b\cdot p_1/\hbar} \, |p_1,p_2\rangle_{\text{in}}\,.
\end{align}
Such an \emph{in} state will evolve to an \emph{out} state in the asymptotic future, $|\text{out}\rangle$, that might contain additional particles created during the interaction. $\Delta O$ is obtained by evaluating the difference of the expectation value of  $\mathbb{O}$ between \emph{in} and \emph{out} states 
\begin{equation}
\label{eq:kmoc_start}
 \Delta O = \langle {\rm out} | \mathbb{O} | {\rm out} \rangle  - \langle {\rm in} | \mathbb{O} | {\rm in} \rangle\,.
\end{equation}
In quantum mechanics, the \emph{out} states are related to the \emph{in} states by the time evolution operator, i.e.\ the S-matrix: $|\text{out}\rangle = S |\text{in} \rangle$ and we can write 
\begin{equation}
 \Delta O = \imath \int  \mathrm{d}\Phi_4(p_1,\cdots\!,p_4)  \, 
					\phi_1(p_1)\phi_2(p_2) \phi^*_2(p_3)\phi^*_1(p_4) \, \hat\delta^{(D)}(\textstyle\sum_i p_i)\,  e^{\imath \, b\cdot (p_1+p_4)/\hbar} \, \, {\cal I}_O
		 \,,
\end{equation}
where we follow the same conventions as Ref.~\cite{Herrmann:2021tct} for the phase-space $d\Phi_n$ and $\hat{\delta}$ factors. The kernel $\calI_O$ is related to the matrix elements via
\begin{equation}
 \widetilde {\cal I}_O \equiv \widehat\delta^{(D)}(\textstyle\sum p_i)\,\calI_O
    =-\imath\, \langle p_4,p_3|S^\dagger[\mathbb{O},S]|p_1,p_2\rangle\,.
\end{equation}
To arrive at this expression, we have used the unitarity of the S-matrix, $S^\dagger S=1$. Following~\cite{Kosower:2018adc}, $\widetilde {\cal I}_O$ can be related to scattering amplitudes by writing $S=1+\imath T$ such that
\begin{equation}
\label{eq:kernel_matrix_element_def}
 \widetilde {\cal I}_O =
        \widetilde {\cal I}_{O,\,{\rm v}} 
        + \widetilde {\cal I}_{O,\,{\rm r}} 
    = \,\langle p_4,p_3|[\mathbb{O},T]|p_1,p_2\rangle 
        -i \, \langle p_4,p_3|T^\dagger[\mathbb{O},T]|p_1,p_2\rangle\,.
\end{equation}
We conveniently separated the kernel $\widetilde {\cal I}_O$ into two contributions $\widetilde {\cal I}_{O, {\rm v}}$ and $\widetilde {\cal I}_{O,\, {\rm r}}$, that we preemptively call \emph{virtual} and \emph{real}, respectively. This nomenclature becomes apparent when one evaluates the expectation values: The \emph{virtual} part of the result is
\begin{align}
\label{eq:kmocvirtual}
    {\cal I}_{O,\,{\rm v}} 
    = 
    \Delta \mathbb{O} \big[{\cal A}(p_1,p_2,p_3,p_4)\big]
    =
    \Delta \mathbb{O} \left[
    \raisebox{-32pt}{
      \!\!\!\scalebox{0.8}{\kmocvirtual}
  } \right] ,
\end{align}
where $\Delta \mathbb{O}$ acts on the scattering amplitude $\calA$. In the \emph{real} kernel
\begin{align}
     {\cal I}_{O,\,{\rm r}}
     &=-\imath\sum_X \int \mathrm{d}\widetilde\Phi_{2+|X|} \,\,  
     \Delta \mathbb{O} \hspace{-.4cm}
     \raisebox{-32pt}{
     \scalebox{0.8}{
       \kmocreal{$r_2$}{$r_1$}}} \,,
    \label{eq:kmocreal}
\end{align}
$\Delta \mathbb{O}$ only acts on the amplitude on the left of the unitarity cut which was introduced by the insertion of a complete sum over states in Eq.~\eqref{eq:kernel_matrix_element_def}. 

While the KMOC formalism can be applied fully quantum mechanically, we are interested in \emph{classical} observables. The classical limit corresponds to the regime where the Compton wavelength of the external particles is the smallest length scale in the problem. KMOC carefully analyze this limit with the net result that all computations ultimately reduce to those of simple plane-wave scattering. In the classical limit, the wavepackets sharply peak about their classical values of the momenta which leads to the appearance of on-shell delta functions and one arrives at a compact expression for the classical change of the observable $O$ in terms of the (asymptotic) impact parameter $b^\mu$, conjugate to the small momentum transfer $q^\mu =p^\mu_1+p^\mu_4\sim \calO(\hbar)$,
\begin{equation}
\label{eq:KMOC_DObs}
    \Delta O = \imath \int \hat{\mathrm{d}}^D q \,
    \hat\delta(-2 p_1 \cdot q)\, \hat\delta(2 p_2 \cdot q) e^{\imath b\cdot q}\,
    \left(
    \mathcal{I}_{O, {\rm v}} + \mathcal{I}_{O,{\rm r}}
    \right)\,.
\end{equation}
The KMOC analysis suggests that we ought to focus on kinematic regions where the massive particle momenta $p_i$ are large and scale like $\calO(1)$ in the classical counting and the four-momentum transfer $q$, as well as graviton loop variables that we will denote by $\ell_i$ below, scale like $\mathcal{O}(\hbar)$. In the effective field theory context, employing terminology from the ``method of regions''~\cite{Beneke:1997zp}, the classical $\hbar$ expansion is therefore equivalent to the so-called soft expansion.

Furthermore, we also expand scattering amplitudes in the coupling $e$ or equivalently $\alpha$
\begin{equation}
   \calA = \calA^{(0)} +\calA^{(1)} +\calA^{(2)} + \cdots =  \!\!
    \raisebox{-22pt}{
      \scalebox{0.8}{
	\kmocvirtualtreenolab	
      }} +\!\!  
      \raisebox{-22pt}{
      \scalebox{0.8}{
	\kmocvirtualnlonolab	
      }} +\!\!
       \raisebox{-22pt}{
      \scalebox{0.8}{
	\kmocvirtualnnlonolab
      }} + \cdots\,,
\end{equation}
where the $L$-loop amplitude is $\calO(\alpha^{L+1})$. The observables (and kernels) have analogous expansions
\begin{align}
\label{eq:KMOC_obs_pert_expansion_schematic}
    \Delta O&= \Delta O^{(0)} + \Delta O^{(1)}  + \Delta O^{(2)}  + \cdots\,,\\
    \calI_{O} &=\calI^{(0)}_{O} +\calI^{(1)}_{O} + \calI^{(2)}_{O} + \cdots\,.
\end{align}
%

%=======================================================================
%\vspace{-0pt}
\subsubsection{Electromagnetic Impulse}
\label{subsec:impulse}
%\vspace{-0pt}
%=======================================================================
%
In this work, we discuss two observables relevant to classical electrodynamic scattering. The first is the impulse, $\Delta p_i^{\mu}$, which is defined as the total change in momentum of one of the particles during the collision. In the KMOC setup this is encoded by the appropriate quantum momentum operator~$\mathbb{P}_i$, which is measured asymptotically far from the collision region as follows
\begin{align}
 \Delta p^\mu_1 = 
 \langle \text{in} | S^\dagger \mathbb{P}^{\mu}_1 S | \text{in} \rangle - \langle \text{in} | \mathbb{P}^\mu_1 |\text{in} \rangle\,.
\end{align}
As summarized above, in the classical limit, this is simply a Fourier transform of the impulse kernel $\calI_{p_1}^\mu$ from momentum transfer $q$ to impact-parameter space $b$
\begin{align}
    \label{eq:classical_impulse}
    \Delta p^\mu_1 & = \imath \int \hat{\mathrm{d}}^Dq 
    \,\hat{\delta}(-2p_1\cdot q)\,\hat{\delta}(2p_2\cdot q)\, e^{\imath b\cdot q}
    \, \calI^\mu_{p_1} \,,
\end{align}
which is separated into virtual and real contributions, given in terms of the amplitude as
\begin{equation}
    \calI_{p_1,\, \text{v}} 
     = 
    q^\mu \hspace{-.4cm}\raisebox{-32pt}{\scalebox{0.8}{\kmocvirtual}}\,, \hskip .4cm  
    \calI_{p_1,\, \text{r}} 
     = - \imath \sum_X \int \mathrm{d}\widetilde{\Phi}_{2+|X|}\  \ell_1^\mu
    \hspace{-.4cm}
     \raisebox{-32pt}{\scalebox{0.8}{\kmocreal{$\ell_2-p_2$}{$\ell_1-p_1$}}}\,,
     \label{eq:classical_impulse_kernel}
\end{equation}
where the numerator insertions $q^\mu$ and $\ell_1$ arise from the measurement function $\Delta \mathbb{P}^\mu_1$ acting on the respective amplitudes, which extracts the momentum change of particle 1. Note that relative to Eq.~\eqref{eq:kmocreal}, we have changed variables in the real contribution by shifting the massive intermediate momenta $r_i = -p_i + \ell_i$, so that all $\ell_i$ are small, $\calO(\hbar)$, in the classical expansion. The impulse on particle $2$ can be obtained by simple relabelling. 

In the following, we often decompose the total impulse into its transverse, $\Delta p_\perp$, and longitudinal, $\Delta p_u$, components
\begin{equation}
\label{eq:impulse_decomp}
    \Delta p^\mu = \Delta p_\perp^\mu + \Delta p_u^\mu\,,
\end{equation}
where $ u_i {\cdot} \Delta p_\perp {=} 0$ and $ q {\cdot} \Delta p_u {=} 0$. The respective kernels get decomposed in a similar fashion
\begin{equation}
\label{eq:impulse_kernel_projection}
  \mathcal{I}^\mu_{p_1} =   {\cal I}_\perp \,q^\mu + 
                            \sum_{i=1,2}{\cal I}_{u_i} \, \check{u}^{\mu}_i \,.
\end{equation}
We define \emph{dual} four-velocities,
\begin{equation}
\label{eq:dual_ui}
   \check{u}^{\mu}_1 = \frac{y u_2^\mu - u_1^\mu}{y^2-1}, \qquad 
   \check{u}^{\mu}_2 = \frac{y u_1^\mu - u_2^\mu}{y^2-1}\,,
\end{equation}
which satisfy $u_i \cdot \check{u}_j = \delta_{ij}$ and remain orthogonal to the momentum transfer $q$. Decomposing the loop momentum dependent impulse numerator
\begin{align}
\label{eq:ell1_decomp}
    \ell^\mu_1  = \frac{\ell_1\cdot q}{q^2} q^\mu 
                + (\ell_1\cdot u_1)\, \check{u}^{\mu}_1
                + (\ell_1\cdot u_2)\, \check{u}^{\mu}_2\,,
\end{align}
exposes that only the transverse part of the impulse has a \emph{virtual} contribution 
\begin{align}
\label{eq:impulse_kernel_perp}
\begin{split}
    {\cal I}_\perp =&
    \hspace{-.3cm}
    \raisebox{-32pt}{\scalebox{0.8}{\kmocvirtual}}
     - \hspace{.3cm} \imath\ \sum_X \int \mathrm{d}\widetilde{\Phi}_{2+|X|} \,\, \frac{\ell_1\cdot q}{q^2} 
    \hspace{-.3cm}
    \raisebox{-35pt}{\scalebox{0.8}{\kmocreal{$\ell_2-p_2$}{$\ell_1-p_1$}}} \,,
\end{split}
\end{align}
whereas the longitudinal part \emph{only} receives contributions from the unitarity cut terms
\begin{align}
\label{eq:impulse_kernel_long}
\begin{split}
   {\cal I}_{u_i} =&-\imath \sum_X \int \mathrm{d}\widetilde{\Phi}_{2+|X|}
   \,\,\ell_1\cdot u_i 
   \hspace{-.3cm}
   \raisebox{-35pt}{\scalebox{0.8}{\kmocreal{$\ell_2-p_2$}{$\ell_1-p_1$}}}\,.
\end{split}
\end{align}
Loop amplitudes generically have real and imaginary parts, so one might wonder how all classical observables end up real-valued, and how various terms in the KMOC setup combine to serve this purpose. Keeping track of factors of `$\imath$', it turns out that the transverse KMOC kernels need to be purely real to yield a real result after the final Fourier transform (Eq.~\eqref{eq:KMOC_DObs}), whereas the longitudinal kernels are purely imaginary. The reality properties of various quantities has been argued abstractly in terms of unitarity cutting rules in Ref.~\cite{Herrmann:2021tct} that later appeared in a slightly different context in Ref.~\cite{Damgaard:2021ipf}. Indeed, it will serve as a nontrivial check of our computation, that all imaginary contributions to the classical observables cancel.

%=======================================================================
\vspace{-0pt}
\subsubsection{Radiated momentum}
\label{subsec:radiated_momentum}\vspace{-0pt}
%=======================================================================
%
Another observable of interest is the total radiated momentum $\Delta R^\mu$ carried away in the form of electromagnetic waves during the scattering of two heavy charged objects. This observable is defined by measuring the momentum operator $\mathbb{R}^\mu$ of the emitted \emph{messenger particles}, here photons. As explained in Ref.~\cite{Kosower:2018adc}, this observable only receives \emph{real} contributions and its respective kernel is
\begin{align}
    \mathcal{I}_{R, {\rm r}}^\mu 
    &=-\imath \sum_X 
    \int\!\mathrm{\mathrm{d}}\widetilde{\Phi}_{2+X}\  \ell^{\mu}_X 
    \raisebox{-32pt}{\scalebox{0.8}{\kmocreal{$\ell_2-p_2$}{$\ell_1-p_1$}}}\,.
    \label{eq:KMOC_kernel_rad}
\end{align}
Like Eqs.~(\ref{eq:classical_impulse}) and (\ref{eq:classical_impulse_kernel}),  Eq.~\eqref{eq:KMOC_kernel_rad} is valid  beyond perturbation theory, however, for explicit calculations we expand it perturbatively in $\alpha$. The first contribution to $\Delta R^\mu$ (obtained from Eq.~\eqref{eq:KMOC_kernel_rad} by performing the Fourier transform to impact-parameter space \eqref{eq:KMOC_DObs}) arises at $\calO(\alpha^3)$. This can be understood from the fact that Bremsstrahlung of finite energy photons only arises once one heavy charged particle is slightly deflected due to its electromagnetic interaction with the other massive charged object.

\medskip

The impulse and radiated momentum are not completely independent observables. As already pointed out in Ref.~\cite{Kosower:2018adc}, their relation goes to the heart of one of the difficulties in traditional approaches to classical field theory with point sources. Two particles that scatter in e.g.~classical electrodynamics exchange momentum via their interaction with the electromagnetic field. In the classical context, this is described by the Lorentz force. However, the energy or the momentum lost by the point particles to radiation is not accounted for by the Lorentz force. In the classical setup, momentum conservation is restored by including the additional Abraham--Lorentz--Dirac (ALD) force~\cite{lorentz:1892theorie,abraham:1912theorie,Dirac:1938a,Landau:1980Classical} and e.g.~Refs.~\cite{Galley:2010es,Forgacs:2012qt,Birnholtz:2013nta} for some recent treatments. In the classical context, the inclusion of the ALD force term is associated with its own problems. In particular, it leads to issues of runaway solutions and causality violations in the description of point charges in classical EM.  In contrast, in the quantum-mechanical description of charged-particle scattering these issues are absent and we will see that our results for the classical two-body dynamics in electrodynamics will conserve energy and momentum automatically. For more discussions on this point, see Sections 3.5 and 5.4 of Ref.~\cite{Kosower:2018adc} for a KMOC integrand level discussion of this point. The main conclusion of their analysis is that the ALD radiation reaction is automatically included in the KMOC setup via the cut-contribution where an on-shell radiation photon is exchanged. This contribution first appears at two-loop order, or $\calO(\alpha^3)$ and was compared to the classical ALD force computation in Section 6.3 of Ref.~\cite{Kosower:2018adc}.   Our agreement below with the $\mathcal O(\alpha^3)$ scattering angle computed 
using the ALD force~\cite{{Saketh:2021sri}} explicitly affirms these conclusions.

%==============================================================
%
%\newpage
\section{Conservative dynamics at $\mathcal{O}(\alpha^3)$}
\label{sec:conservative}
%
%==============================================================

%================================================
%
\subsection{Conservative scattering amplitudes}
\label{subsec:amps_cons}
%
%================================================
In this section we give the results for the tree level, one-loop and two-loop scattering amplitudes in scalar QED in the conservative sector. We combine the integrands derived in section \ref{subsec:integrands} with the integration techniques sketched in section \ref{subsec:integration}. A detailed account of the potential region integrals is found in Refs.~\cite{Bern:2019crd,Bern:2019nnu,Parra-Martinez:2020dzs}. All potential region $L$-loop amplitudes will henceforth be denoted by $\mathcal{A}^{(L)}_{4,\pot}$.

The tree-level amplitude can be obtained from Eq.~\eqref{eq:4s_tree_amp_soft_vars} by switching from the soft variables $y$, and $\mb_1$ to $\sigma$ and $m_i$ via the relations from Subsection \ref{subsec:kinematics} 
\begin{equation}
  \mathcal{A}_{4,\pot}^{(0)} = -(4\pi\alpha\,  q_1q_2)\frac{4 m_1m_2\sigma}{-q^2}\,,
  \label{eq:treeAmplitude}
\end{equation}
where we use the charge normalization $\alpha = e^2/4\pi$ and denote the multiple of the elementary charge $e$ of the massive objects by $q_i$.
  
The conservative one-loop amplitude can be obtained from the integrand in Eq.~\eqref{eq:one-loop-QED_integrand} by taking the small-$q$ expansion of the covariant integrand. Upon reducing the $q$-expanded integrals to a basis of master integrals, it is given in terms of the sum of the box and crossed box integrals, as well as two triangle integrals evaluated in the potential region. (Bubble integrals are zero in the potential region.)
\begin{equation}
\mathcal{A}_{4,\pot}^{(1)} = -\imath
c_{\rm II}\, 
\big(\, I^{\pot}_{\rm II} + I^{\pot}_{\rm X}\,\big)
-\imath c_{\triangleleft}\, I^{\pot}_{\triangleleft}
-\imath c_{\triangleright}\, I^{\pot}_{\triangleright}.
  \label{eq:1lintegrandpot}
\end{equation}
The coefficients are
\begin{align}
c_{\rm II} = (16 \pi\,  \alpha q_1 q_2\, m_1 m_2 \sigma)^2 \,,
\quad
c_{\triangleleft} = -\frac{1}{2}(16 \pi\,  \alpha q_1 q_2\, m_1)^2 \,, 
\quad
c_{\triangleright} = -\frac{1}{2}(16 \pi\,  \alpha q_1 q_2\, m_2)^2\,. 
\end{align}
Inserting the explicit values of the integrals in the potential region (see e.g.~Ref.~\cite{Parra-Martinez:2020dzs}) yields
\begin{align}
\begin{split}
\mathcal{A}_{4,\pot}^{(1)} =(4\pi\alpha\,  q_1q_2)^2\frac{1}{(4\pi)^2} \left(\frac{-q^2}{\bar\mu^2}\right)^{-\epsilon}
\bigg\{&
 \frac{1}{(-q^2)} \frac {\imath \pi (\sigma^2 m_1m_2)} {2 \sqrt{\relf^2-1}}
\frac{ e^{\epsilon \EulerGamma} \Gamma (-\epsilon )^2 \Gamma (1 + \epsilon)}{\Gamma (-2 \epsilon )} 
\\
& \hspace{-2cm}+ 8\frac{1}{\sqrtmQSq} \sqrt{\pi} (m_1{+}m_2)
\frac{ e^{\epsilon \EulerGamma} \Gamma \left(\frac{1}{2}-\epsilon \right)^2 \Gamma \left(\epsilon+\frac{1}{2}\right)} { 2\Gamma(1-2 \epsilon )} 
\\
& \hspace{-2cm} -  \epsilon \frac{1}{\sqrtmQSq} \frac{ \sqrt{\pi}(m_1 {+} m_2)}{(\relf^2-1) }
\frac{ e^{\epsilon \EulerGamma} \Gamma \left(\frac{1}{2}-\epsilon \right)^2 \Gamma \left(\epsilon+\frac{1}{2}\right)} { \Gamma(1-2 \epsilon )}
\\
& \hspace{-2cm} -  \epsilon\frac{ \imath \pi  \left(m_1^2{+}m_2^2{+}2 m_1 m_2 \relf\right)}{8 m_1 m_2 \left(\relf^2-1\right)^{3/2}}
\frac{ e^{\epsilon \EulerGamma} \Gamma (-\epsilon )^2 \Gamma (1 + \epsilon)}{\Gamma (-2 \epsilon )}
\bigg\}  +\ldots \, , 
\end{split}
\label{eq:boxplusxbox}
\end{align}
where $\bar{\mu}^2=4\pi e^{-\EulerGamma}\mu^2$, while $\mu$ and $\EulerGamma$ are the dimensional regularization scale, and the Euler-Mascheroni constant, respectively. The ellipsis stand for terms with polynomial (including constant) dependence on $q^2$, with or without poles in $\epsilon$. Such terms give rise to contact interactions after Fourier transform to impact-parameter space, and are irrelevant for long-range classical physics.

\noindent
Finally,  the two-loop amplitude in the potential region is given by
\begin{align}
\begin{split}
  \mathcal{A}_{4,\pot}^{(2)}={} &
(4\pi\alpha\,  q_1q_2)^3
\left(\frac{\imath}{(4\pi)^2}\right)^2
\left(\frac{-q^2}{\bar\mu^2}\right)^{-2\epsilon}
\bigg\{
{-}\frac{1}{(-q^2)} \frac {32\pi^2 m^2 \nu\, \sigma^3} { (\relf^2-1)}
\left[ \frac 1 {\epsilon^2} - \frac{\pi^2} 6 \right] 
 \\
&\quad\qquad + \frac{1}{\sqrtmQSq} \frac{16\imath \pi \, m}{(\relf^2-1)^{1/2} }\left[
\frac{1}{\epsilon}-2\log(2)-\frac{4\sigma^3}{\sigma^2-1}+\calO(\epsilon) \right]
 \\
&\quad\qquad +\frac{8\pi^2}{\epsilon}\left[ \frac {2\nu(1-\sigma^2-\sigma^4)+(1-2\nu)(\sigma-2\sigma^3)} {\nu (\relf^2-1)^2}
+\calO(\epsilon)
\right]
\bigg\}+ \ldots  
\end{split}
\label{eq:2loopAmplitude}
\end{align}
where we have used the total mass, $m$, and symmetric mass ratio, $\nu$,
\begin{align}
  m &= m_1 + m_2\,,\quad \nu = \frac{m_1m_2}{(m_1 + m_2)^2}\,.
\end{align}

%================================================
%
\subsection{Eikonal approach to classical conservative scattering}
\label{subsec:eikonal_phase_cons}
%
%================================================

Armed with the conservative amplitude through two loops,  we may compute the eikonal phase. Traditionally,  one Fourier transforms the amplitudes to impact-parameter space in order to extract the eikonal phase. Here, following Ref.~\cite{Parra-Martinez:2020dzs}, we instead use the eikonal exponentiation directly in momentum space.  This comes at the cost of products in impact-parameter space becoming convolutions in momentum space. However, all needed convolutions have already been evaluated in Ref.~\cite{Parra-Martinez:2020dzs}.  We summarize the result of our calculation of the eikonal phase (Our convention for eikonal phase differs by a factor of 2 from that of \cite{Parra-Martinez:2020dzs} and is consistent with the definition of the eikonal in the soft region of section \ref{subsec:eikonal_phase_rad}.)
\begin{align}
2\delta^{(0)}_{\pot} (\sigma, \vect q_\perp) &= -(4\pi\alpha\,  q_1q_2)4 m_1m_2\sigma\frac{1}{\vect q^2_\perp}\,, \\
2\delta^{(1)}_{\pot} (\sigma, \vect q_\perp) &=(4\pi\alpha\,  q_1q_2)^2\frac{(m_1+m_2)}{4} \frac{1}{|\vect q_\perp|} \,,\\
2\delta^{(2)}_{\pot} (\sigma, \vect q_\perp) &=(4\pi\alpha\,  q_1q_2)^3 \frac{2\nu+(1-2\nu)\sigma}{16(\sigma^2-1)\nu} \log(\vect q_\perp^2)\,,
\end{align}
where we use the $D-2$-dimensional spacelike vector, $\vect q_{\perp}$, transverse to the scattering plane, satisfying $\vect q_{\perp}^2 = - q^2$. 
Finally, we can perform the Fourier transform to obtain the more familiar eikonal phase in impact-parameter space
\begin{equation}
  \delta(\sigma, \vect b_\mathrm{e}) = \frac{1}{4m_1 m_2 \sqrt{\sigma^2-1}}\int \frac{\mathrm{d}^{D-2}\vect q_\perp}{(2\pi)^{D-2}}\, e^{\imath\vect b_\mathrm{e} \cdot \vect q_\perp}\, \delta(\sigma, \vect q_\perp)\,,
\end{equation}
with the result
\begin{align}
2\delta^{(0)}_{\pot} (\sigma, \vect b_\mathrm{e}) &=(4\pi \alpha\,  q_1q_2)\frac{\sigma}{4\pi \sqrt{\sigma^2-1}}\log(\vect b_\mathrm{e}^2)\,, \\
2\delta^{(1)}_{\pot} (\sigma, \vect b_\mathrm{e}) &=(4\pi\alpha\,  q_1q_2)^2\frac{1}{32\pi m \nu\sqrt{\sigma^2-1}} \frac{1}{|\vect b_\mathrm{e}|} \,,\\
2\delta^{(2)}_{\pot} (\sigma, \vect b_\mathrm{e}) &=-(4\pi \alpha\,  q_1q_2)^3\frac{
\left(2\nu+(1-2\nu)\sigma\right)
}{64\pi^3 m^2\nu^2\sqrt{\sigma^2-1}} \frac{1}{\vect b_\mathrm{e}^2}\,,
\end{align}
where we have dropped terms that do not contribute to the classical scattering angle.
As discussed below (see Eq.~\eqref{bevsb}), the eikonal imapact parameter $\vect b_\mathrm{e}$ is distinct from the geometric impact parameter $\vect b$.

%================================================
%
\subsection{Conservative eikonal phase, scattering angle, and two-body Hamiltonian}
\label{subsec:conservative_Hamiltonian}
%
%================================================

The stationary phase approximation of the Fourier transform of the exponentiated impact-parameter amplitude back to momentum space 
\begin{align}
 \calA(\sigma, -q^2) = \int d^{D-2} \vect b_\mathrm{e} \left(e^{2 \imath  \delta(\sigma,\vect b_\mathrm{e} )} -1 \right) e^{-\imath \,\vect q  \cdot  \vect b_\mathrm{e}}
\end{align}
yields the relation
\begin{equation}
  \vect q   =-  \frac{\partial }{\partial \vect b_\mathrm{e}} 2\delta(\sigma,\vect b_\mathrm{e})\,.
  \label{eq:qdelta}
\end{equation}
The magnitude of $\vect q$ is related to the scattering angle $\chi$ and the magnitude of the three-momentum $\vect p$ in the center-of-mass  frame by
\begin{equation}
  |\vect q| = 2 |\vect p| \sin\frac\chi 2\,.
\label{eq:qangle}
\end{equation}
From this, we may now calculate the gravitational scattering angle from the eikonal phase using the formula 
\begin{equation}
  \sin\frac\chi 2 = - \frac{1}{2 |\vect p|} \frac{\partial }{\partial |\vect b_\mathrm{e}|}2 \delta(\sigma,\vect b_\mathrm{e})\,.
\end{equation}
where in terms of the center of mass energy $E=\sqrt{s}$ and/or $\sigma$
\begin{align}
\label{eq:pinf}
p_\infty \equiv  |\vect p| =& \frac{m_1m_2\sqrt{\sigma^2-1}}{E} 
 = \frac{m_1m_2\sqrt{\sigma^2-1}}{\sqrt{m_1^2 + m_2^2 + 2m_1m_2\sigma}}
= \frac{m\nu\sqrt{\sigma^2-1}}{\sqrt{1+2\nu(\sigma-1)}}\,.
\end{align}
Using this formula we find the following result for the scattering angle
\begin{align}
\label{Anglebe1}  \chi^{(0)}_{\pot} &= - \frac{\alpha\, q_1 q_2}{|\vect b_e|}\frac{2\sigma}{|\vect p| \sqrt{\sigma^2-1}} \,,\\
\label{Anglebe2}  \chi^{(1)}_{\pot} &= \frac{(\alpha\, q_1 q_2)^2}{|\vect b_e|^2}\frac{1}{2 |\vect p|m\nu\sqrt{\sigma^2-1}}  \,,\\
  \chi^{(2)}_{\pot} &=- \frac{(\alpha\, q_1 q_2)^3}{|\vect b_e|^3}
   \left[ \frac{\sigma^3}{3|\vect p|^3(\sigma^2-1)^{3/2}}+ \frac{ 
2 \left(2\nu+(1-2\nu)\sigma\right) }{|\vect p| m^2\nu^2\sqrt{\sigma^2-1}} \right] \,.
\end{align}
Comparing the result in Eqs.~(\ref{Anglebe1})-(\ref{Anglebe2}), we find agreement (up to conventions) with Westpfahl~\cite{Westpfahl:1985tsl}. It is useful to write the formulae in terms of the angular momentum, $J$, as defined via
\begin{equation}
  J = | \vect b \times \vect p | =|\vect b| |\vect p|~,
  \label{eq:lb}
\end{equation}
where $\vect b$ is the asymptotic impact parameter perpendicular to the incoming center of mass momentum $\vect p$. As noted earlier, this is not the impact parameter, $\vect b_\mathrm{e}$, relevant to the eikonal phase and pointing in the direction of the momentum transfer $\vect q$. The magnitude of $\vect b$ and $\vect b_\mathrm{e}$ are then related by 
\begin{equation}
  |\vect b| = |\vect b_\mathrm{e}| \cos\frac{\chi}{2}\,,
\label{bevsb}
\end{equation}
so that the angular momentum is
\begin{equation}
  J =|\vect b_\mathrm{e}| |\vect p|  \cos\frac{\chi}{2}\,.
  \label{eq:lbe}
\end{equation}
This difference is unimportant at leading orders, and it will only matter at order $J^{-3}$. Using the relation \eqref{eq:lbe} we find the scattering angle in terms of the angular momentum
\begin{align}
\begin{split}
\label{eq:cons_angles_J}
  \chi^{(0)}_{\pot} &=-\frac{\alpha\, q_1 q_2}{J}\frac{2 \sigma}{\sqrt{\sigma^2-1}}  \,,\\
  \chi^{(1)}_{\pot} &= \frac{(\alpha\, q_1 q_2)^2}{J^2}\frac{\pi}{2 \sqrt{1+2\nu(\sigma-1)}}  ,\\
  \chi^{(2)}_{\pot} &=\frac{(\alpha\, q_1 q_2)^3}{J^3}\frac{4\nu(3-3\sigma^2+\sigma^4)
+(1-2\nu)(6\sigma-4\sigma^3)}{3(1+2\nu(\sigma-1))(\sigma^2-1)^{3/2}}  \,.
\end{split}
\end{align}
One can write the conservative Hamiltonian for a system of two spinless charges in an expansion in powers of the electromagnetic coupling as
\begin{align}
\label{Hamiltonian}
H({\bm r}^2, {\bm p}^2) =& \sqrt{{\bm p}^2+m_1^2}+  \sqrt{{\bm p}^2+m_2^2} 
\\ &
+ c_1({\bm p}^2) \frac{\tilde{\alpha}}{|\bm r|}
+c_2({\bm p}^2) \left(\frac{\tilde{\alpha}}{|\bm r|}\right)^2
+c_3({\bm p}^2) \left(\frac{\tilde{\alpha}}{|\bm r|}\right)^3+\dots \,,
\nonumber
\end{align}
where $\tilde{\alpha}{=}4\pi\alpha q_1q_2$ and the $c_i({\bm p}^2)$ are yet to be determined coefficients. 
Such a Hamiltonian was used in Ref.~\cite{Bern:2019crd}, to obtain a related expansion of the classical scattering angle
\begin{align}
\chi = & \frac{ P_1}{p_\infty} \left(\frac{\tilde{\alpha}}{J} \right)
                + \frac{\pi}{2}  P_2  \left(\frac{\tilde{\alpha}}{J} \right)^2
                - \frac{  P_1^3 - 12 p_\infty^2 P_1 P_2 - 24 p_\infty^4 P_3}{12p_\infty^3}  \left(\frac{\tilde{\alpha}}{J} \right)^3 
                +\calO \left((\tilde{\alpha}/J)^4\right)
               \, ,
\label{theangle}
\end{align}
where $p_\infty$ was defined in Eq.~\eqref{eq:pinf}. The $P_i$ coefficients in the scattering angle are the natural expansion coefficients for the radial momentum $p^2_r(r)$ (see Sec.~11 of \cite{Bern:2019crd} for details in the context of GR), but are also related to the $c_i$ coefficients in the Hamiltonian via~\cite{Bern:2019crd}
\begin{align}
\label{P1}
P_1 = \null &  -2 E \xi\,  \bar{c}{}_1 \,,
\\[2pt]
 \label{P2}
P_2 =\null & -2 E \xi\,  \bar{c}{}_2 +(1-3 \xi ) \, \bar{c}{}_1^2+4 E^2 \xi ^2\,  \bar{c}{}_1 \bar{c}'_1 \,,
\\[2pt]
\label{P3}
P_3 = \null & -2 E \xi \bar{c}_3 +2 (1-3 \xi )\bar{c}_1 \bar{c}_2
   -4E^3 \xi ^3  \bar{c}{}_1 \left(2  \bar{c}'^{\,2}_1+ \bar{c}{}_1 \bar{c}''_1\right)
   +4 E^2 \xi ^2 \left(\bar{c}{}_2
   \bar{c}'_1+\bar{c}_1 \bar{c}'_2\right) 
   \nonumber\\[0pt]
  & -6 E (1-3 \xi ) \xi 
   \bar{c}{}_1^2 \bar{c}'_1
   +\frac{(1 - 4 \xi) \bar{c}_1^3}{E} \,,
 \end{align}  
where $\bar{c}_i {\equiv} c_i(p_\infty^2)$, primes denote derivatives with respect to the argument, $E {=} E_1 {+} E_2 = m \sqrt{1{+}2 \nu  (\sigma {-}1)}$ and $\xi {=} E_1E_2/E^2 {=} \frac{\nu  \left(\nu  (\sigma {-}1)^2{+}\sigma \right)}{(1{+}2 \nu  (\sigma {-}1))^2}$. Conversely, one may start from our scattering angles in Eq.~(\ref{eq:cons_angles_J}), deduce 
the $P_i$ coefficients in Eq.~\eqref{theangle} up to $\calO(\alpha^3)$
\begin{align}
P_1  =  -\frac{2 m\, \nu\,  \sigma }{(4 \pi)  \, \Gamma}\,, \qquad
P_2  =  \frac{1}{(4\pi)^2} \frac{1}{\Gamma}\,,\qquad 
P_3 = \frac{(\Gamma{-}1)\sigma + 2\nu (\sigma{-}1)}{(4\pi)^3\, \Gamma \, m \, \nu\, (\sigma^2-1)}\,, 
\end{align}
where we have defined $\Gamma = E/m = \sqrt{1+2\nu(\sigma{-}1)}$. Relatedly, the coefficients in the expansion of the scattering angle can also be expressed via suitably defined finite parts of scattering amplitudes (see discussion around Eq.~(12) of \cite{Bern:2019nnu}) in terms of slightly rescaled coefficients $\tilde{d}_i$ which differ from $P_i$ by powers of $p_\infty$, $\tilde{d}_i = p^{i-2}_\infty\, P_i$:\footnote{We use $\tilde{d}_i$ rather than $d_i$ to signal slight normalization differences by factors of $\pi$ and the coupling constant compared to \cite{Bern:2019nnu}. }
\begin{align}
\label{eq:angle_ds}
 \chi = \tilde{d}_1 \left(\frac{\tilde \alpha}{J}\right) 
          + \frac{\pi}{2} \tilde{d}_2  \left(\frac{\tilde \alpha}{J}\right)^2
          - \left(\frac{\tilde{d}_1^3}{12} - \tilde{d}_1 \, \tilde{d}_2 + 2 \tilde{d}_3\right) \left(\frac{\tilde \alpha}{J}\right)^3  
          + \calO\big((\tilde{\alpha}/J)^4\big)\,,
\end{align}
where the coefficients now read
\begin{align}
 \tilde{d}_1 = - \frac{2 \sigma}{(4\pi) \sqrt{\sigma^2-1}}\,, \qquad
 \tilde{d}_2 =  \frac{1}{(4\pi)^2 \, \Gamma }\,, \qquad
 \tilde{d}_3 = \frac{(\Gamma{-}1) (1+\Gamma + \sigma)}{(4\pi)^3 \Gamma^2 \sqrt{\sigma ^2-1}}\,.
\end{align}
An interesting feature is that two-loop function $\tilde{d}_3$ vanishes in the test mass limit $\nu {\to} 0$ ($ \Gamma{\to}1$), i.e.~$ \tilde{d}_3 \stackrel{\nu {\to} 0}{\longrightarrow} 0$.  This is due to the fact that $\Gamma{-}1$ starts at $\calO(\nu)$ in the test mass expansion. Crucially, $\tilde{d}_3=0$ implies that the $\calO(\alpha^3)$ test-body Hamiltonian is fully determined by lower-loop information. In fact, in the test-body limit, the scattering angle
is computable exactly to all orders in the coupling (see
e.g. Ref.~\cite{Kol:2021jjc}\footnote{In comparison to Ref.~\cite{Kol:2021jjc}, our scattering angle is defined to go to zero in the absence of interactions ($\alpha{\to}0$) which explains the additional $-\pi$. Our conventions for the charges lead to an additional sign in the argument of the arctan.}  and references therein)
\begin{align}
\label{eq:testbody_angle}
\begin{split}
\chi_{{\rm test}} & =  \frac{J}{\sqrt{J^2-(\alpha q_1 q_2)^2}} 
\left(\pi + 2 \, \text{arctan} \left[\frac{-(\alpha q_1 q_2)}{\beta \sqrt{J^2-(\alpha q_1 q_2)^2}}\right]\right)-\pi \,, \\[3pt]
& = 
-\frac{2 \alpha  q_1 q_2}{\beta  J}
+\frac{\pi  \alpha ^2 q_1^2 q_2^2}{2 J^2} 
-\frac{2 \alpha ^3 \left(3 \beta ^2-1\right) q_1^3 q_2^3}{3 \beta ^3 J^3}
+\frac{3 \pi  \alpha ^4 q_1^4 q_2^4}{8 J^4}+ \calO(\alpha^5)\,,
\end{split}
\end{align}
where $\beta = \sqrt{\sigma^2 -1}/\sigma$ is the velocity.  With the test-body angle available to all 
orders in the coupling constant, we can investigate the angle relation of Eq.~\eqref{theangle} for 
the $P_i$ to higher orders in perturbation theory. At $\calO(\alpha^4)$, 
we would find (see Eq.~(11.25) of \cite{Bern:2019crd})
\begin{align}
\label{eq:angle_in_P_higher_order}
\begin{split}
\chi & = \eqref{theangle} + \frac{3\pi}{8} \left( P^2_2 + 2 P_1 P_3 + 2 p^2_\infty P_4\right)  \left(\frac{\tilde{\alpha}}{J} \right)^4 + \calO\big((\tilde{\alpha}/J)^5\big)\,, 
\\
& = \eqref{eq:angle_ds} +  \frac{3\pi}{8} \left(\tilde{d}^2_2 + 2 \tilde{d}_1 \tilde{d}_3 + 2 \tilde{d}_4\right)  \left(\frac{\tilde{\alpha}}{J} \right)^4 + \calO\big((\tilde{\alpha}/J)^5\big)
\end{split}
\end{align}
Comparing the explicit angle in the test-body limit \eqref{eq:testbody_angle} to the expansion in 
Eq.~\eqref{eq:angle_in_P_higher_order}, we see that $ \tilde{d}_3= \tilde{d}_4 =0$. Indeed, all $ \tilde{d}_{i>2}=0$ 
vanish in the test-body limit which in turn implies that the higher Hamiltonian coefficients $c_{i>2}$ 
are equally determined by at most one-loop data. In light of this discussion, the reason there is a simple closed form solution for the scattering angle 
is connected to the simplicity of the test mass Hamiltonian.

For the sake of completeness, we also tabulate the $c_i$ coefficients in the Hamiltonian 
or general mass dependence up to $\calO(\alpha^3)$. This yields
\begin{align}
c_1&=\frac{1}{4\pi(2\Gamma^2\xi^2)}(2\Gamma^2\xi^2\tilde{\nu}) \,,
\label{eq:c1Explicit}
\\
c_2&=\frac{1}{(4\pi)^2m(2\Gamma^2\xi^2)}
\left(-\xi+2\Gamma\xi\tilde{\nu}-\Gamma\xi(1-\xi)\tilde{\nu}^2\right)\,,
\label{eq:c2Explicit}
\\
\label{eq:c3Explicit}
c_3&=\frac{1}{(4\pi)^3m^2(2\Gamma^2\xi^2)}
\left\lbrace
\frac{\sigma^2}{1-\sigma^2}\left(\frac{\Gamma-1}{\Gamma}\right)\frac{1}{\tilde{\nu}}
-\frac{1+\sigma+2\Gamma\xi}{(1+\sigma)\Gamma}
\right.\\
\nonumber 
& 	\qquad\qquad\qquad\qquad\qquad\qquad\qquad 
\left.
+\frac{2\Gamma+1-\xi}{\Gamma}\tilde{\nu}
-(3-4\xi)\tilde{\nu}^2+(1-2\xi)\tilde{\nu}^3
\right\rbrace \,,
\end{align}
where $\tilde{\nu} =\frac{\sigma\nu}{\Gamma^2\xi}= \frac{\sigma  (1+2 \nu  (\sigma -1))}{\nu  (\sigma -1)^2+\sigma }$.

%================================================
%
\subsection{Conservative impulse and scattering angle via KMOC}
\label{subsec:KMOC_cons_results}
%
%================================================
Besides extracting the classical observables via the eikonal approach discussed in Sec.~\ref{subsec:eikonal_phase_cons} or in terms of a classical Hamiltonian (see Sec.~\ref{subsec:conservative_Hamiltonian}), we have also computed the conservative electromagnetic impulse within the KMOC framework, generally discussed in Sec. \ref{subsec:KMOC_general}. In the conservative setting, energy is conserved in the scattering process, so that there is no radiated momentum or energy loss to consider. In terms of the method of regions, this is encoded in the fact that all photons are purely potential and can never go on-shell, so that the radiation cuts in Eq.~(\ref{eq:KMOC_kernel_rad}) are always absent. Likewise, for the impulse, we only have to consider elastic processes without exchanged messenger particles in the \emph{real} contribution in Eq.~(\ref{eq:classical_impulse_kernel}). 

Due to energy conservation in the conservative sector, it suffices to determine the transverse impulse only and the longitudinal part is completely determined by the on-shell conditions and momentum conservation of the process 
\begin{align}
 (p_i + \Delta p_i)^2 = p^2_i\, \qquad \Delta p_1 + \Delta p_2 =0\,.
\end{align}
Employing generalized unitarity, we constructed the relevant conservative integrand to extract the transverse impulse kernel from Eq.~(\ref{eq:impulse_kernel_perp}) where only the first two diagrams of Fig.~(\ref{fig:cubic_graphs}) are relevant at $\calO(\alpha^3)$. Using the same integration and assembly pipeline that has already produced the conservative gravitational results in Ref.~\cite{Herrmann:2021tct} together with the potential region values of the master integrals from Ref.~\cite{Parra-Martinez:2020dzs}, we obtain the transverse impulse on particle 1 at two loops\footnote{We drop the explicit label of the perturbative order of the observable (see Eq.~\eqref{eq:KMOC_obs_pert_expansion_schematic}) for brevity. The same information is encoded in the order of $\alpha$. }
\begin{align}
\Delta p^{\mu,{\rm cons.}}_{1,\perp} = (\alpha q_1 q_2)^3 \frac{2 \left(2 m_1 m_2 \left(\sigma^4{-}\sigma^2 + 1\right)+(m_1^2  + m_2^2) \sigma \right)}{m_1^2 m_2^2  \left(\sigma ^2-1\right)^{5/2}}  \frac{b^{\mu }}{|b|^4}\,.
\end{align}
The conservative longitudinal impulse at $\calO(\alpha^3)$ is 
\begin{align}
\label{eq:impulse_cons_long_p1}
\Delta p^{\mu, {\rm cons.}}_{1,u} = (\alpha q_1 q_2)^3 \frac{\pi \left(m_1 + m_2\right)  \sigma }{m_1 m_2\, |b|^3  \left(\sigma^2-1\right)}
 \left(\frac{\check{u}^\mu_2}{m_2} - \frac{\check{u}^\mu_1}{m_1}\right)\,.
\end{align}
Both the longitudinal and transverse impulse agree with the ones obtained from the scattering angles \eqref{eq:cons_angles_J} and the help of the parametrization of the impulse in terms of the conservative scattering angle of appendix B in Ref.~\cite{Herrmann:2021tct}. We also agree with the recent computation of the impulse obtained by solving classical equations of motion~\cite{Saketh:2021sri}. 

In comparing to the gravitational result~\cite{Herrmann:2021tct}, in electrodynamics there is no $\text{arcsinh} \sqrt{\frac{\sigma-1}{2}}$ high-energy singularity in the transverse impulse, which also renders this feature absent from the scattering angle. At high energies, i.e. large $\sigma \gg 1$, and fixed impact parameter $b$, the leading large sigma behavior of the transverse and longitudinal impulse scale like $1/\sigma$ and $1/\sigma^2$ respectively. (For the longitudinal impulse, one has to take the definition of the $\check u_i \stackrel{\sigma \gg 1}{\sim} 1/\sigma$ in Eq.~\eqref{eq:dual_ui} into account.)

%==============================================================
%
%\newpage
\section{Radiative dynamics at $\mathcal{O}(\alpha^3)$}
\label{sec:radiative}
%
%==============================================================

%================================================
%
\subsection{Radiative scattering amplitudes}
\label{subsec:amps_soft}
%
%================================================
In this section we give the results for the scattering amplitudes in the soft region that includes radiation effects by combining the integrands from section \ref{subsec:integrands} with the integration techniques sketched in section \ref{subsec:integration}.  Detailed results of all soft-region integrals can be found in Ref.~\cite{Herrmann:2021tct}.  The tree-level amplitude remains
\begin{align}
\label{eq:tree_amp_soft}
\mathcal{A}^{(0)}_4(p_1,p_2,p_3,p_4)=
 \vcenter{\hbox{\scalebox{1}{\treetPhoton}}} = - (4\pi \alpha\, q_1 q_2)\frac{4 m_1 m_2\, \sigma}{-q^2}\,,
\end{align}

\medskip

\noindent
The one-loop amplitude in the soft region is 
\begin{align}
\label{eq:1loop_amp_soft}
\begin{split}
\hspace{-.5cm}
\mathcal{A}^{(1)}_4(p_1,p_2,p_3,p_4) 
= & (4\pi \alpha\, q_1 q_2)^2\!\! \left[\imath \left(\!\frac{1 }{4 \pi}\!\right)^{2-\epsilon} \!\!\!\!\frac{e^{-\gamma  \epsilon}}{(-q^2)^\epsilon}\!\right]\!\!\! 
\Bigg[
\frac{4 \pi ^2 m_1 m_2 \left(z^2 + 1\right)^2 e^{\gamma  \epsilon } \csc (\pi  \epsilon ) \Gamma (-\epsilon )}{(-q^2)z \left(z^2-1\right) \Gamma (-2 \epsilon )}
\\&
+ \frac{\imath \pi ^2 \left(m_1+m_2\right) 4^{\epsilon + 1} e^{\gamma  \epsilon } \left(4 \left(z^2+1\right)^2 \epsilon -\left(z^2-1\right)^2\right) \sec (\pi  \epsilon )}{\sqrt{-q^2} \left(z^2-1\right)^2 \Gamma (1-\epsilon )}
\\&
+ \frac{\pi ^{3/2} 2^{2 \epsilon +1} e^{\gamma  \epsilon } \csc (\pi  \epsilon )}{\Gamma \left(\frac{3}{2}-\epsilon \right)m_1 m_2 \left(z^2-1\right)^3 }\ f(z,\epsilon)
+ \calO(|q|)
\Bigg] \,,
\end{split}
\end{align}
where $\gamma$ is the Euler-Mascharoni constant and
\begin{align}
%\begin{split}
f(z,\epsilon) {=} 
\Big[
& 2 \pi  \left(m_1^2 + m_2^2\right) z \left(z^2 + 1\right)^2 \epsilon  (2 \epsilon{-}1)
+m_1 m_2 \Big\{
4 \left(z^2 + 1\right)^2 \left((\pi + 2 i) z^2{-}2 i + \pi \right) \epsilon^2 
\nn \\& \hspace{-1.5cm}
-2 \left(\pi  \left(z^2 + 1\right)^3 + 2 i \left(z^2{-}1\right)^3\right) \epsilon 
+2 i (z^2 + 1)(2 \epsilon{-}1) \left(2 \left(z^2 + 1\right)^2\! \epsilon {-}z^4 + 6 z^2{-}1\right) \log (z)
\nn \\& \hspace{-1.5cm}
-i \left(z^6+5 z^4-5 z^2-1\right)
\Big\}
\Big] \,,
%\end{split}
\end{align}
is written in terms of $z$, related to $\sigma=\frac{1+z^2}{2z}$ in order to rationalize square roots such as $\sqrt{\sigma^2-1} = \frac{1-z^2}{2z}$.  When converting the computations in terms of the soft-variables $y,\, x,\, \overline{m}_i$ to the original masses $m_i$ and $\sigma, z$, one has to be careful to take into account subleading $\calO(q^2)$ terms in the conversion in order to obtain the correct expression for the one-loop quantum piece. 

\medskip

\noindent
The relevant parts of the two-loop amplitude are
\begin{align}
\label{eq:2loop_amp_soft}
\begin{split}
\hspace{-.7cm}
\frac{\mathcal{A}^{(2)}_4\!(p_1,p_2,p_3,p_4)}{(4\pi \alpha)^3}&{=}
 \left[\imath \left(\!\frac{1 }{4 \pi}\!\right)^{2-\epsilon}\!\! \!\!\!\frac{e^{-\gamma  \epsilon}}{(-q^2)^\epsilon}\!\right]^2 \!\!
 \left[\!
 	\frac{a^{(2)}_{-1,-2}}{(-q^2)\, \epsilon^2}  +  \frac{a^{(2)}_{-1,-1}}{(-q^2)\, \epsilon} 
   +  \frac{a^{(2)}_{-\frac12,-1}}{\sqrt{-q^2}\, \epsilon} 
   +   \frac{a^{(2)}_{0,-2}}{\epsilon^2}  +  \frac{a^{(2)}_{0,-1}}{\epsilon}\!
 \right]\!,
 \hspace{-.2cm}
\end{split}
\end{align}
where the two-loop amplitude coefficients $a^{(2)}_{i,j}$ are labeled by the order in the $q$ and $\epsilon$ expansion. In principle, we also have access to higher-order terms in the $\epsilon$ and $q$ expansion, but they do not play a role for the purpose of our discussion here. The expression for the classical term $a^{(2)}_{0,-1}$ is rather lengthy, so we do not display it here. (We supply all expressions in computer readable form as an ancillary file.) The classical term $a^{(2)}_{0,-1}$ contains contributions from both the charge sector proportional to $q^3_1 q^3_2$ as well as from the mushroom topologies $q^4_i q^2_j$. In contrast, the classically divergent terms as well as the $1/\epsilon^2$ coefficient of the classical piece are entirely within the $q^3_1 q^3_2$ sub-sector, 
\begin{align}
a^{(2)}_{-1,-2} & = -(q_1 q_2)^3 \frac{16 \pi ^2 m_1 m_2 \left(z^2 + 1\right)^3}{z \left(z^2{-}1\right)^2 }
\,, \qquad \qquad \qquad \qquad \qquad
a^{(2)}_{-1,-1}   = 0\,,
\\
a^{(2)}_{-\frac12,-1} & =  (q_1 q_2)^3  \frac{16 \imath \pi^3\, (m_1 + m_2)\left(z^2+1\right) }{ \left(z^2-1\right)}\,,
\\
a^{(2)}_{0,-2} & = (q_1 q_2)^3 \frac{8 \imath \pi  \left(z^2 + 1\right) \left(3 z^6 + 7 z^4{-}7 z^2{-}3+2 \left(z^6{-}9 z^4{-}9 z^2 + 1\right) \log (z)\right)}{\left(z^2{-}1\right)^4}\,.
\end{align}
Again, all results are written in the terms of the $z$ variable that rationalizes $\sqrt{\sigma^2-1}$.

%================================================
%
\subsection{Eikonal approach to classical scattering including radiation effects}
\label{subsec:eikonal_phase_rad}
%
%================================================

In a first implementation of radiative effects in scalar QED at $\calO(\alpha^3)$, we follow the eikonal approach that has been successfully used in Ref.~\cite{DiVecchia:2021bdo} to extract the classical scattering angle in gravity. The steps in their derivation are independent of the theory under consideration as long as they admit a suitable classical limit. It is therefore natural to expect that we can follow Sec.~2.2 of Ref.~\cite{DiVecchia:2021bdo} for scalar QED. As we will show, this expectation is indeed correct. Since the physical intuition behind the eikonal approach has already been discussed extensively by Di Vecchia et al.~\cite{DiVecchia:2021bdo}, we restrict ourselves to only the most important formulae. At the heart of the eikonal approach is the expected exponentiation of certain parts of the scattering amplitude in impact-parameter space
\begin{align}
\label{eq:eikonal_ansatz}
1+ i \widetilde{\mathcal{A}}(s,b_e) = \big(1+ 2 i \Delta(s,b_e)\big) e^{2i \delta(s,b_e)} \,,
\end{align}
where $\delta(s,b_e)$ is the classical eikonal phase and $\Delta(s,b_e)$ is a quantum remainder that starts with one-loop contributions. (We follow the notation of Ref.~\cite{DiVecchia:2021bdo} but keep in mind that we express $s$ in terms of $\sigma$ or $z$.) Let us note, that it is a priori unclear what part of the amplitude exponentiates beyond leading order. In the conservative sector discussed in Sec.~\ref{subsec:eikonal_phase_cons}, there is an good argument based on elastic unitarity, that the amplitude (including quantum terms) exponentiates. Here, we follow the prescription of Ref.~\cite{DiVecchia:2021bdo} that exponentiate strictly classical terms and all quantum corrections are assigned to $\Delta(s,b_e)$. However, from the conservative perspective, one could expect that certain quantum terms that originate from the potential region should still exponentiate and lead to quantum corrections of the classical eikonal phase $\delta(s,b_e)$ and only quantum radiation pieces need not resum to a phase. We do not pursue this separation further and leave a detailed investigation to future work. The momentum-space amplitude $\mathcal{A}(s,-q^2)$ is Fourier-transformed to the $D-2$-dimensional impact-parameter space, conjugate to the (space-like) momentum-transfer $q$
\begin{align}
\label{eq:eik_amp_fourier_transform}
\widetilde{\mathcal{A}}(s,b_e) = \int \frac{d^{D-2}q}{(2\pi)^{D-2}} \frac{\mathcal{A}(s,-q^2)}{4 m_1 m_2 \sqrt{\sigma^2-1}} e^{i b_e\cdot q}\,.
\end{align}
Here, we distinguish the eikonal impact parameter $b_e$ from the asymptotic impact parameter $b$ related by the geometric identity
\begin{align}
\label{eq:impact_parameter_mapping}
 b_e = \frac{b}{\cos \frac{\chi}{2}} \,,
\end{align}
where $\chi$ is the scattering angle. The momentum-space scattering amplitude $\mathcal{A}(s,-q^2)$ entering in Eq.~\eqref{eq:eik_amp_fourier_transform} has an expansion in terms of the coupling constant (loop-expansion) as well as an expansion in powers of $-q^2$. Via standard Fourier integrals, the power-series expansion in $-q^2$ gets converted to a similar expansion in impact-parameter space
\begin{align}
\label{eq:abstract_mom_amplitudes}
\hspace{-.5cm}
 \mathcal{A}^{(0)}_4 & {=} \frac{a^{(0)}_{-1}}{-q^2} 
&&
\hspace{-.3cm}
 \to
 \quad
 \widetilde{\mathcal{A}}^{(0)}_4   {=} \frac{\widetilde{a}^{(0)}_{-1}}{\left(-b^2_e\right)^{-\epsilon}} \,,\hspace{-.2cm}
 \\
 \hspace{-.5cm}
 \mathcal{A}^{(1)}_4 &  {=} \frac{a^{(1)}_{-1}}{\left(-q^2\right)^{1+\epsilon}} 
 				    +  \frac{a^{(1)}_{-1/2}}{\left(-q^2\right)^{\frac12+\epsilon}} 
				    +   \frac{a^{(1)}_{0}}{\left(-q^2\right)^{\epsilon}} 
&&
\hspace{-.3cm}
 \to
 \quad
 \widetilde{\mathcal{A}}^{(1)}_4   {=} \frac{\widetilde{a}^{(1)}_{-1}}{\left(-b^2_e\right)^{-2\epsilon}} 
 						   +   \frac{\widetilde{a}^{(1)}_{-1/2}}{\left(-b^2_e\right)^{\frac12-2\epsilon}} 
						   +   \frac{\widetilde{a}^{(1)}_{0}}{\left(-b^2_e\right)^{1-2\epsilon}} \,,\hspace{-.2cm}
\\
\hspace{-.5cm}
\mathcal{A}^{(2)}_4 &  {=} \frac{a^{(2)}_{-1}}{\left(-q^2\right)^{1+2\epsilon}} 
				    +   \frac{a^{(2)}_{-1/2}}{\left(-q^2\right)^{\frac12+2\epsilon}} 
				    +   \frac{a^{(2)}_{0}}{\left(-q^2\right)^{2\epsilon}} 
&&
\hspace{-.3cm}
 \to
 \quad
 \widetilde{\mathcal{A}}^{(2)}_4   {=} \frac{\widetilde{a}^{(2)}_{-1}}{\left(-b^2_e\right)^{-3\epsilon}} 
 						   +   \frac{\widetilde{a}^{(2)}_{-1/2}}{\left(-b^2_e\right)^{\frac12-3\epsilon}} 
						   +   \frac{\widetilde{a}^{(2)}_{0}}{\left(-b^2_e\right)^{1-3\epsilon}} \,, \hspace{-.2cm}
\end{align} 
where the impact-parameter space coefficients $\widetilde{a}^{(\ell)}_{j}$ are directly related to the momentum-space expressions around $D=4-2\epsilon$ dimensions via 
\begin{align}
\begin{split}
   \int\! \frac{\mathrm{d}^{D-2}q }{(2\pi)^{D-2}}\,  \frac{e^{\imath  b_e\cdot q}}{(-q^2)^{\alpha} }
       =\frac{ \Gamma\left(D/2-1-\alpha\right)}
          {2^{2\alpha+2} (\pi)^{\frac{D-2}{2}}\Gamma\left(\alpha\right)} \frac{1}{(-b^2_e)^{\frac{D-2-2\alpha}{2}}}\,.
    \label{eq:basicFT}
\end{split}
\end{align}
Furthermore, the $a^{(\ell)}_{i}$ are simply related to the $a^{(\ell)}_{i,j}$ of Eqs.~\eqref{eq:tree_amp_soft}, \eqref{eq:1loop_amp_soft}, and \eqref{eq:2loop_amp_soft}. Expanding the right-hand-side of the eikonal ansatz in Eq.~\eqref{eq:eikonal_ansatz} perturbatively in the coupling $\alpha$
\begin{align}
\delta = \delta^{(0)}+\delta^{(1)} + \delta^{(2)} + \cdots \,, 
\qquad 
\Delta = \Delta^{(1)} + \cdots\,,
\end{align}
one can match both sides of Eq.~\eqref{eq:eikonal_ansatz} to the desired order in $\alpha$ to determine the eikonal quantities in terms of fixed-order amplitude results. Note that all classically divergent (or `super-classical') terms are completely determined by lower order data and serve as nontrivial cross-check of the computation. Equipped with the eikonal phase, one can Fourier-transform back to momentum space which allows for the identification of the \emph{classical} momentum transfer $Q^\mu$ from a stationary-phase approximation 
\begin{align}
Q^\mu = - \frac{\partial \, \text{Re }2\, \delta(s,b_e)}{\partial b^\mu_e} \,,
\end{align}
which is related to the scattering angle $\chi$ via
\begin{align}
 |Q| = 2 p \sin \frac{\chi}{2} \,,
\end{align}
where $p=\frac{\sqrt{\sigma^2-1} m_1m_2}{\sqrt{m^2_1 + m^2_2 + 2 m_1 m_2 \sigma}}$ is the asymptotic center of mass momentum.  

\noindent
In terms of the asymptotic impact parameter that is tied to the center-of-mass angular momentum $J = p \ b$ (see Eq.~\eqref{eq:impact_parameter_mapping}), the scattering angle is
\begin{align}
\label{eq:scatt_angle_bJ}
 \tan \frac{\chi}{2} = - \frac{1}{2p} \frac{\partial \, 2\, \text{Re } \delta}{\partial b}\,.
\end{align}
We can now take the perturbative expansion of Eq.~(\ref{eq:eikonal_ansatz}) together with the explicit values of the tree-, one-loop, and two-loop amplitudes in momentum space of Eqs.~\eqref{eq:tree_amp_soft}, \eqref{eq:1loop_amp_soft}, and \eqref{eq:2loop_amp_soft} to convert the variables to the normal masses $m_i$ and to $\sigma$ in order to obtain the explicit values for the $a^{(\ell)}_{i}(s)$ coefficients in Eq.~\eqref{eq:abstract_mom_amplitudes} from which we can extract the eikonal parameters
\begin{align}
\label{eqs:delta_softs}
 \delta^{(0)}(s,b_e) & = \frac{1}{(-b^2_e)^{-\epsilon}} \frac{\pi^{\epsilon}\, \Gamma (-\epsilon)}{32\pi\, m_1 m_2 \sqrt{\sigma ^2-1}}\  a^{(0)}_{-1} \,,
 \\
 \delta^{(1)}(s,b_e) & = \frac{1}{(-b^2_e)^{\frac12-2\epsilon}} \frac{2^{-2 \epsilon} \pi ^{\epsilon}\, \Gamma \left(\frac{1}{2}-2 \epsilon\right)}{16 \pi \, m_1m_2 \sqrt{\sigma ^2-1}\,  \Gamma \left(\epsilon+\frac{1}{2}\right)}\ a^{(1)}_{-1/2} \,,\\
 \delta^{(2)}(s,b_e) & = \frac{1}{(-b^2_e)^{1-3\epsilon}} \frac{1}{2 m_1 m_2} 
 \Bigg[
  \frac{4^{-2 \epsilon} \pi ^{\epsilon}\, \Gamma (1-3 \epsilon)}{4 \pi \, \sqrt{\sigma ^2-1} \Gamma (2 \epsilon)}\ a^{(2)}_0
  \\
  \label{eq:delta_soft_2}
  &\hspace{3.7cm}
 -\imath \frac{2^{-2 \epsilon} \pi^{2 \epsilon}\,  \Gamma (1-2 \epsilon) \Gamma (-\epsilon)}{64 \pi^2\, m_1 m_2\left(\sigma ^2-1\right) \Gamma (\epsilon)}\ a^{(0)}_{-1}\, a^{(1)}_{0}
 \Bigg] \,, \nn
\end{align}
and the one-loop quantum remainder,
\begin{align}
\Delta^{(1)}(s,b_e) = \frac{1}{(-b^2_e)^{1-2\epsilon}} \frac{2^{-2 \epsilon} \pi ^{\epsilon} \, \Gamma (1-2 \epsilon)}{8\pi \, m_1 m_2 \sqrt{\sigma ^2-1} \Gamma (\epsilon)}\ a^{(1)}_{0}\,.
\end{align}
We note that the imaginary part of the two-loop eikonal, $\delta^{(2)}$, still contains a $1/\epsilon$ infrared singularity (due to the $1/\epsilon^2$ contribution of $a^{(2)}_0$ in Eq.~\eqref{eq:delta_soft_2}), consistent with a similar feature in gravity~\cite{DiVecchia:2021bdo}, whereas the real part of the phase which is relevant for the physical scattering angle is IR finite. 

With the explicit values of the eikonal parameters at hand, we can take Eq.~\eqref{eq:scatt_angle_bJ}, convert the eikonal impact parameter in Eq.~\eqref{eqs:delta_softs} with the help of Eq.~\eqref{eq:impact_parameter_mapping}, and perturbatively expand the angle $\chi = \chi^{(0)} + \chi^{(1)} + \chi^{(2)} + \cdots$. This allows us to solve  Eq.~\eqref{eq:scatt_angle_bJ} order by order in the coupling in terms of expressions that we know from the scattering ampliutudes. In particular, we find
\begin{align}
 \chi^{(0)} & =   - (\alpha q_1 q_2) \frac{2 \sigma }{J \sqrt{\sigma^2-1}}\,,
 \\
  \chi^{(1)} & =  (\alpha q_1 q_2)^2  \frac{\pi (m_1+m_2) }{2 J^2 \sqrt{m_1^2+m_2^2+2 m_1 m_2\, \sigma}}\,,
  \\
  \label{eq:soft_angle_2loop_eik}
  \chi^{(2)} & = (\alpha q_1 q_2)^3  \frac{(m_1^2 {+} m_2^2) \left(6 \sigma{-}4 \sigma^3\right)
    +4 m_1 m_2 \left(3 +  (\sigma ^2{-}3  \sigma \sqrt{\sigma^2{-}1} {-}3) \sigma ^2
    +6 \sigma ^2 \text{arcsinh}\sqrt{\frac{\sigma-1}{2}}\right)}
  {3 J^3 \left(\sigma ^2{-}1\right)^{3/2} \left(m_1^2 + m_2^2 + 2 m_1 m_2\, \sigma\right)} \nn \\
   & \null \hskip .4 cm 
    +\alpha^3 q^2_1 q^2_2  \frac{4  \sigma ^2 \left(m_2^2 q_1^2+m_1^2 q_2^2\right)}{3 J^3 \left(m_1^2 + m_2^2 + 2 m_1 m_2\, \sigma\right)}\,,
\end{align}
where the second line of Eq.~\eqref{eq:soft_angle_2loop_eik} originates from the mushroom sector. Our results are in complete agreement with those of Saketh et al.~\cite{Saketh:2021sri}, once we take into account the different conventions $\chi^{(L)}_{{\rm us}}=-\chi^{(L)}_{\text{\cite{Saketh:2021sri}}}$. Before discussing these results in a bit more detail, it is beneficial to split the $\calO(\alpha^3)$ scattering angle into conservative and radiative parts,
\begin{align}
 \chi^{(2)} = \chi^{(2)}_{{\rm cons.}} + \chi^{(2)}_{{\rm rad.}}\,,
\end{align}
where
\begin{align}
 \chi^{(2)}_{{\rm cons.}}  & =  \left(\alpha q_1 q_2\right)^3 
 \frac{2 \left( \sigma  \left(3-2 \sigma ^2\right)(m_1^2  +  m_2^2)+2 m_1 m_2 \left(\sigma^4 - 3 \sigma^2 + 3\right)\right)}
 {3 J^3 \left(\sigma^2 - 1\right)^{3/2} \left(m_1^2 + m_2^2 + 2 m_1 m_2\, \sigma\right)} \,,
 \\
  \chi^{(2)}_{{\rm rad.}} & =  
   -  \left(\alpha q_1 q_2\right)^3 \frac{4  m_1 m_2 \, \sigma ^2 \left(\sigma  \sqrt{\sigma^2 - 1} - 2\, \text{arcsinh}\sqrt{\frac{\sigma - 1}{2}}\right)}
  {J^3 \left(\sigma^2 - 1\right)^{3/2} \left(m_1^2 + m_2^2 + 2 m_1 m_2\, \sigma\right)}
\nonumber \\
& \hskip 3 cm 
  +\alpha^3 q_1^2 q_2^2 \frac{4 \sigma ^2 \left(m_2^2 q_1^2+m_1^2 q_2^2\right)}
  {3 J^3 \left(m_1^2 + m_2^2 + 2 m_1 m_2\, \sigma\right)} \,.
 \end{align}
 The radiative part of the angle can also be used to extract or compare to the radiated angular momentum~\cite{Bini:2012ji,Damour:2020tta} at one order lower via the linear response relation
 \begin{align}
  \chi_{{\rm rad. }} = \frac12 \frac{\partial \chi_{{\rm cons.}}}{\partial J} J_{{\rm rad.}} + \calO(\alpha^4)\,.
 \end{align}
where $\chi_{{\rm cons.}}$ starts at $\calO(\alpha)$.

%================================================
%
\subsubsection*{High-energy limit of scattering angle}
%
%================================================
%
Similar to the discussion in Ref.~\cite{Saketh:2021sri}, we can investigate the behavior of our generic scattering angles of Eq.~\eqref{eq:soft_angle_2loop_eik} in special kinematic limits. One interesting regime is the high-energy limit which corresponds to the center-of-mass energy $E{=}\sqrt{m^2_1 + m^2_2 + 2m_1m_2\, \sigma} \gg (m_1 + m_2)$ for $m_1$ and $m_2$ held fixed. This is achieved by taking $\sigma \gg 1$. Note that this limit is \emph{not} equivalent to the massless limit where $m_1, m_2 {\to} 0$, which is outside the range of validity of our classical approximation $m^2_i \gg |q|^2$, since the Compton wavelength of the particles $\lambda_c \sim 1/m$ becomes large compared to the impact parameter $|b| \sim 1/|q|$, whereas classical physics requires $\lambda_c \ll |b|$. The exact details of the high-energy limit depend on the parameters that are held fixed in the scattering experiment. In particular, so far we have opted to write the scattering angle in terms of the large angular momentum $J$, related to the impact parameter $b$, 
\begin{align}
 J = \frac{m_1 m_2 \sqrt{\sigma^2 - 1} |b| }{\sqrt{m_1^2 + m_2^2 + 2 m_1 m_2\, \sigma}}
 \quad \stackrel{{\rm HE}}{\longrightarrow} \quad  
 \frac{|b| E}{2} +\calO(1/E) \,,
\end{align}
by an additional factor of $E$ in the high energy limit which shifts the overall energy dependence of the answer by overall powers of $E$ in the conversion of $L$-loop results which are proportional to $1/J^{L+1}$. 

We are now in the position to discuss the high energy limit of the scattering angles, both conservative and radiative. Up to $\calO(\alpha^3)$ and to leading order in the high-energy limit (in each individual charge sector), we find
\begin{align}
\begin{split}
  \chi_{{\rm cons.}} \Big|_{{\rm HE}} = &  
  \chi^{(0)}\Big|_{{\rm HE}}  +  \chi^{(1)}\Big|_{{\rm HE}} +\chi^{(2)}_{{\rm cons.}} \Big|_{{\rm HE}}  + \calO(\alpha^4)\,,
\end{split} 
\end{align}
where we have the explicit high-energy values of the scattering angles
\begin{align}
 \label{eq:chi_cons_tree_high_energy}
  \chi^{(0)}\Big|_{{\rm HE}}& = - \frac{4 \alpha  m_1 m_2}{E\, |b|} \! \left[\!\frac{q_1}{m_1}\!\right] \left[\!\frac{q_2}{m_2}\!\right] ,
  \\
   \label{eq:chi_cons_1loop_high_energy}
  \chi^{(1)}\Big|_{{\rm HE}}& =\frac{2 \pi  \alpha ^2 m_1^2 m_2^2 \left(m_1 + m_2\right)}{E^3\,|b| ^2} \! \left[\!\frac{q_1}{m_1}\!\right]^2 \!\left[\!\frac{q_2}{m_2}\!\right]^2 ,
  \\
 \label{eq:chi_cons_2loop_high_energy}
 \chi^{(2)}_{{\rm cons.}} \Big|_{{\rm HE}}& =\frac{16 \alpha^3 m_1^3 m_2^3}{3 E^3\, | b| ^3}\! \left[\!\frac{q_1}{m_1}\!\right]^3\!\left[\!\frac{q_2}{m_2}\!\right]^3,
\end{align}
in terms of the charge-to-mass ratios $r_i {\equiv} q_i/m_i$. The radiative angle is more interesting,
\begin{align}
\begin{split}
\label{eq:chi_rad_high_energy}
\hspace{-.5cm}
\chi^{(2)}_{{\rm rad.}} \Big|_{{\rm HE}}  & {=} 
 - \frac{16 \alpha ^3 m_1^3 m_2^3}{E^3\, |b|^3} \left[\!\frac{q_1}{m_1}\!\right]^3 \!\left[\!\frac{q_2}{m_2}\!\right]^3
 + \frac{8 \alpha ^3 m_1^2 m_2^2}{3\, E\, |b| ^3} \left(
\left[\!\frac{q_1}{m_1}\!\right]^4 \!\left[\!\frac{q_2}{m_2}\!\right]^2
 + 
\left[\!\frac{q_1}{m_1}\!\right]^2 \!\left[\!\frac{q_2}{m_2}\!\right]^4
\right) ,
\hspace{-.4cm}
\end{split}
\end{align}
which, as has been pointed out in Ref.~\cite{Saketh:2021sri}, develops a mass singularity in the mushroom sector due to uncanceled $1/m^2_i$ in the second term of Eq.~\eqref{eq:chi_rad_high_energy}. From our perspective, one should not take this as a true singularity of the result but rather as a breakdown of the classical approximation which requires $|q|^2\ll m^2_i$, or equivalently $m^2_i |b|^2 \gg 1$, which prevents us from setting $m_i{\to}0$. Presumably, one could take into account coherence effects to discuss the classical scattering of light states \cite{Cristofoli:2021vyo}. There, however, the particle interpretation breaks down and one should instead talk about the scattering of classical waves described by coherent states. 

Note that the mushroom contribution is dominant in the high-energy limit and behaves like $1/E$ at $\calO(\alpha^3)$ whereas the $q^3_1 q^3_2$ sector scales like $1/E^3$ at large $E$. Combining the high-energy limit of the conservative angle in Eq.~\eqref{eq:chi_cons_2loop_high_energy} with the high-energy limit of the $q^3_1 q^3_2$-sector of the radiative angle, we find
\begin{align}
\label{eq:chi_qed_iteration_2loops}
\chi^{(2)}_{q^3_1 q^3_2} \Big|_{{\rm HE}}  & {=} 
 - \frac{32 \alpha^3 m_1^3 m_2^3}{3 E^3\, | b| ^3}\! \left[\!\frac{q_1}{m_1}\!\right]^3\!\left[\!\frac{q_2}{m_2}\!\right]^3 
= \frac{\left[  \chi^{(0)}\Big|_{{\rm HE}} \right]^3}{3!}\,,
\end{align}
which is given in terms of the third power of the high-energy limit of the tree-level angle. The contribution from the mushroom diagrams on the other hand does not follow such an iterative structure at high energies. 

As noted in the introduction, we view QED as a useful model for gravitational calculations that not only lets us test computational setups in a simpler context, but also sheds light on important physical questions. To this end, we would like to compare the results of some classical observables in electrodynamics to the ones in GR. One of the simplest places of comparison is the high-energy limit of the scattering angle. First, at tree- and one-loop level, the leading order behavior of the gravitational scattering angle is~\cite{Westpfahl:1985tsl}
\begin{align}
  \chi^{\phantom{(0)}}_{{\rm GR}} \Big|_{{\rm HE}}  =  
  \chi^{(0)}_{{\rm GR}}\Big|_{{\rm HE}}  +  \chi^{(1)}_{{\rm GR}}\Big|_{{\rm HE}} + \calO(G^3)
  =
  \frac{4 GE }{|b|} + \frac{15 \pi  G^2 E (m_1 + m_2)}{4\, |b|^2}+\calO(G^3) \,.
\end{align}
The high-energy-limit of the tree-level gravitational scattering angle, $\chi^{(0)}_{{\rm GR}}\Big|_{{\rm HE}}$ is the double-copy of the QED angle with the replacement $q_i \to E$ and $\alpha \to G$. This is a direct consequence of the fact that the tree-level amplitude is dominated by the $t$-channel exchange diagram which is obtained as the product of two three-particle amplitudes. In the high-energy limit, these have the double-copy property. At one-loop, there is still some residual double-copy property present that is exposed by the same replacement as above. However, the numerical prefactor does not directly match anymore. 

At two-loop order, it is worthwhile to recall that the conservative scattering angle in GR~\cite{Bern:2019nnu,Bern:2019crd} contains a logarithmic high-energy singularity that is canceled by radiative contributions~\cite{DiVecchia:2020ymx,Damour:2020tta}
\begin{align}
\chi^{(2),{\rm GR}}_{{\rm rad.}} \Big|_{{\rm HE}} & =\frac{8 G^3 E^3  \left(6 \log \left(\frac{E^2}{m_1 m_2}\right)-5\right)}
										     {3\, |b|^{3}} \,,
										     \\
\chi^{(2),{\rm GR}}_{{\rm cons.}} \Big|_{{\rm HE}} & =  -\frac{8 G^3 E^3 \left(6 \log \left(\frac{E^2}{m_1 m_2}\right)-9\right)}{3\,|b|^{3}} \,,
\\
\left(\chi^{(2),{\rm GR}}_{{\rm cons.}} +\chi^{(2),{\rm GR}}_{{\rm rad.}}\right) \!\Big|_{{\rm HE}} 
		& = \frac{32 G^3 E^3}{3|b|^3} 
		   =  \frac{\left[  \chi^{(0)}_{{\rm GR}}\Big|_{{\rm HE}} \right]^3}{3!}\,.
\end{align}
The fact that the leading high-energy limit of the full scattering angle in gravity is given by the third power of the tree-level angle is furthermore responsible for the observed high-energy universality of gravitational scattering at $\calO(G^3)$~\cite{DiVecchia:2020ymx,Damour:2020tta}. In comparison, for QED, only the box-sector ($\sim q^3_1 q^3_2$) angle follows the same iteration structure (see Eq.~\eqref{eq:chi_qed_iteration_2loops}), whereas the mushroom terms (that are actually leading in the high-energy limit) do not.

Nonetheless, as was the case at one-loop order, the structure of the high-energy limit of the angle still follows a double-copy-like relation between QED and gravity under the replacement $q_i/m_i\to 1$, $m_i \to E$, and $\alpha \to G$. Notably, at two-loops in gravity, the logarithmic high-energy singularity of the individual conservative and radiative contributions to the angle is directly entangled with the mass singularity. This is due to the fact that both quantities appear linked as the argument of a logarithm $\log \frac{E^2}{m_1 m_2}$ so that one can think about mass or high-energy divergences interchangeably. In QED, this is no longer true, as we have explicitly seen in Eqs.~\eqref{eq:chi_rad_high_energy} and \eqref{eq:chi_cons_2loop_high_energy} above.  

Interestingly, in the conservative sector of GR at $\calO(G^4)$, dimensional analysis and explicit computation exposed a $1/m$ singularity~\cite{Bern:2021dqo} whose fate is unclear once radiation effects will ultimately be taken into account. Ref.~\cite{Bern:2021dqo} speculated that the $1/m$ is canceled by radiative contributions similar to the cancellation of the $\log m$ at $\calO(G^3)$; however, it is also possible that this term remains and merely signals the breakdown of the classical approximation which requires that $m^2\gg |q|^2$. As such, QED seems to be an ideal toy example that exposes these features already at lower-loop order than gravity which is a consequence of the different mass dimensions of the electromagnetic ($\alpha$) and gravitational ($G$) coupling constants. An explicit investigation of this aspect in gravity is an interesting open problem.

%================================================
%
\subsection{Radiative impulse, energy loss, and scattering angle via KMOC}
%
%================================================
Besides the eikonal analysis of the previous subsection, we have furthermore analyzed the classical two-loop scalar QED observables in the KMOC framework in complete analogy to the gravitational case that has been obtained by two of the authors~\cite{Herrmann:2021tct}. Due to the linearity of QED, the present analysis is a lot simpler due to the limited number of diagrams appearing in the computation. In fact, we were able to use exactly the same analysis pipeline from before, just substituting in the new QED integrands from section \ref{subsec:integrands}. 

We first assemble all relevant terms for the transverse part of the impulse, which receives both \emph{virtual} and \emph{real} contributions according to \eqref{eq:impulse_kernel_perp}. Now that we are in the soft region, we do have situations where internal photons are allowed to go on-shell, so that we have to both consider two- and three-particle cuts at $\calO(\alpha^3)$ to find
\begin{align}
\Delta p^\mu_{1,\perp} =& \frac{(\alpha q_1 q_2)^3}{m_1 m_2}
\!\!  
\left[
\frac{8 \sigma^2 \text{arcsinh}\sqrt{\frac{\sigma-1}{2}}}{\left(\sigma^2 - 1\right)^{5/2}}
 - \frac{4 \sigma ^3}{ \left(\sigma^2 - 1\right)^2}
 + \frac{2 \left(2 m_1 m_2 \left(\sigma^4 {-} \sigma^2 {+} 1\right)+(m_1^2 {+} m_2^2)\sigma \right)}{m_1 m_2 \left(\sigma^2 - 1\right)^{5/2}}
\right] \!\!
\frac{b^{\mu }}{|b|^4} 
\nonumber\\
&\hspace{2cm} +\alpha^3\,q_1^2 q_2^2 \, \frac{4\sigma ^2 \left(m_2^2\, q_1^2 + m_1^2 \,q_2^2\right)}{3 m_1^2 m_2^2 \left(\sigma^2 - 1\right)} \frac{b^{\mu }}{|b|^4}\,.
\end{align}
In comparison to the conservative sector, we now have two different charge sectors, the one with $(q_1 q_2)^3$ dependence and the other with $q^2_i q^4_j$. In QED, these sectors are independently gauge-invariant and do not interfer. This is distinct from the gravitational setting, where all results were proportional to $G^3$ which can be roughly understood from a double-copy point of view where the charges in gauge theory $q_i$ get replaced by the gravitational charges, i.e. the masses $m_i$. Unlike in the conservative sector, now the $\text{arcsinh}\sqrt{\frac{\sigma-1}{2}}$ appears, but is not proportional to any leading power of $\sigma$ so that the ultrarelativistic limit is well-behaved. From the above result, we can subtract the conservative contribution to single out the radiative part of the impulse
\begin{align}
\Delta p^{\mu,{\rm rad.}}_{1,\perp} =& \frac{\alpha ^3 q_1^2 q_2^2}{m_1 m_2} \frac{b^\mu }{| b| ^4} 
\Bigg[
q_1q_2\left(\frac{8 \sigma^2 \text{arcsinh}\sqrt{\frac{\sigma-1}{2}}}{\left(\sigma^2 - 1\right)^{5/2}}  -  \frac{4 \sigma ^3}{ \left(\sigma^2 - 1\right)^2} \right)
+ \frac{4  \sigma ^2 \left(m_2^2 q_1^2+m_1^2 q_2^2\right)}{3 m_1 m_2 \left(\sigma ^2-1\right)}
\Bigg]\,. \nonumber \\
\end{align}
From the radiative transverse impulse, one can extract the radiative correction to the scattering angle~\cite{Herrmann:2021tct,Saketh:2021sri} which equally splits into different charge sectors
\begin{align}
\hspace{-.4cm}
\chi^{(2)}_{{\rm rad}}{=}\alpha^3 q_1^2 q_2^2
\frac{4\sigma ^2\!\! \left[3 m_1 m_2 q_1 q_2 \left(2 \text{arcsinh}\sqrt{\frac{\sigma-1}{2}}  - \sigma  \sqrt{\sigma ^2 - 1}\right)  +  (m_2^2 q_1^2  +  m_1^2 q_2^2) \left(\sigma ^2 - 1\right)^{\frac32} \right]}
{3 J^3 \left(\sigma ^2-1\right)^{3/2} \left(m_1^2 + m_2^2 + 2 m_1 m_2\, \sigma\right)}\,.
\end{align}
For the longitudinal part, we only give the additional contribution due to radiation effects. In order to obtain the full longitudinal part, one would have to add the conservative result from Eq.~(\ref{eq:impulse_cons_long_p1})
\begin{align}
\Delta p^{\mu, {\rm rad.}}_{1,u}  = \frac{\alpha^3 q^2_1 q^2_2 \pi}{m_1 m_2\, | b| ^3}  \check{u}^\mu_2 \Bigg[ & 
q_1 q_2
\frac{ \left(3 \sigma^3 {-} 4 \sigma ^2 {+} 9 \sigma {-} 4\right)(\sigma^2 - 1) 
	- 2 \left(3 \sigma ^2 {+} 1\right) \sqrt{\sigma^2 {-} 1}\, \text{arcsinh}\sqrt{\frac{\sigma-1}{2}}}
     {4  \left(\sigma^2 - 1\right)^{5/2}} 
\nonumber \\
& -\frac{\left(3 \sigma ^2+1\right)  \left(m_2^2 \, \sigma\, q_1^2  + m_1^2\, q_2^2\right) }{12 m_1 m_2 \sqrt{\sigma^2-1}}
\Bigg]
\,,
\end{align}
where the radiative part of the momentum change on particle 1 is purely in the direction of $\check{u}^\mu_2$. We now can compute the radiated momentum in the scattering of two charged scalars in QED in two different ways. First, we simply take the impulse on particle 1 plus the impulse on particle 2 (which can be obtained from the above results by simple relabeling) together with momentum conservation $\Delta p^\mu_1 + \Delta p^\mu_2 + \Delta R^\mu =0$. Alternatively, we can directly compute the radiated momentum from KMOC, see Eq.~\eqref{eq:KMOC_kernel_rad} to find 
\begin{align}
\hspace{-.8cm}
 \Delta R^\mu {=} \frac{\alpha^3 q^2_1 q^2_2 \pi}{|b|^3}\!  \Bigg[ & \! 
 \frac{q_1 q_2}{m_1 m_2} \frac{u_1^{\mu } + u_2^{\mu }}{\sigma+1} 
   \Bigg\{\!\frac{2 \left(3 \sigma^2 + 1\right)  \text{arcsinh}\sqrt{\frac{\sigma-1}{2}} -\sqrt{\sigma^2 - 1} \left(3 \sigma^3 - 4 \sigma ^2 + 9 \sigma - 4\right)}{4(\sigma^2-1)^2}\!\Bigg\} 
 \nn \\
 &
 +
 \frac{\left(3 \sigma^2 + 1\right) \left(m_2^2 q_1^2 u_1^{\mu }+m_1^2 q_2^2 u_2^{\mu }\right)}{12 m_1^2 m_2^2 \sqrt{\sigma ^2-1}}
 \Bigg]\,,
\hspace{-.8cm}
\end{align}
which points entirely along the longitudinal directions $u^\mu_{1}$ and $u^\mu_{2}$ with appropriate strength proportional to the charge-to-mass ratios $q_i/m_i$ of the scattering particles. In comparison to the GR results obtained previously~\cite{Herrmann:2021lqe}, where the masses are always positive, the electric charges can change sign. For the $q^3_1 q^2_2$ charge sector, this implies that the contribution to the radiated energy can be positive or negative. However, the combined energy loss e.g. in the center-of-mass system 
\begin{align}
\Delta E_{{\rm c.m}} = \Delta R \cdot (u_1 + u_2)\,,
\end{align}
is numerically positive for arbitrary values of the charge $q_1/q_2$ and mass $m_1 / m_2$ ratios.

%================================================
%
\subsection{Radiative radial action}
%
%================================================

For completeness, we give the explicit results for the \emph{soft radial action} for which we found an empirical shortcut in terms of explicit subtractions of master integrals that has been described in section~\ref{subsec:soft_radial_action_subtraction_scheme}. Upon Fourier-transforming to impact-parameter space $b$, the two-loop soft radial action is 
\begin{align}
\begin{split}
\hspace{-.2cm}
 I^{(2)}_{r} (J,\sigma) =\null & \frac{(\alpha q_1 q_2)^3} {3 J^2 \left(\sigma^2 - 1\right)^{3/2} \left(m_1^2 + m_2^2 + 2 m_1 m_2 \sigma \right)} \Bigg[
 \left(m_1^2 + m_2^2\right) \sigma  \left(2 \sigma ^2 - 3\right)   
\\
& \hskip .7 cm
  - 2 m_1 m_2  \left\{ 3 + \!\left(\sigma^2 - 3 \sigma \sqrt{\sigma^2 - 1} - 3\right)\! \sigma^2 + 6 \sigma^2 \text{arcsinh}\sqrt{\frac{\sigma - 1}{2}} \, \right \}
 \Bigg]
 \\[5pt]
 & \hskip .1 cm \null
 -\frac{2 \alpha ^3 q_1^2 q_2^2 \ \sigma ^2 \left(m_2^2 q_1^2+m_1^2 q_2^2\right)}
         {3 J^2  \left(m_1^2 + m_2^2 + 2 m_1 m_2 \sigma \right)} \,,
\end{split}         
\end{align}
%
\iffalse
\begin{align}
 I^{(2)}_{r} (J,\sigma) = & \frac{(\alpha q_1 q_2)^3 \left[
 \left(m_1^2 + m_2^2\right) \sigma  \left(2 \sigma ^2 - 3\right)
  - 2 m_1 m_2 \left[3 + \!\left(\sigma^2 - 3 \sqrt{\sigma^2 - 1} \sigma - 3\right)\! \sigma^2 + 6 \sigma^2 \text{arcsinh}\sqrt{\frac{\sigma - 1}{2}}\right]
 \right]}
 {3 J^2 \left(\sigma^2 - 1\right)^{3/2} \left(m_1^2 + m_2^2 + 2 m_1 m_2 \sigma \right)} 
 \nn \\
 & 
 -\frac{2 \alpha ^3 q_1^2 q_2^2 \ \sigma ^2 \left(m_2^2 q_1^2+m_1^2 q_2^2\right)}
 	 {3\,  J^2  \left(m_1^2 + m_2^2 + 2 m_1 m_2 \sigma \right)} \,,
\end{align}
\fi
%
from which we find the scattering angle 
\begin{align}
 \chi^{(2)} = \frac{\partial I^{(2)}_r(J,\sigma)}{\partial J}\,,
\end{align}
which exactly agrees with our eikonal result in Eq.~\eqref{eq:soft_angle_2loop_eik} and the ones that can be obtained from the impulse via the KMOC setup. At this point, our prescription for the modification of the boundary conditions for the soft master integrals was inspired by an analogous procedure in the conservative sector and it would be interesting to understand this method more systematically so that it can be applied at higher orders. It would be very interesting to find a direct computational method for the phase of the S-matrix in the representation of Ref.~\cite{Damgaard:2021ipf} and the relation to the velocity cuts of Ref.~\cite{Bjerrum-Bohr:2021wwt} as well as the heavy-particle effective theory implementation of related ideas of Ref.~\cite{Brandhuber:2021eyq}.

%================================================
%
\vspace{-.3cm}
\section{Conclusions}
\label{sec:conclusions}
\vspace{-.3cm}
%
%================================================

In this paper we applied scattering amplitude techniques to study the
classical two-body scattering problem in scalar electrodynamics in a
regime similar to the post-Minkowskian expansion of general
relativity.  Scalar QED has many similarities to GR, 
except that it is far simpler because the underlying
interactions are linear, making it useful as a toy model of the
gravitational problem~\cite{Westpfahl:1985tsl,Buonanno:2000qq}.  The
recent calculation of the scattering angle through $\calO (\alpha^3)$,
including radiative effects, by iteratively solving the classical
equations of motion~\cite{Saketh:2021sri} motivated the computation of
the corresponding results from amplitudes based approaches.  We
derived the classical scattering angle from both the eikonal
phase~\cite{Glauber:1956Lecture} of the amplitude, and from the KMOC
formalism~\cite{Kosower:2018adc}.  We also noted a simplified
computation of the radial action including radiative effects. The
`soft radial action' was obtained by a change in the boundary
conditions of the differential equations for the soft master integrals.
This automatically removes classically singular iterations and
directly yields the radial action (including radiative contributions)
up to two-loop order from which we obtain the classical
scattering angle.  All three approaches give the same scattering angle
which furthermore agrees with the one found by solving the classical
equations of motion~\cite{Saketh:2021sri}.

An interesting feature of the scattering angle in QED is that, at
$\calO(\alpha^3)$, it contains factors of $1/m_i$ which are singular
for $m_i \rightarrow 0$, despite having included both conservative and
radiative effects consistently.  Of course, this region is outside the
validity of our setup since it violates the classical requirement that
the Compton wavelength is smaller than the inter-particle separation.
Translated to momentum space this amounts to the classical hierarchy
of scales $-q^2\ll m^2_i$ where the masses are larger than the
momentum transfer which is violated for $m_i{\to}0$.  The same
breakdown of the approximation was found in
Ref.~\cite{Saketh:2021sri}, where it was tied to terms generated via
the ALD force~\cite{lorentz:1892theorie, abraham:1912theorie,
  Dirac:1938a}.  In the classical setup, the ALD force has to be added
`by hand' to account for radiation loss during the scattering event.
In contrast, in amplitudes-based frameworks, the radiation loss is
taken into account automatically and is on an equal footing with other
terms in the amplitudes. The appearance of the mass singularities in
QED may be contrasted with the corresponding $\calO(G^3)$ scattering
angle in GR: once radiative effects are included the mass singularity
cancels~\cite{DiVecchia:2020ymx,Damour:2020tta}.  On the other hand,
at $\calO (G^4)$ the pure potential contributions to the scattering
angle contain a $1/m$ singularity~\cite{Bern:2021dqo}.  Our results
for the electrodynamics suggest that these may not cancel even after
including gravitational radiation.

Following Refs.~\cite{Kalin:2019rwq, Kalin:2019inp,
  Saketh:2021sri,Cho:2021arx} many physical observables can be
analytically continued from the scattering problem to the bound-state
problem.  In particular, the scattering angle is directly tied to the
periastron advance. In gravity, at $\calO (G^4)$ this is greatly
complicated by the nonlocal-in-time tail effect~\cite{Thorne:1980ru, Blanchet:1987wq, Glauber:1956Lecture, Blanchet:1993ec,Cho:2021arx}.
Since the tail effect arises from gravitational radiation
interacting with potential gravitons (due to nonlinear graviton
couplings in GR), electrodynamics should be immune from this
complication, given that photons do not self interact classically in 
the absence of nonlinear higher derivative corrections to QED. 

In summary, we expect electrodynamics to continue to serve as  
a useful toy model for the classical gravitational binary dynamics.

%===============================================
\subsubsection*{Acknowledgements}
%===============================================
%
We thank Clifford Cheung, Michael Ruf, and Mikhail Solon for very enlightening discussions.
We also thank Alessandra Buonanno, Justin Vines, Jan Steinhoff, and
Muddu Saketh for very helpful discussions and for sharing the results
of Ref.~\cite{Saketh:2021sri} prior to publication, where the
scattering angle is derived from the classical equations of motion. We thank 
Julio Parra-Martinez for discussions and comments on the manuscript.
This work was supported in part by the U.S. Department of Energy (DOE)
under Award Number \mbox{DE-SC0009937}.  
This project has received funding from the European Union’s Horizon 2020 research and innovation program under the Marie Sklodowska-Curie grant agreement No. 847523 ‘INTERACTIONS’. M.Z.'s work is supported by the U.K.\ Royal Society through Grant URF{\textbackslash}R1{\textbackslash}20109.
We are also grateful for support from the Mani L. Bhaumik Institute for Theoretical Physics.

%================================================

%================================================
%    References (& /Document)
%================================================
\bibliographystyle{JHEP}
\bibliography{jhep_refs.bib}

\providecommand{\href}[2]{#2}\begingroup\raggedright\begin{thebibliography}{100}

\bibitem{Abbott:2016blz}
{\scshape LIGO Scientific, Virgo} collaboration, B.~Abbott et~al.,
  \emph{{Observation of Gravitational Waves from a Binary Black Hole Merger}},
  \href{https://doi.org/10.1103/PhysRevLett.116.061102}{\emph{Phys. Rev. Lett.}
  {\bfseries 116} (2016) 061102}
  [\href{https://arxiv.org/abs/1602.03837}{{\ttfamily 1602.03837}}].

\bibitem{TheLIGOScientific:2017qsa}
{\scshape LIGO Scientific, Virgo} collaboration, B.~Abbott et~al.,
  \emph{{GW170817: Observation of Gravitational Waves from a Binary Neutron
  Star Inspiral}},
  \href{https://doi.org/10.1103/PhysRevLett.119.161101}{\emph{Phys. Rev. Lett.}
  {\bfseries 119} (2017) 161101}
  [\href{https://arxiv.org/abs/1710.05832}{{\ttfamily 1710.05832}}].

\bibitem{Punturo:2010zz}
M.~Punturo et~al., \emph{{The Einstein Telescope: A third-generation
  gravitational wave observatory}},
  \href{https://doi.org/10.1088/0264-9381/27/19/194002}{\emph{Class. Quant.
  Grav.} {\bfseries 27} (2010) 194002}.

\bibitem{Dwyer:2014fpa}
S.~Dwyer, D.~Sigg, S.~W. Ballmer, L.~Barsotti, N.~Mavalvala and M.~Evans,
  \emph{{Gravitational wave detector with cosmological reach}},
  \href{https://doi.org/10.1103/PhysRevD.91.082001}{\emph{Phys. Rev. D}
  {\bfseries 91} (2015) 082001}
  [\href{https://arxiv.org/abs/1410.0612}{{\ttfamily 1410.0612}}].

\bibitem{LISA:2017pwj}
{\scshape LISA} collaboration, P.~Amaro-Seoane et~al., \emph{{Laser
  Interferometer Space Antenna}},
  \href{https://arxiv.org/abs/1702.00786}{{\ttfamily 1702.00786}}.

\bibitem{Reitze:2019iox}
D.~Reitze et~al., \emph{{Cosmic Explorer: The U.S. Contribution to
  Gravitational-Wave Astronomy beyond LIGO}}, {\emph{Bull. Am. Astron. Soc.}
  {\bfseries 51} (2019) 035}
  [\href{https://arxiv.org/abs/1907.04833}{{\ttfamily 1907.04833}}].

\bibitem{Buonanno:1998gg}
A.~Buonanno and T.~Damour, \emph{{Effective one-body approach to general
  relativistic two-body dynamics}},
  \href{https://doi.org/10.1103/PhysRevD.59.084006}{\emph{Phys. Rev. D}
  {\bfseries 59} (1999) 084006}
  [\href{https://arxiv.org/abs/gr-qc/9811091}{{\ttfamily gr-qc/9811091}}].

\bibitem{Pretorius:2005gq}
F.~Pretorius, \emph{{Evolution of binary black hole spacetimes}},
  \href{https://doi.org/10.1103/PhysRevLett.95.121101}{\emph{Phys. Rev. Lett.}
  {\bfseries 95} (2005) 121101}
  [\href{https://arxiv.org/abs/gr-qc/0507014}{{\ttfamily gr-qc/0507014}}].

\bibitem{Campanelli:2005dd}
M.~Campanelli, C.~O. Lousto, P.~Marronetti and Y.~Zlochower, \emph{{Accurate
  evolutions of orbiting black-hole binaries without excision}},
  \href{https://doi.org/10.1103/PhysRevLett.96.111101}{\emph{Phys. Rev. Lett.}
  {\bfseries 96} (2006) 111101}
  [\href{https://arxiv.org/abs/gr-qc/0511048}{{\ttfamily gr-qc/0511048}}].

\bibitem{Baker:2005vv}
J.~G. Baker, J.~Centrella, D.-I. Choi, M.~Koppitz and J.~van Meter,
  \emph{{Gravitational wave extraction from an inspiraling configuration of
  merging black holes}},
  \href{https://doi.org/10.1103/PhysRevLett.96.111102}{\emph{Phys. Rev. Lett.}
  {\bfseries 96} (2006) 111102}
  [\href{https://arxiv.org/abs/gr-qc/0511103}{{\ttfamily gr-qc/0511103}}].

\bibitem{Mino:1996nk}
Y.~Mino, M.~Sasaki and T.~Tanaka, \emph{{Gravitational radiation reaction to a
  particle motion}},
  \href{https://doi.org/10.1103/PhysRevD.55.3457}{\emph{Phys. Rev. D}
  {\bfseries 55} (1997) 3457}
  [\href{https://arxiv.org/abs/gr-qc/9606018}{{\ttfamily gr-qc/9606018}}].

\bibitem{Quinn:1996am}
T.~C. Quinn and R.~M. Wald, \emph{{An Axiomatic approach to electromagnetic and
  gravitational radiation reaction of particles in curved space-time}},
  \href{https://doi.org/10.1103/PhysRevD.56.3381}{\emph{Phys. Rev. D}
  {\bfseries 56} (1997) 3381}
  [\href{https://arxiv.org/abs/gr-qc/9610053}{{\ttfamily gr-qc/9610053}}].

\bibitem{PNDroste}
J.~Droste, \emph{{The field of $n$ moving centres in Einstein's theory of
  gravitation}}, {\emph{Proc. Acad. Sci. Amst.} {\bfseries 26} (1916) 447}.

\bibitem{Einstein:1938yz}
A.~Einstein, L.~Infeld and B.~Hoffmann, \emph{{The Gravitational equations and
  the problem of motion}}, \href{https://doi.org/10.2307/1968714}{\emph{Annals
  Math.} {\bfseries 39} (1938) 65}.

\bibitem{Goldberger:2004jt}
W.~D. Goldberger and I.~Z. Rothstein, \emph{{An Effective field theory of
  gravity for extended objects}},
  \href{https://doi.org/10.1103/PhysRevD.73.104029}{\emph{Phys. Rev. D}
  {\bfseries 73} (2006) 104029}
  [\href{https://arxiv.org/abs/hep-th/0409156}{{\ttfamily hep-th/0409156}}].

\bibitem{Kol:2007rx}
B.~Kol and M.~Smolkin, \emph{{Classical Effective Field Theory and Caged Black
  Holes}}, \href{https://doi.org/10.1103/PhysRevD.77.064033}{\emph{Phys. Rev.
  D} {\bfseries 77} (2008) 064033}
  [\href{https://arxiv.org/abs/0712.2822}{{\ttfamily 0712.2822}}].

\bibitem{Kol:2007bc}
B.~Kol and M.~Smolkin, \emph{{Non-Relativistic Gravitation: From Newton to
  Einstein and Back}},
  \href{https://doi.org/10.1088/0264-9381/25/14/145011}{\emph{Class. Quant.
  Grav.} {\bfseries 25} (2008) 145011}
  [\href{https://arxiv.org/abs/0712.4116}{{\ttfamily 0712.4116}}].

\bibitem{Gilmore:2008gq}
J.~B. Gilmore and A.~Ross, \emph{{Effective field theory calculation of second
  post-Newtonian binary dynamics}},
  \href{https://doi.org/10.1103/PhysRevD.78.124021}{\emph{Phys. Rev. D}
  {\bfseries 78} (2008) 124021}
  [\href{https://arxiv.org/abs/0810.1328}{{\ttfamily 0810.1328}}].

\bibitem{Foffa:2011ub}
S.~Foffa and R.~Sturani, \emph{{Effective field theory calculation of
  conservative binary dynamics at third post-Newtonian order}},
  \href{https://doi.org/10.1103/PhysRevD.84.044031}{\emph{Phys. Rev. D}
  {\bfseries 84} (2011) 044031}
  [\href{https://arxiv.org/abs/1104.1122}{{\ttfamily 1104.1122}}].

\bibitem{Foffa:2016rgu}
S.~Foffa, P.~Mastrolia, R.~Sturani and C.~Sturm, \emph{{Effective field theory
  approach to the gravitational two-body dynamics, at fourth post-Newtonian
  order and quintic in the Newton constant}},
  \href{https://doi.org/10.1103/PhysRevD.95.104009}{\emph{Phys. Rev. D}
  {\bfseries 95} (2017) 104009}
  [\href{https://arxiv.org/abs/1612.00482}{{\ttfamily 1612.00482}}].

\bibitem{Porto:2017dgs}
R.~A. Porto and I.~Z. Rothstein, \emph{{Apparent ambiguities in the
  post-Newtonian expansion for binary systems}},
  \href{https://doi.org/10.1103/PhysRevD.96.024062}{\emph{Phys. Rev. D}
  {\bfseries 96} (2017) 024062}
  [\href{https://arxiv.org/abs/1703.06433}{{\ttfamily 1703.06433}}].

\bibitem{Foffa:2019hrb}
S.~Foffa, P.~Mastrolia, R.~Sturani, C.~Sturm and W.~J. Torres~Bobadilla,
  \emph{{Static two-body potential at fifth post-Newtonian order}},
  \href{https://doi.org/10.1103/PhysRevLett.122.241605}{\emph{Phys. Rev. Lett.}
  {\bfseries 122} (2019) 241605}
  [\href{https://arxiv.org/abs/1902.10571}{{\ttfamily 1902.10571}}].

\bibitem{Blumlein:2019zku}
J.~Bl\"umlein, A.~Maier and P.~Marquard, \emph{{Five-Loop Static Contribution
  to the Gravitational Interaction Potential of Two Point Masses}},
  \href{https://doi.org/10.1016/j.physletb.2019.135100}{\emph{Phys. Lett. B}
  {\bfseries 800} (2020) 135100}
  [\href{https://arxiv.org/abs/1902.11180}{{\ttfamily 1902.11180}}].

\bibitem{Foffa:2019rdf}
S.~Foffa and R.~Sturani, \emph{{Conservative dynamics of binary systems to
  fourth Post-Newtonian order in the EFT approach I: Regularized Lagrangian}},
  \href{https://doi.org/10.1103/PhysRevD.100.024047}{\emph{Phys. Rev. D}
  {\bfseries 100} (2019) 024047}
  [\href{https://arxiv.org/abs/1903.05113}{{\ttfamily 1903.05113}}].

\bibitem{Foffa:2019yfl}
S.~Foffa, R.~A. Porto, I.~Rothstein and R.~Sturani, \emph{{Conservative
  dynamics of binary systems to fourth Post-Newtonian order in the EFT approach
  II: Renormalized Lagrangian}},
  \href{https://doi.org/10.1103/PhysRevD.100.024048}{\emph{Phys. Rev. D}
  {\bfseries 100} (2019) 024048}
  [\href{https://arxiv.org/abs/1903.05118}{{\ttfamily 1903.05118}}].

\bibitem{Blumlein:2020pog}
J.~Bl\"umlein, A.~Maier, P.~Marquard and G.~Sch\"afer, \emph{{Fourth
  post-Newtonian Hamiltonian dynamics of two-body systems from an effective
  field theory approach}},
  \href{https://doi.org/10.1016/j.nuclphysb.2020.115041}{\emph{Nucl. Phys. B}
  {\bfseries 955} (2020) 115041}
  [\href{https://arxiv.org/abs/2003.01692}{{\ttfamily 2003.01692}}].

\bibitem{Bertotti:1956}
B.~Bertotti, \emph{{On gravitational motion}},
  \href{https://doi.org/10.1007/BF02746175}{\emph{Nuovo Cimento} {\bfseries 4}
  (1956) 898}.

\bibitem{Kerr:1959zlt}
R.~P. Kerr, \emph{{The Lorentz-covariant approximation method in general
  relativity I}}, \href{https://doi.org/10.1007/bf02732767}{\emph{Nuovo Cim.}
  {\bfseries 13} (1959) 469}.

\bibitem{Bertotti:1960wuq}
B.~Bertotti and J.~Plebanski, \emph{{Theory of gravitational perturbations in
  the fast motion approximation}},
  \href{https://doi.org/10.1016/0003-4916(60)90132-9}{\emph{Annals Phys.}
  {\bfseries 11} (1960) 169}.

\bibitem{Portilla:1979xx}
M.~Portilla, \emph{{Momentum and angular momentum of two gravitating
  particles}}, \href{https://doi.org/10.1088/0305-4470/12/7/025}{\emph{J. Phys.
  A} {\bfseries 12} (1979) 1075}.

\bibitem{Westpfahl:1979gu}
K.~Westpfahl and M.~Goller, \emph{{Gravitational scattering of two relativistic
  particles in postlinear approximation}},
  \href{https://doi.org/10.1007/BF02817047}{\emph{Lett. Nuovo Cim.} {\bfseries
  26} (1979) 573}.

\bibitem{Bel:1981be}
L.~Bel, T.~Damour, N.~Deruelle, J.~Ibanez and J.~Martin,
  \emph{{Poincar\'e-invariant gravitational field and equations of motion of
  two pointlike objects: The postlinear approximation of general relativity}},
  \href{https://doi.org/10.1007/BF00756073}{\emph{Gen. Rel. Grav.} {\bfseries
  13} (1981) 963}.

\bibitem{Westpfahl:1985tsl}
K.~Westpfahl, \emph{{High-Speed Scattering of Charged and Uncharged Particles
  in General Relativity}},
  \href{https://doi.org/10.1002/prop.2190330802}{\emph{Fortsch. Phys.}
  {\bfseries 33} (1985) 417}.

\bibitem{Ledvinka:2008tk}
T.~Ledvinka, G.~Schaefer and J.~Bicak, \emph{{Relativistic Closed-Form
  Hamiltonian for Many-Body Gravitating Systems in the Post-Minkowskian
  Approximation}},
  \href{https://doi.org/10.1103/PhysRevLett.100.251101}{\emph{Phys. Rev. Lett.}
  {\bfseries 100} (2008) 251101}
  [\href{https://arxiv.org/abs/0807.0214}{{\ttfamily 0807.0214}}].

\bibitem{Damour:2017zjx}
T.~Damour, \emph{{High-energy gravitational scattering and the general
  relativistic two-body problem}},
  \href{https://doi.org/10.1103/PhysRevD.97.044038}{\emph{Phys. Rev. D}
  {\bfseries 97} (2018) 044038}
  [\href{https://arxiv.org/abs/1710.10599}{{\ttfamily 1710.10599}}].

\bibitem{Bjerrum-Bohr:2013bxa}
N.~E.~J. Bjerrum-Bohr, J.~F. Donoghue and P.~Vanhove, \emph{{On-shell
  Techniques and Universal Results in Quantum Gravity}},
  \href{https://doi.org/10.1007/JHEP02(2014)111}{\emph{JHEP} {\bfseries 02}
  (2014) 111} [\href{https://arxiv.org/abs/1309.0804}{{\ttfamily 1309.0804}}].

\bibitem{Bjerrum-Bohr:2018xdl}
N.~E.~J. Bjerrum-Bohr, P.~H. Damgaard, G.~Festuccia, L.~Plant\'e and
  P.~Vanhove, \emph{{General Relativity from Scattering Amplitudes}},
  \href{https://doi.org/10.1103/PhysRevLett.121.171601}{\emph{Phys. Rev. Lett.}
  {\bfseries 121} (2018) 171601}
  [\href{https://arxiv.org/abs/1806.04920}{{\ttfamily 1806.04920}}].

\bibitem{Cheung:2018wkq}
C.~Cheung, I.~Z. Rothstein and M.~P. Solon, \emph{{From Scattering Amplitudes
  to Classical Potentials in the Post-Minkowskian Expansion}},
  \href{https://doi.org/10.1103/PhysRevLett.121.251101}{\emph{Phys. Rev. Lett.}
  {\bfseries 121} (2018) 251101}
  [\href{https://arxiv.org/abs/1808.02489}{{\ttfamily 1808.02489}}].

\bibitem{Kosower:2018adc}
D.~A. Kosower, B.~Maybee and D.~O'Connell, \emph{{Amplitudes, Observables, and
  Classical Scattering}},
  \href{https://doi.org/10.1007/JHEP02(2019)137}{\emph{JHEP} {\bfseries 02}
  (2019) 137} [\href{https://arxiv.org/abs/1811.10950}{{\ttfamily
  1811.10950}}].

\bibitem{Maybee:2019jus}
B.~Maybee, D.~O'Connell and J.~Vines, \emph{{Observables and amplitudes for
  spinning particles and black holes}},
  \href{https://doi.org/10.1007/JHEP12(2019)156}{\emph{JHEP} {\bfseries 12}
  (2019) 156} [\href{https://arxiv.org/abs/1906.09260}{{\ttfamily
  1906.09260}}].

\bibitem{Bern:2019nnu}
Z.~Bern, C.~Cheung, R.~Roiban, C.-H. Shen, M.~P. Solon and M.~Zeng,
  \emph{{Scattering Amplitudes and the Conservative Hamiltonian for Binary
  Systems at Third Post-Minkowskian Order}},
  \href{https://doi.org/10.1103/PhysRevLett.122.201603}{\emph{Phys. Rev. Lett.}
  {\bfseries 122} (2019) 201603}
  [\href{https://arxiv.org/abs/1901.04424}{{\ttfamily 1901.04424}}].

\bibitem{Bern:2019crd}
Z.~Bern, C.~Cheung, R.~Roiban, C.-H. Shen, M.~P. Solon and M.~Zeng,
  \emph{{Black Hole Binary Dynamics from the Double Copy and Effective
  Theory}}, \href{https://doi.org/10.1007/JHEP10(2019)206}{\emph{JHEP}
  {\bfseries 10} (2019) 206}
  [\href{https://arxiv.org/abs/1908.01493}{{\ttfamily 1908.01493}}].

\bibitem{Antonelli:2019ytb}
A.~Antonelli, A.~Buonanno, J.~Steinhoff, M.~van~de Meent and J.~Vines,
  \emph{{Energetics of two-body Hamiltonians in post-Minkowskian gravity}},
  \href{https://doi.org/10.1103/PhysRevD.99.104004}{\emph{Phys. Rev. D}
  {\bfseries 99} (2019) 104004}
  [\href{https://arxiv.org/abs/1901.07102}{{\ttfamily 1901.07102}}].

\bibitem{KoemansCollado:2019ggb}
A.~Koemans~Collado, P.~Di~Vecchia and R.~Russo, \emph{{Revisiting the second
  post-Minkowskian eikonal and the dynamics of binary black holes}},
  \href{https://doi.org/10.1103/PhysRevD.100.066028}{\emph{Phys. Rev. D}
  {\bfseries 100} (2019) 066028}
  [\href{https://arxiv.org/abs/1904.02667}{{\ttfamily 1904.02667}}].

\bibitem{Cristofoli:2020uzm}
A.~Cristofoli, P.~H. Damgaard, P.~Di~Vecchia and C.~Heissenberg,
  \emph{{Second-order Post-Minkowskian scattering in arbitrary dimensions}},
  \href{https://doi.org/10.1007/JHEP07(2020)122}{\emph{JHEP} {\bfseries 07}
  (2020) 122} [\href{https://arxiv.org/abs/2003.10274}{{\ttfamily
  2003.10274}}].

\bibitem{Blumlein:2020znm}
J.~Bl\"umlein, A.~Maier, P.~Marquard and G.~Sch\"afer, \emph{{Testing binary
  dynamics in gravity at the sixth post-Newtonian level}},
  \href{https://doi.org/10.1016/j.physletb.2020.135496}{\emph{Phys. Lett. B}
  {\bfseries 807} (2020) 135496}
  [\href{https://arxiv.org/abs/2003.07145}{{\ttfamily 2003.07145}}].

\bibitem{Cheung:2020gyp}
C.~Cheung and M.~P. Solon, \emph{{Classical gravitational scattering at $
  \mathcal{O} $(G$^{3}$) from Feynman diagrams}},
  \href{https://doi.org/10.1007/JHEP06(2020)144}{\emph{JHEP} {\bfseries 06}
  (2020) 144} [\href{https://arxiv.org/abs/2003.08351}{{\ttfamily
  2003.08351}}].

\bibitem{Bini:2020wpo}
D.~Bini, T.~Damour and A.~Geralico, \emph{{Binary dynamics at the fifth and
  fifth-and-a-half post-Newtonian orders}},
  \href{https://doi.org/10.1103/PhysRevD.102.024062}{\emph{Phys. Rev. D}
  {\bfseries 102} (2020) 024062}
  [\href{https://arxiv.org/abs/2003.11891}{{\ttfamily 2003.11891}}].

\bibitem{Bini:2020nsb}
D.~Bini, T.~Damour and A.~Geralico, \emph{{Sixth post-Newtonian local-in-time
  dynamics of binary systems}},
  \href{https://doi.org/10.1103/PhysRevD.102.024061}{\emph{Phys. Rev. D}
  {\bfseries 102} (2020) 024061}
  [\href{https://arxiv.org/abs/2004.05407}{{\ttfamily 2004.05407}}].

\bibitem{Bini:2019nra}
D.~Bini, T.~Damour and A.~Geralico, \emph{{Novel approach to binary dynamics:
  application to the fifth post-Newtonian level}},
  \href{https://doi.org/10.1103/PhysRevLett.123.231104}{\emph{Phys. Rev. Lett.}
  {\bfseries 123} (2019) 231104}
  [\href{https://arxiv.org/abs/1909.02375}{{\ttfamily 1909.02375}}].

\bibitem{Siemonsen:2019dsu}
N.~Siemonsen and J.~Vines, \emph{{Test black holes, scattering amplitudes and
  perturbations of Kerr spacetime}},
  \href{https://doi.org/10.1103/PhysRevD.101.064066}{\emph{Phys. Rev. D}
  {\bfseries 101} (2020) 064066}
  [\href{https://arxiv.org/abs/1909.07361}{{\ttfamily 1909.07361}}].

\bibitem{Antonelli:2020aeb}
A.~Antonelli, C.~Kavanagh, M.~Khalil, J.~Steinhoff and J.~Vines,
  \emph{{Gravitational spin-orbit coupling through third-subleading
  post-Newtonian order: from first-order self-force to arbitrary mass ratios}},
  \href{https://doi.org/10.1103/PhysRevLett.125.011103}{\emph{Phys. Rev. Lett.}
  {\bfseries 125} (2020) 011103}
  [\href{https://arxiv.org/abs/2003.11391}{{\ttfamily 2003.11391}}].

\bibitem{Bern:2021dqo}
Z.~Bern, J.~Parra-Martinez, R.~Roiban, M.~S. Ruf, C.-H. Shen, M.~P. Solon
  et~al., \emph{{Scattering Amplitudes and Conservative Binary Dynamics at
  ${\cal O}(G^4)$}},
  \href{https://doi.org/10.1103/PhysRevLett.126.171601}{\emph{Phys. Rev. Lett.}
  {\bfseries 126} (2021) 171601}
  [\href{https://arxiv.org/abs/2101.07254}{{\ttfamily 2101.07254}}].

\bibitem{Bini:2021gat}
D.~Bini, T.~Damour and A.~Geralico, \emph{{Radiative contributions to
  gravitational scattering}},
  \href{https://doi.org/10.1103/PhysRevD.104.084031}{\emph{Phys. Rev. D}
  {\bfseries 104} (2021) 084031}
  [\href{https://arxiv.org/abs/2107.08896}{{\ttfamily 2107.08896}}].

\bibitem{Blumlein:2021txe}
J.~Bl\"umlein, A.~Maier, P.~Marquard and G.~Sch\"afer, \emph{{The fifth-order
  post-Newtonian Hamiltonian dynamics of two-body systems from an effective
  field theory approach}},  \href{https://arxiv.org/abs/2110.13822}{{\ttfamily
  2110.13822}}.

\bibitem{Bern:2021yeh}
Z.~Bern, J.~Parra-Martinez, R.~Roiban, M.~S. Ruf, C.-H. Shen, M.~P. Solon
  et~al., \emph{{Scattering Amplitudes, the Tail Effect, and Conservative
  Binary Dynamics at $O(G^4)$}},
  \href{https://arxiv.org/abs/2112.10750}{{\ttfamily 2112.10750}}.

\bibitem{Neill:2013wsa}
D.~Neill and I.~Z. Rothstein, \emph{{Classical Space-Times from the S Matrix}},
  \href{https://doi.org/10.1016/j.nuclphysb.2013.09.007}{\emph{Nucl. Phys. B}
  {\bfseries 877} (2013) 177}
  [\href{https://arxiv.org/abs/1304.7263}{{\ttfamily 1304.7263}}].

\bibitem{Cristofoli:2021vyo}
A.~Cristofoli, R.~Gonzo, D.~A. Kosower and D.~O'Connell, \emph{{Waveforms from
  Amplitudes}},  \href{https://arxiv.org/abs/2107.10193}{{\ttfamily
  2107.10193}}.

\bibitem{Bern:1994zx}
Z.~Bern, L.~J. Dixon, D.~C. Dunbar and D.~A. Kosower, \emph{{One loop n point
  gauge theory amplitudes, unitarity and collinear limits}},
  \href{https://doi.org/10.1016/0550-3213(94)90179-1}{\emph{Nucl. Phys. B}
  {\bfseries 425} (1994) 217}
  [\href{https://arxiv.org/abs/hep-ph/9403226}{{\ttfamily hep-ph/9403226}}].

\bibitem{Bern:1994cg}
Z.~Bern, L.~J. Dixon, D.~C. Dunbar and D.~A. Kosower, \emph{{Fusing gauge
  theory tree amplitudes into loop amplitudes}},
  \href{https://doi.org/10.1016/0550-3213(94)00488-Z}{\emph{Nucl. Phys. B}
  {\bfseries 435} (1995) 59}
  [\href{https://arxiv.org/abs/hep-ph/9409265}{{\ttfamily hep-ph/9409265}}].

\bibitem{Britto:2004nc}
R.~Britto, F.~Cachazo and B.~Feng, \emph{{Generalized unitarity and one-loop
  amplitudes in N=4 super-Yang-Mills}},
  \href{https://doi.org/10.1016/j.nuclphysb.2005.07.014}{\emph{Nucl. Phys. B}
  {\bfseries 725} (2005) 275}
  [\href{https://arxiv.org/abs/hep-th/0412103}{{\ttfamily hep-th/0412103}}].

\bibitem{KLT}
H.~Kawai, D.~C. Lewellen and S.~H.~H. Tye, \emph{{A Relation Between Tree
  Amplitudes of Closed and Open Strings}},
  \href{https://doi.org/10.1016/0550-3213(86)90362-7}{\emph{Nucl. Phys. B}
  {\bfseries 269} (1986) 1}.

\bibitem{Bern:2008qj}
Z.~Bern, J.~Carrasco and H.~Johansson, \emph{{New Relations for Gauge-Theory
  Amplitudes}}, \href{https://doi.org/10.1103/PhysRevD.78.085011}{\emph{Phys.
  Rev. D} {\bfseries 78} (2008) 085011}
  [\href{https://arxiv.org/abs/0805.3993}{{\ttfamily 0805.3993}}].

\bibitem{Bern:2010ue}
Z.~Bern, J.~J.~M. Carrasco and H.~Johansson, \emph{{Perturbative Quantum
  Gravity as a Double Copy of Gauge Theory}},
  \href{https://doi.org/10.1103/PhysRevLett.105.061602}{\emph{Phys. Rev. Lett.}
  {\bfseries 105} (2010) 061602}
  [\href{https://arxiv.org/abs/1004.0476}{{\ttfamily 1004.0476}}].

\bibitem{Bern:2012uf}
Z.~Bern, J.~Carrasco, L.~Dixon, H.~Johansson and R.~Roiban, \emph{{Simplifying
  Multiloop Integrands and Ultraviolet Divergences of Gauge Theory and Gravity
  Amplitudes}}, \href{https://doi.org/10.1103/PhysRevD.85.105014}{\emph{Phys.
  Rev. D} {\bfseries 85} (2012) 105014}
  [\href{https://arxiv.org/abs/1201.5366}{{\ttfamily 1201.5366}}].

\bibitem{Bern:2019prr}
Z.~Bern, J.~J. Carrasco, M.~Chiodaroli, H.~Johansson and R.~Roiban, \emph{{The
  Duality Between Color and Kinematics and its Applications}},
  \href{https://arxiv.org/abs/1909.01358}{{\ttfamily 1909.01358}}.

\bibitem{Parra-Martinez:2020dzs}
J.~Parra-Martinez, M.~S. Ruf and M.~Zeng, \emph{{Extremal black hole scattering
  at $\mathcal{O}(G^3)$: graviton dominance, eikonal exponentiation, and
  differential equations}},
  \href{https://doi.org/10.1007/JHEP11(2020)023}{\emph{JHEP} {\bfseries 11}
  (2020) 023} [\href{https://arxiv.org/abs/2005.04236}{{\ttfamily
  2005.04236}}].

\bibitem{Tkachov:1981wb}
F.~V. Tkachov, \emph{{A Theorem on Analytical Calculability of Four Loop
  Renormalization Group Functions}},
  \href{https://doi.org/10.1016/0370-2693(81)90288-4}{\emph{Phys. Lett. B}
  {\bfseries 100} (1981) 65}.

\bibitem{Chetyrkin:1981qh}
K.~Chetyrkin and F.~Tkachov, \emph{{Integration by Parts: The Algorithm to
  Calculate beta Functions in 4 Loops}},
  \href{https://doi.org/10.1016/0550-3213(81)90199-1}{\emph{Nucl. Phys. B}
  {\bfseries 192} (1981) 159}.

\bibitem{Laporta:2001dd}
S.~Laporta, \emph{{High precision calculation of multiloop Feynman integrals by
  difference equations}},
  \href{https://doi.org/10.1016/S0217-751X(00)00215-7}{\emph{Int. J. Mod. Phys.
  A} {\bfseries 15} (2000) 5087}
  [\href{https://arxiv.org/abs/hep-ph/0102033}{{\ttfamily hep-ph/0102033}}].

\bibitem{Kotikov:1990kg}
A.~Kotikov, \emph{{Differential equations method: New technique for massive
  Feynman diagrams calculation}},
  \href{https://doi.org/10.1016/0370-2693(91)90413-K}{\emph{Phys. Lett. B}
  {\bfseries 254} (1991) 158}.

\bibitem{Bern:1992em}
Z.~Bern, L.~J. Dixon and D.~A. Kosower, \emph{{Dimensionally regulated one loop
  integrals}}, \href{https://doi.org/10.1016/0370-2693(93)90400-C}{\emph{Phys.
  Lett. B} {\bfseries 302} (1993) 299}
  [\href{https://arxiv.org/abs/hep-ph/9212308}{{\ttfamily hep-ph/9212308}}].

\bibitem{Remiddi:1997ny}
E.~Remiddi, \emph{{Differential equations for Feynman graph amplitudes}},
  {\emph{Nuovo Cim. A} {\bfseries 110} (1997) 1435}
  [\href{https://arxiv.org/abs/hep-th/9711188}{{\ttfamily hep-th/9711188}}].

\bibitem{Gehrmann:1999as}
T.~Gehrmann and E.~Remiddi, \emph{{Differential equations for two loop four
  point functions}},
  \href{https://doi.org/10.1016/S0550-3213(00)00223-6}{\emph{Nucl. Phys. B}
  {\bfseries 580} (2000) 485}
  [\href{https://arxiv.org/abs/hep-ph/9912329}{{\ttfamily hep-ph/9912329}}].

\bibitem{Henn:2013pwa}
J.~M. Henn, \emph{{Multiloop integrals in dimensional regularization made
  simple}}, \href{https://doi.org/10.1103/PhysRevLett.110.251601}{\emph{Phys.
  Rev. Lett.} {\bfseries 110} (2013) 251601}
  [\href{https://arxiv.org/abs/1304.1806}{{\ttfamily 1304.1806}}].

\bibitem{Henn:2014qga}
J.~M. Henn, \emph{{Lectures on differential equations for Feynman integrals}},
  \href{https://doi.org/10.1088/1751-8113/48/15/153001}{\emph{J. Phys. A}
  {\bfseries 48} (2015) 153001}
  [\href{https://arxiv.org/abs/1412.2296}{{\ttfamily 1412.2296}}].

\bibitem{Anastasiou:2002yz}
C.~Anastasiou and K.~Melnikov, \emph{{Higgs boson production at hadron
  colliders in NNLO QCD}},
  \href{https://doi.org/10.1016/S0550-3213(02)00837-4}{\emph{Nucl. Phys. B}
  {\bfseries 646} (2002) 220}
  [\href{https://arxiv.org/abs/hep-ph/0207004}{{\ttfamily hep-ph/0207004}}].

\bibitem{Anastasiou:2002qz}
C.~Anastasiou, L.~J. Dixon and K.~Melnikov, \emph{{NLO Higgs boson rapidity
  distributions at hadron colliders}},
  \href{https://doi.org/10.1016/S0920-5632(03)80168-8}{\emph{Nucl. Phys. B
  Proc. Suppl.} {\bfseries 116} (2003) 193}
  [\href{https://arxiv.org/abs/hep-ph/0211141}{{\ttfamily hep-ph/0211141}}].

\bibitem{Anastasiou:2003yy}
C.~Anastasiou, L.~J. Dixon, K.~Melnikov and F.~Petriello, \emph{{Dilepton
  rapidity distribution in the Drell-Yan process at NNLO in QCD}},
  \href{https://doi.org/10.1103/PhysRevLett.91.182002}{\emph{Phys. Rev. Lett.}
  {\bfseries 91} (2003) 182002}
  [\href{https://arxiv.org/abs/hep-ph/0306192}{{\ttfamily hep-ph/0306192}}].

\bibitem{Anastasiou:2015yha}
C.~Anastasiou, C.~Duhr, F.~Dulat, E.~Furlan, F.~Herzog and B.~Mistlberger,
  \emph{{Soft expansion of double-real-virtual corrections to Higgs production
  at N$^{3}$LO}}, \href{https://doi.org/10.1007/JHEP08(2015)051}{\emph{JHEP}
  {\bfseries 08} (2015) 051}
  [\href{https://arxiv.org/abs/1505.04110}{{\ttfamily 1505.04110}}].

\bibitem{Bonnor:1959}
W.~Bonnor, \emph{{Spherical gravitational waves}}, {\emph{Philos. Trans. R.
  Soc. London} {\bfseries A 251} (1959) 233}.

\bibitem{BonnorRotenberg:1966}
W.~Bonnor and M.~Rotenberg, \emph{{Gravitational waves from isolated sources}},
  {\emph{Proc. R. Soc. London} {\bfseries Ser A 289} (1966) 247}.

\bibitem{Thorne:1980ru}
K.~S. Thorne, \emph{{Multipole Expansions of Gravitational Radiation}},
  \href{https://doi.org/10.1103/RevModPhys.52.299}{\emph{Rev. Mod. Phys.}
  {\bfseries 52} (1980) 299}.

\bibitem{Blanchet:1987wq}
L.~Blanchet and T.~Damour, \emph{{Tail Transported Temporal Correlations in the
  Dynamics of a Gravitating System}},
  \href{https://doi.org/10.1103/PhysRevD.37.1410}{\emph{Phys. Rev. D}
  {\bfseries 37} (1988) 1410}.

\bibitem{Blanchet:1992br}
L.~Blanchet and T.~Damour, \emph{{Hereditary effects in gravitational
  radiation}}, \href{https://doi.org/10.1103/PhysRevD.46.4304}{\emph{Phys. Rev.
  D} {\bfseries 46} (1992) 4304}.

\bibitem{Blanchet:1993ec}
L.~Blanchet and G.~Schaefer, \emph{{Gravitational wave tails and binary star
  systems}}, \href{https://doi.org/10.1088/0264-9381/10/12/026}{\emph{Class.
  Quant. Grav.} {\bfseries 10} (1993) 2699}.

\bibitem{Kalin:2019rwq}
G.~K\"alin and R.~A. Porto, \emph{{From Boundary Data to Bound States}},
  \href{https://doi.org/10.1007/JHEP01(2020)072}{\emph{JHEP} {\bfseries 01}
  (2020) 072} [\href{https://arxiv.org/abs/1910.03008}{{\ttfamily
  1910.03008}}].

\bibitem{Kalin:2019inp}
G.~K\"alin and R.~A. Porto, \emph{{From boundary data to bound states. Part II.
  Scattering angle to dynamical invariants (with twist)}},
  \href{https://doi.org/10.1007/JHEP02(2020)120}{\emph{JHEP} {\bfseries 02}
  (2020) 120} [\href{https://arxiv.org/abs/1911.09130}{{\ttfamily
  1911.09130}}].

\bibitem{Bini:2020hmy}
D.~Bini, T.~Damour and A.~Geralico, \emph{{Sixth post-Newtonian
  nonlocal-in-time dynamics of binary systems}},
  \href{https://doi.org/10.1103/PhysRevD.102.084047}{\emph{Phys. Rev. D}
  {\bfseries 102} (2020) 084047}
  [\href{https://arxiv.org/abs/2007.11239}{{\ttfamily 2007.11239}}].

\bibitem{Saketh:2021sri}
M.~V.~S. Saketh, J.~Vines, J.~Steinhoff and A.~Buonanno, \emph{{Conservative
  and radiative dynamics in classical relativistic scattering and bound
  systems}},  \href{https://arxiv.org/abs/2109.05994}{{\ttfamily 2109.05994}}.

\bibitem{Bini:2017wfr}
D.~Bini and T.~Damour, \emph{{Gravitational scattering of two black holes at
  the fourth post-Newtonian approximation}},
  \href{https://doi.org/10.1103/PhysRevD.96.064021}{\emph{Phys. Rev. D}
  {\bfseries 96} (2017) 064021}
  [\href{https://arxiv.org/abs/1706.06877}{{\ttfamily 1706.06877}}].

\bibitem{Cho:2021arx}
G.~Cho, G.~K\"alin and R.~A. Porto, \emph{{From Boundary Data to Bound States
  III: Radiative Effects}},  \href{https://arxiv.org/abs/2112.03976}{{\ttfamily
  2112.03976}}.

\bibitem{Damgaard:2021ipf}
P.~H. Damgaard, L.~Plante and P.~Vanhove, \emph{{On an exponential
  representation of the gravitational S-matrix}},
  \href{https://doi.org/10.1007/JHEP11(2021)213}{\emph{JHEP} {\bfseries 11}
  (2021) 213} [\href{https://arxiv.org/abs/2107.12891}{{\ttfamily
  2107.12891}}].

\bibitem{Kol:2021jjc}
U.~Kol, D.~O'Connell and O.~Telem, \emph{{The Radial Action from Probe
  Amplitudes to All Orders}},
  \href{https://arxiv.org/abs/2109.12092}{{\ttfamily 2109.12092}}.

\bibitem{Glauber:1956Lecture}
R.~J. Glauber, \emph{{Lectures in theoretical physics}}, {\emph{Interscience
  Publishers Inc., New York} {\bfseries I} (1959) 315}.

\bibitem{Amati:1990xe}
D.~Amati, M.~Ciafaloni and G.~Veneziano, \emph{{Higher Order Gravitational
  Deflection and Soft Bremsstrahlung in Planckian Energy Superstring
  Collisions}}, \href{https://doi.org/10.1016/0550-3213(90)90375-N}{\emph{Nucl.
  Phys. B} {\bfseries 347} (1990) 550}.

\bibitem{Laenen:2008gt}
E.~Laenen, G.~Stavenga and C.~D. White, \emph{{Path integral approach to
  eikonal and next-to-eikonal exponentiation}},
  \href{https://doi.org/10.1088/1126-6708/2009/03/054}{\emph{JHEP} {\bfseries
  03} (2009) 054} [\href{https://arxiv.org/abs/0811.2067}{{\ttfamily
  0811.2067}}].

\bibitem{Akhoury:2013yua}
R.~Akhoury, R.~Saotome and G.~Sterman, \emph{{High Energy Scattering in
  Perturbative Quantum Gravity at Next to Leading Power}},
  \href{https://doi.org/10.1103/PhysRevD.103.064036}{\emph{Phys. Rev. D}
  {\bfseries 103} (2021) 064036}
  [\href{https://arxiv.org/abs/1308.5204}{{\ttfamily 1308.5204}}].

\bibitem{Bern:2020gjj}
Z.~Bern, H.~Ita, J.~Parra-Martinez and M.~S. Ruf, \emph{{Universality in the
  classical limit of massless gravitational scattering}},
  \href{https://doi.org/10.1103/PhysRevLett.125.031601}{\emph{Phys. Rev. Lett.}
  {\bfseries 125} (2020) 031601}
  [\href{https://arxiv.org/abs/2002.02459}{{\ttfamily 2002.02459}}].

\bibitem{delaCruz:2021gjp}
L.~de~la Cruz, A.~Luna and T.~Scheopner, \emph{{Yang-Mills observables: from
  KMOC to eikonal through EFT}},
  \href{https://arxiv.org/abs/2108.02178}{{\ttfamily 2108.02178}}.

\bibitem{Vaidya:2014kza}
V.~Vaidya, \emph{{Gravitational spin Hamiltonians from the S matrix}},
  \href{https://doi.org/10.1103/PhysRevD.91.024017}{\emph{Phys. Rev. D}
  {\bfseries 91} (2015) 024017}
  [\href{https://arxiv.org/abs/1410.5348}{{\ttfamily 1410.5348}}].

\bibitem{Vines:2017hyw}
J.~Vines, \emph{{Scattering of two spinning black holes in post-Minkowskian
  gravity, to all orders in spin, and effective-one-body mappings}},
  \href{https://doi.org/10.1088/1361-6382/aaa3a8}{\emph{Class. Quant. Grav.}
  {\bfseries 35} (2018) 084002}
  [\href{https://arxiv.org/abs/1709.06016}{{\ttfamily 1709.06016}}].

\bibitem{Guevara:2017csg}
A.~Guevara, \emph{{Holomorphic Classical Limit for Spin Effects in
  Gravitational and Electromagnetic Scattering}},
  \href{https://doi.org/10.1007/JHEP04(2019)033}{\emph{JHEP} {\bfseries 04}
  (2019) 033} [\href{https://arxiv.org/abs/1706.02314}{{\ttfamily
  1706.02314}}].

\bibitem{Vines:2018gqi}
J.~Vines, J.~Steinhoff and A.~Buonanno, \emph{{Spinning-black-hole scattering
  and the test-black-hole limit at second post-Minkowskian order}},
  \href{https://doi.org/10.1103/PhysRevD.99.064054}{\emph{Phys. Rev. D}
  {\bfseries 99} (2019) 064054}
  [\href{https://arxiv.org/abs/1812.00956}{{\ttfamily 1812.00956}}].

\bibitem{Guevara:2018wpp}
A.~Guevara, A.~Ochirov and J.~Vines, \emph{{Scattering of Spinning Black Holes
  from Exponentiated Soft Factors}},
  \href{https://doi.org/10.1007/JHEP09(2019)056}{\emph{JHEP} {\bfseries 09}
  (2019) 056} [\href{https://arxiv.org/abs/1812.06895}{{\ttfamily
  1812.06895}}].

\bibitem{Chung:2018kqs}
M.-Z. Chung, Y.-T. Huang, J.-W. Kim and S.~Lee, \emph{{The simplest massive
  S-matrix: from minimal coupling to Black Holes}},
  \href{https://doi.org/10.1007/JHEP04(2019)156}{\emph{JHEP} {\bfseries 04}
  (2019) 156} [\href{https://arxiv.org/abs/1812.08752}{{\ttfamily
  1812.08752}}].

\bibitem{Guevara:2019fsj}
A.~Guevara, A.~Ochirov and J.~Vines, \emph{{Black-hole scattering with general
  spin directions from minimal-coupling amplitudes}},
  \href{https://doi.org/10.1103/PhysRevD.100.104024}{\emph{Phys. Rev. D}
  {\bfseries 100} (2019) 104024}
  [\href{https://arxiv.org/abs/1906.10071}{{\ttfamily 1906.10071}}].

\bibitem{Chung:2019duq}
M.-Z. Chung, Y.-T. Huang and J.-W. Kim, \emph{{Classical potential for general
  spinning bodies}}, \href{https://doi.org/10.1007/JHEP09(2020)074}{\emph{JHEP}
  {\bfseries 09} (2020) 074}
  [\href{https://arxiv.org/abs/1908.08463}{{\ttfamily 1908.08463}}].

\bibitem{Damgaard:2019lfh}
P.~H. Damgaard, K.~Haddad and A.~Helset, \emph{{Heavy Black Hole Effective
  Theory}}, \href{https://doi.org/10.1007/JHEP11(2019)070}{\emph{JHEP}
  {\bfseries 11} (2019) 070}
  [\href{https://arxiv.org/abs/1908.10308}{{\ttfamily 1908.10308}}].

\bibitem{Aoude:2020onz}
R.~Aoude, K.~Haddad and A.~Helset, \emph{{On-shell heavy particle effective
  theories}}, \href{https://doi.org/10.1007/JHEP05(2020)051}{\emph{JHEP}
  {\bfseries 05} (2020) 051}
  [\href{https://arxiv.org/abs/2001.09164}{{\ttfamily 2001.09164}}].

\bibitem{Bern:2020buy}
Z.~Bern, A.~Luna, R.~Roiban, C.-H. Shen and M.~Zeng, \emph{{Spinning black hole
  binary dynamics, scattering amplitudes, and effective field theory}},
  \href{https://doi.org/10.1103/PhysRevD.104.065014}{\emph{Phys. Rev. D}
  {\bfseries 104} (2021) 065014}
  [\href{https://arxiv.org/abs/2005.03071}{{\ttfamily 2005.03071}}].

\bibitem{Kosmopoulos:2021zoq}
D.~Kosmopoulos and A.~Luna, \emph{{Quadratic-in-spin Hamiltonian at $
  \mathcal{O} $(G$^{2}$) from scattering amplitudes}},
  \href{https://doi.org/10.1007/JHEP07(2021)037}{\emph{JHEP} {\bfseries 07}
  (2021) 037} [\href{https://arxiv.org/abs/2102.10137}{{\ttfamily
  2102.10137}}].

\bibitem{Guevara:2020xjx}
A.~Guevara, B.~Maybee, A.~Ochirov, D.~O'connell and J.~Vines, \emph{{A
  worldsheet for Kerr}},
  \href{https://doi.org/10.1007/JHEP03(2021)201}{\emph{JHEP} {\bfseries 03}
  (2021) 201} [\href{https://arxiv.org/abs/2012.11570}{{\ttfamily
  2012.11570}}].

\bibitem{Levi:2020uwu}
M.~Levi, A.~J. Mcleod and M.~Von~Hippel, \emph{{N$^{3}$LO gravitational
  quadratic-in-spin interactions at G$^{4}$}},
  \href{https://doi.org/10.1007/JHEP07(2021)116}{\emph{JHEP} {\bfseries 07}
  (2021) 116} [\href{https://arxiv.org/abs/2003.07890}{{\ttfamily
  2003.07890}}].

\bibitem{Levi:2020kvb}
M.~Levi, A.~J. Mcleod and M.~Von~Hippel, \emph{{N$^3$LO gravitational
  spin-orbit coupling at order $G^4$}},
  \href{https://doi.org/10.1007/JHEP07(2021)115}{\emph{JHEP} {\bfseries 07}
  (2021) 115} [\href{https://arxiv.org/abs/2003.02827}{{\ttfamily
  2003.02827}}].

\bibitem{Chen:2021qkk}
W.-M. Chen, M.-Z. Chung, Y.-t. Huang and J.-W. Kim, \emph{{The 2PM Hamiltonian
  for binary Kerr to quartic in spin}},
  \href{https://arxiv.org/abs/2111.13639}{{\ttfamily 2111.13639}}.

\bibitem{Brandhuber:2019qpg}
A.~Brandhuber and G.~Travaglini, \emph{{On higher-derivative effects on the
  gravitational potential and particle bending}},
  \href{https://doi.org/10.1007/JHEP01(2020)010}{\emph{JHEP} {\bfseries 01}
  (2020) 010} [\href{https://arxiv.org/abs/1905.05657}{{\ttfamily
  1905.05657}}].

\bibitem{Huber:2019ugz}
M.~Accettulli~Huber, A.~Brandhuber, S.~De~Angelis and G.~Travaglini,
  \emph{{Note on the absence of $R^2$ corrections to Newton\textquoteright{}s
  potential}}, \href{https://doi.org/10.1103/PhysRevD.101.046011}{\emph{Phys.
  Rev. D} {\bfseries 101} (2020) 046011}
  [\href{https://arxiv.org/abs/1911.10108}{{\ttfamily 1911.10108}}].

\bibitem{Cheung:2020sdj}
C.~Cheung and M.~P. Solon, \emph{{Tidal Effects in the Post-Minkowskian
  Expansion}},
  \href{https://doi.org/10.1103/PhysRevLett.125.191601}{\emph{Phys. Rev. Lett.}
  {\bfseries 125} (2020) 191601}
  [\href{https://arxiv.org/abs/2006.06665}{{\ttfamily 2006.06665}}].

\bibitem{Kalin:2020lmz}
G.~K\"alin, Z.~Liu and R.~A. Porto, \emph{{Conservative Tidal Effects in
  Compact Binary Systems to Next-to-Leading Post-Minkowskian Order}},
  \href{https://doi.org/10.1103/PhysRevD.102.124025}{\emph{Phys. Rev. D}
  {\bfseries 102} (2020) 124025}
  [\href{https://arxiv.org/abs/2008.06047}{{\ttfamily 2008.06047}}].

\bibitem{Haddad:2020que}
K.~Haddad and A.~Helset, \emph{{Tidal effects in quantum field theory}},
  \href{https://doi.org/10.1007/JHEP12(2020)024}{\emph{JHEP} {\bfseries 12}
  (2020) 024} [\href{https://arxiv.org/abs/2008.04920}{{\ttfamily
  2008.04920}}].

\bibitem{Aoude:2020ygw}
R.~Aoude, K.~Haddad and A.~Helset, \emph{{Tidal effects for spinning
  particles}}, \href{https://doi.org/10.1007/JHEP03(2021)097}{\emph{JHEP}
  {\bfseries 03} (2021) 097}
  [\href{https://arxiv.org/abs/2012.05256}{{\ttfamily 2012.05256}}].

\bibitem{AccettulliHuber:2020oou}
M.~Accettulli~Huber, A.~Brandhuber, S.~De~Angelis and G.~Travaglini,
  \emph{{Eikonal phase matrix, deflection angle and time delay in effective
  field theories of gravity}},
  \href{https://doi.org/10.1103/PhysRevD.102.046014}{\emph{Phys. Rev. D}
  {\bfseries 102} (2020) 046014}
  [\href{https://arxiv.org/abs/2006.02375}{{\ttfamily 2006.02375}}].

\bibitem{Huber:2020xny}
M.~Accettulli~Huber, A.~Brandhuber, S.~De~Angelis and G.~Travaglini,
  \emph{{From amplitudes to gravitational radiation with cubic interactions and
  tidal effects}},  \href{https://arxiv.org/abs/2012.06548}{{\ttfamily
  2012.06548}}.

\bibitem{Cheung:2020gbf}
C.~Cheung, N.~Shah and M.~P. Solon, \emph{{Mining the Geodesic Equation for
  Scattering Data}},
  \href{https://doi.org/10.1103/PhysRevD.103.024030}{\emph{Phys. Rev. D}
  {\bfseries 103} (2021) 024030}
  [\href{https://arxiv.org/abs/2010.08568}{{\ttfamily 2010.08568}}].

\bibitem{Bern:2020uwk}
Z.~Bern, J.~Parra-Martinez, R.~Roiban, E.~Sawyer and C.-H. Shen, \emph{{Leading
  Nonlinear Tidal Effects and Scattering Amplitudes}},
  \href{https://doi.org/10.1007/JHEP05(2021)188}{\emph{JHEP} {\bfseries 05}
  (2021) 188} [\href{https://arxiv.org/abs/2010.08559}{{\ttfamily
  2010.08559}}].

\bibitem{DiVecchia:2020ymx}
P.~Di~Vecchia, C.~Heissenberg, R.~Russo and G.~Veneziano, \emph{{Universality
  of ultra-relativistic gravitational scattering}},
  \href{https://doi.org/10.1016/j.physletb.2020.135924}{\emph{Phys. Lett. B}
  {\bfseries 811} (2020) 135924}
  [\href{https://arxiv.org/abs/2008.12743}{{\ttfamily 2008.12743}}].

\bibitem{Damour:2020tta}
T.~Damour, \emph{{Radiative contribution to classical gravitational scattering
  at the third order in $G$}},
  \href{https://doi.org/10.1103/PhysRevD.102.124008}{\emph{Phys. Rev. D}
  {\bfseries 102} (2020) 124008}
  [\href{https://arxiv.org/abs/2010.01641}{{\ttfamily 2010.01641}}].

\bibitem{Buonanno:2000qq}
A.~Buonanno, \emph{{Reduction of the two-body dynamics to a one-body
  description in classical electrodynamics}},
  \href{https://doi.org/10.1103/PhysRevD.62.104022}{\emph{Phys. Rev. D}
  {\bfseries 62} (2000) 104022}
  [\href{https://arxiv.org/abs/hep-th/0004042}{{\ttfamily hep-th/0004042}}].

\bibitem{lorentz:1892theorie}
H.~A. Lorentz, \emph{La th{\'e}orie {\'e}lectromagn{\'e}tique de Maxwell et son
  application aux corps mouvants}, vol.~25. EJ Brill, 1892.

\bibitem{abraham:1912theorie}
M.~Abraham, \emph{Theorie der Elektrizit{\"a}t}. BG Teubner, 1912.

\bibitem{Dirac:1938a}
P.~A.~M. Dirac, \emph{Classical theory of radiating electrons},
  \href{https://doi.org/10.1098/rspa.1938.0124}{\emph{Proceedings of the Royal
  Society of London. Series A. Mathematical and Physical Sciences} {\bfseries
  167} (1938) 148}
  [\href{https://arxiv.org/abs/https://royalsocietypublishing.org/doi/pdf/10.1098/rspa.1938.0124}{{\ttfamily
  https://royalsocietypublishing.org/doi/pdf/10.1098/rspa.1938.0124}}].

\bibitem{DiVecchia:2021bdo}
P.~Di~Vecchia, C.~Heissenberg, R.~Russo and G.~Veneziano, \emph{{The eikonal
  approach to gravitational scattering and radiation at $ \mathcal{O}
  $(G$^{3}$)}}, \href{https://doi.org/10.1007/JHEP07(2021)169}{\emph{JHEP}
  {\bfseries 07} (2021) 169}
  [\href{https://arxiv.org/abs/2104.03256}{{\ttfamily 2104.03256}}].

\bibitem{Beneke:1997zp}
M.~Beneke and V.~A. Smirnov, \emph{{Asymptotic expansion of Feynman integrals
  near threshold}},
  \href{https://doi.org/10.1016/S0550-3213(98)00138-2}{\emph{Nucl. Phys. B}
  {\bfseries 522} (1998) 321}
  [\href{https://arxiv.org/abs/hep-ph/9711391}{{\ttfamily hep-ph/9711391}}].

\bibitem{Herrmann:2021tct}
E.~Herrmann, J.~Parra-Martinez, M.~S. Ruf and M.~Zeng, \emph{{Radiative
  classical gravitational observables at $ \mathcal{O} $(G$^{3}$) from
  scattering amplitudes}},
  \href{https://doi.org/10.1007/JHEP10(2021)148}{\emph{JHEP} {\bfseries 10}
  (2021) 148} [\href{https://arxiv.org/abs/2104.03957}{{\ttfamily
  2104.03957}}].

\bibitem{Kosmopoulos:2020pcd}
D.~Kosmopoulos, \emph{{Simplifying $D$-Dimensional Physical-State Sums in Gauge
  Theory and Gravity}},  \href{https://arxiv.org/abs/2009.00141}{{\ttfamily
  2009.00141}}.

\bibitem{Bourjaily:2017wjl}
J.~L. Bourjaily, E.~Herrmann and J.~Trnka, \emph{{Prescriptive Unitarity}},
  \href{https://doi.org/10.1007/JHEP06(2017)059}{\emph{JHEP} {\bfseries 06}
  (2017) 059} [\href{https://arxiv.org/abs/1704.05460}{{\ttfamily
  1704.05460}}].

\bibitem{Herrmann:2021lqe}
E.~Herrmann, J.~Parra-Martinez, M.~S. Ruf and M.~Zeng, \emph{{Gravitational
  Bremsstrahlung from Reverse Unitarity}},
  \href{https://doi.org/10.1103/PhysRevLett.126.201602}{\emph{Phys. Rev. Lett.}
  {\bfseries 126} (2021) 201602}
  [\href{https://arxiv.org/abs/2101.07255}{{\ttfamily 2101.07255}}].

\bibitem{Landau:1976mech}
L.~D. Landau and E.~M. Lifshitz, \emph{Mechanics, Third Edition: Volume 1
  (Course of Theoretical Physics)}. Butterworth-Heinemann, 3~ed., Jan., 1976.

\bibitem{Bjerrum-Bohr:2021wwt}
N.~E.~J. Bjerrum-Bohr, L.~Plante and P.~Vanhove, \emph{{Post-Minkowskian Radial
  Action from Soft Limits and Velocity Cuts}},
  \href{https://arxiv.org/abs/2111.02976}{{\ttfamily 2111.02976}}.

\bibitem{Brandhuber:2021eyq}
A.~Brandhuber, G.~Chen, G.~Travaglini and C.~Wen, \emph{{Classical
  gravitational scattering from a gauge-invariant double copy}},
  \href{https://doi.org/10.1007/JHEP10(2021)118}{\emph{JHEP} {\bfseries 10}
  (2021) 118} [\href{https://arxiv.org/abs/2108.04216}{{\ttfamily
  2108.04216}}].

\bibitem{Landau:1980Classical}
L.~D. Landau and E.~M. Lifshitz, \emph{The Classical Theory of Fields}.
  Butterworth-Heinemann, 4~ed., Jan., 1980.

\bibitem{Galley:2010es}
C.~R. Galley, A.~K. Leibovich and I.~Z. Rothstein, \emph{{Finite size
  corrections to the radiation reaction force in classical electrodynamics}},
  \href{https://doi.org/10.1103/PhysRevLett.105.094802}{\emph{Phys. Rev. Lett.}
  {\bfseries 105} (2010) 094802}
  [\href{https://arxiv.org/abs/1005.2617}{{\ttfamily 1005.2617}}].

\bibitem{Forgacs:2012qt}
P.~Forgacs, T.~Herpay and P.~Kovacs, \emph{{Comment on `Finite Size Corrections
  to the Radiation Reaction Force in Classical Electrodynamics'}},
  \href{https://doi.org/10.1103/PhysRevLett.109.029501}{\emph{Phys. Rev. Lett.}
  {\bfseries 109} (2012) 029501}
  [\href{https://arxiv.org/abs/1202.6289}{{\ttfamily 1202.6289}}].

\bibitem{Birnholtz:2013nta}
O.~Birnholtz, S.~Hadar and B.~Kol, \emph{{Theory of post-Newtonian radiation
  and reaction}}, \href{https://doi.org/10.1103/PhysRevD.88.104037}{\emph{Phys.
  Rev. D} {\bfseries 88} (2013) 104037}
  [\href{https://arxiv.org/abs/1305.6930}{{\ttfamily 1305.6930}}].

\bibitem{Bini:2012ji}
D.~Bini and T.~Damour, \emph{{Gravitational radiation reaction along general
  orbits in the effective one-body formalism}},
  \href{https://doi.org/10.1103/PhysRevD.86.124012}{\emph{Phys. Rev. D}
  {\bfseries 86} (2012) 124012}
  [\href{https://arxiv.org/abs/1210.2834}{{\ttfamily 1210.2834}}].

\end{thebibliography}\endgroup

\end{document}